\documentclass[11pt]{article}

\usepackage{graphicx,epsfig}
\usepackage{multicol}
\usepackage{amsmath,amssymb,cite,color,hyperref}
\newcommand{\be}{\begin{equation}}
\newcommand{\ee}{\end{equation}}
\newcommand{\bea}{\begin{eqnarray}}
\newcommand{\eea}{\end{eqnarray}}
\newcommand{\bi}{\begin{itemize}}
\newcommand{\ei}{\end{itemize}}
\newcommand{\ben}{\begin{enumerate}}
\newcommand{\een}{\end{enumerate}}

\newcommand{\lc}{\left[}
\newcommand{\rc}{\right]}
\newcommand{\lp}{\left(}
\newcommand{\rp}{\right)}

\def\frac#1#2{{{#1}\over {#2}}}
\def\gsim{\mathrel{\rlap{\lower4pt\hbox{\hskip1pt$\sim$}}
    \raise1pt\hbox{$>$}}}         %greater than or approx. symbol
\def\lsim{\mathrel{\rlap{\lower4pt\hbox{\hskip1pt$\sim$}}
    \raise1pt\hbox{$<$}}}         %less than or approx. symbol

\newcommand{\draft}[1]{}

\definecolor{grey}{rgb}{0.5,0.5,0.5}

\begin{document}
\begin{flushright}
CERN-PH-TH/2012-263\\
Edinburgh 2012/21\\
SMU-HEP-12-16\\
LCTS/2012-26\\
IFUM-1003-FT\\
\end{flushright}

\begin{center}
{\large\bf Parton distribution benchmarking with LHC data}
\vspace{0.6cm}

Richard~D.~Ball$^1$, Stefano~Carrazza$^{2,3}$, Luigi Del Debbio$^1$, Stefano Forte$^{2,3}$, Jun Gao$^4$, Nathan
Hartland$^1$, Joey Huston$^5$, 
Pavel Nadolsky$^4$,  Juan Rojo$^{6}$, Daniel Stump$^5$, Robert~S.~Thorne$^7$,  C.-P. Yuan$^5$

\vspace{1.cm}
{\it 
  ~$^1$  Tait Institute, University of Edinburgh,\\
JCMB, KB, Mayfield Rd, Edinburgh EH9 3JZ, Scotland\\
 ~$^2$ Dipartimento di Fisica, Universit\`a di Milano and\\
~$^3$ INFN, Sezione di Milano,\\ Via Celoria 16, I-20133 Milano,
Italy\\
~$^4$ Department of Physics, Southern Methodist University, 
Dallas, TX 75275, USA \\
~$^5$ Department of Physics \& Astronomy, Michigan State University, \\
 East Lansing, MI 48824, USA \\
~$^6$ PH Department, TH Unit, CERN, CH-1211 Geneva 23, Switzerland \\
~$^7$ Department of Physics and Astronomy, University College London, WC1E 6BT, UK \\}
\end{center}

\vspace{0.5cm}

\begin{center}
{\bf \large Abstract:}
\end{center}

We present a detailed comparison of the most recent sets of NNLO PDFs
from the ABM, CT, HERAPDF, MSTW and NNPDF collaborations. 
We compare parton distributions at low and high scales
and parton luminosities relevant for LHC phenomenology.
We study the PDF dependence of LHC benchmark inclusive cross sections
and differential distributions for electroweak boson and jet production
in the cases in which the experimental covariance matrix is available.
We quantify the agreement between data and theory by computing the $\chi^2$ for
each data set with all the various PDFs. PDF comparisons are performed
consistently for common values of the strong coupling.
We also present a benchmark comparison of jet production at the LHC, comparing
the results from various available codes and scale settings.
Finally, we discuss the implications of the
updated NNLO PDF sets for the combined PDF+$\alpha_s$
uncertainty in the gluon fusion Higgs production cross section. 

\clearpage

\tableofcontents

\clearpage

\section{Introduction}
\label{sec:intro}

Parton distribution functions (PDFs) are one of the dominant sources 
of systematic uncertainty 
in many of the LHC cross sections relevant for Standard Model precision physics,
Higgs boson characterization and new physics searches. 
The dependence of benchmark total cross sections on PDFs at 
the $7$ TeV LHC was discussed
in Refs.~\cite{Watt:2011kp,Watt:2012np}. The purpose of the present  
paper is on the one hand to update these benchmark comparisons by  
including the most recent PDF sets from the various 
 collaborations, and on the other hand to perform quantitative
comparisons with $7$ TeV data for differential distributions,
and with $8$ TeV data for inclusive cross sections.

There have been several new NNLO PDF releases since the previous 
benchmark studies~\cite{Watt:2011kp}. The ABM collaboration have released  
ABM11~\cite{Alekhin:2012ig},
which supersedes ABKM09~\cite{Alekhin:2009ni}. It uses the combined
HERA-I data,  $\overline{\mbox{MS}}$  running heavy quark
masses for DIS structure functions~\cite{Alekhin:2010sv}, 
and provides PDF sets for a range of values of $\alpha_s$ in a fixed
flavor number scheme with 
$N_f=5$. The CT collaboration have recently released a CT10 NNLO
PDF set~\cite{Nadolsky:2012ia,cteqnnlo}, based 
on the same global dataset as CT10 NLO~\cite{Lai:2010vv},
and using a NNLO implementation of the S-ACOT-$\chi$ variable
flavor number scheme
for heavy quark structure functions \cite{Guzzi:2011ew}. 
The HERAPDF collaboration have  
released the HERAPDF1.5 NNLO PDF set~\cite{Radescu:2010zz,CooperSarkar:2011aa}, 
which in addition to the combined HERA-I dataset uses 
the inclusive HERA-II data 
from H1~\cite{Aaron:2012qi} and 
ZEUS~\cite{ZEUS:2012bx}\footnote{Note however that the fit uses preliminary 
data which are not exactly the same as the final published data.}. 
The latest release from NNPDF is the NNPDF2.3~\cite{Ball:2012cx} set.
Like the previous NNPDF2.1 release this uses the 
FONLL VFNS at NNLO~\cite{Forte:2010ta}, and
now also includes relevant LHC data for which the experimental 
correlation matrix is available. This is currently the only set which 
include LHC data in the fit.

As in previous benchmarks, we also use the MSTW08 NNLO 
PDFs~\cite{Martin:2009iq}. 
Although no new public release has been provided, several
partial updates have been presented, discussing the
impact on the MSTW08 PDFs of the combined HERA-I data
and the Tevatron $W$ lepton asymmetry~\cite{Thorne:2010kj} and of the
LHC $W$ lepton asymmetry data~\cite{Watt:2012tq}, and additionally
the ATLAS $W,Z$ and inclusive jet data in \cite{Martin:2012xx}.
We do not include in this benchmark study the JR09 PDF 
set~\cite{JimenezDelgado:2009tv} because it is available only 
for a single value of $\alpha_s(M_Z)$.

PDF sets  will be compared consistently for a common value
of $\alpha_s$.  All the PDF sets included in this
benchmark comparison provide $\alpha_s(M_Z)$ variations
in a relatively wide range, as summarized in Table~\ref{tab:as}. 
Unless otherwise specified, in the rest of the paper we will
always quote $\alpha_s$ at a scale $Q=M_Z$.
We will show results for PDFs,
parton luminosities, physical cross sections and 
$\chi^2$ values for $\alpha_s(M_Z)=0.118$ as a baseline,
and whenever we want to study the effect of varying
$\alpha_s$ we will provide results for two values of  $\alpha_s(M_Z)$,
$\alpha_s=0.117$ and $0.119$. The motivation for this choice is that
these  values approximately bracket the current 2012 PDG best
fit value~\cite{Beringer:1900zz}, $\alpha_s(M_Z)=0.1184\pm 0.0007$. 
They also include the preferred or best-fit $\alpha_s$ values of CT,
MSTW and NNPDF at
NNLO~\cite{Nadolsky:2012ia,Lionetti:2011pw,Ball:2011us,Martin:2009bu}.  
When error sets are only provided at a single value of $\alpha_s$
we will determine uncertainties at other values of $\alpha_s$ by
computing percentage uncertainties  at the value of $\alpha_s$ at
which error sets are provided, and then applying the same percentage
uncertainty to the central value computed for other $\alpha_s$ values.
For the PDF plots of Sect~\ref{sec:pdfs} only (but not for
luminosities)  the uncertainty shown on
the plot for values of $\alpha_s$  for which error sets are not
available will be taken as the {\it absolute}
PDF uncertainty computed at the $\alpha_s$ value at which error sets
are provided: this is because relative uncertainties on PDFs become
meaningless in regions where the PDF is very close to zero.

%%%%%%%%%%%%%%%%%%%%

%%%%%%%%%%%%%%%%%%%%%%%%%%%%%%%
\begin{table}[h]
\centering
\small
\begin{tabular}{c||c|c|c|c|c}
\hline
PDF set  & Reference & $\alpha_s^{(0)}$ (NLO) & $\alpha_s$ range (NLO)
& $\alpha_s^{(0)}$ (NNLO) & $\alpha_s$ range (NNLO) \\
\hline
\hline
ABM11 $N_f=5$      & \cite{Alekhin:2012ig}  & 0.1181 & $\lc 0.110,0.130 \rc$ & 0.1134   & $\lc 0.104,0.120 \rc$    \\
CT10       & \cite{cteqnnlo} & 0.118  & $\lc 0.112,0.127 \rc$  & 0.118   & 
$\lc 0.112,0.127 \rc$  \\
HERAPDF1.5 & \cite{Radescu:2010zz,CooperSarkar:2011aa}  & 0.1176 & $\lc 0.114,0.122 \rc$  & 0.1176   & $\lc 0.114,0.122 \rc$   \\
MSTW08     &  \cite{Martin:2009iq} & 0.1202 & $\lc 0.110,0.130 \rc$  & 0.1171   & $\lc 0.107,0.127 \rc$   \\
NNPDF2.3   & \cite{Ball:2012cx} & all  & $\lc 0.114,0.124 \rc$  & all   &   $\lc 0.114,0.124 \rc$ \\
\hline
\end{tabular}
\caption{\small PDF sets used in this paper. We quote the value
$\alpha_s^{(0)}\lp M_Z\rp$ for which PDF uncertainties are provided, 
and the range in $\alpha_s(M_Z)$ in which PDF central
values are available (in steps of $0.001$). For ABM11 the $\alpha_s(M_Z)$ varying PDF
sets are only available for the $N_f=5$ PDF set.  \label{tab:as}}
\end{table}
%%%%%%%%%%%%%%%%%%%%%%%%%%%%

The structure of this paper is the following: in Sect.~\ref{sec:pdfs} 
we begin by comparing the various sets
of NNLO PDFs and the associated parton luminosities, 
and discuss the similarities and differences between each of the sets.
In Sect.~\ref{sec:LHCincl} we compute predictions for LHC inclusive
cross sections at 8 TeV, including Higgs cross sections. Finally in
Sect.~\ref{sec:LHCdist} we compare PDF predictions for all available
LHC data at 7 TeV with experimental covariance matrix, and quantify the
data theory agreement for each of the PDF sets. 
Then we turn
to discuss in more detail the case of the ATLAS inclusive jet data
in Sect.~\ref{sec:atlasjets}, where we compare different codes
and theory scale settings
for jet production.
Finally in Sect.~\ref{sec:higgs} we discuss the 
implications of this benchmarking for
the particular case of the Higgs cross section in
gluon fusion and examine
possible extensions of the current (2010) PDF4LHC recommendation.
Then we conclude and
discuss the prospects for future benchmarking studies in 
Sect.~\ref{sec:conclusions}. A more technical appendix summarizes
the issue of the dependence on the
$\chi^2$ definition.

All the above groups provide versions of the respective PDF
sets both at NLO and at NNLO. In this paper
we will show only the NNLO
PDFs, for the particular values of $\alpha_s$ mentioned
above. We 
have however produced the results presented here
also at NLO and for a wider range of $\alpha_s$ values.
The complete catalog of plots can be obtained online from {\tt HepForge}: 
\begin{center} \bf
\url{http://nnpdf.hepforge.org/html/pdfbench/catalog}.
\end{center}

\clearpage

\section{Parton distributions and parton luminosities}
\label{sec:pdfs}

In this section we 
compare PDFs and then parton luminosities between
the various groups.
For definiteness we show here comparisons only between 
PDFs and luminosities at NNLO for $\alpha_s=0.118$.
Results for several other values of $\alpha_s$ and at NLO
can be obtained from the catalog of plots on the {\tt HepForge}
website.

\subsection{Parton distributions}

We  compare  parton distributions at $Q^2=$ 25 GeV$^2$, above the $b$ quark
threshold since ABM11 only provide their $N_f=5$ PDFs for a range 
of values of $\alpha_s$.\footnote{The ABM11 PDFs are provided
as FFN sets with different numbers of active flavours: $N_f$=3,
4 and 5. For scales $Q^2$ below the charm threshold the  $N_f$=3
set must be used, between the charm and bottom threshold the  $N_f$=4
set should be used and above the bottom threshold it is the $N_f$=5
set to be used. Of all these various FFN sets, only those with $N_f$=5
are provided for a variety of $\alpha_s$ values. } 
For each  PDF we compare first NNPDF2.3, CT10 and MSTW08,
and then NNPDF2.3, ABM11 and HERAPDF1.5 (with NNPDF2.3 thus being 
used as a common reference).
We consider PDF uncertainties only and not 
the $\alpha_s$ uncertainty, except for the ABM11 PDFs, where the $\alpha_s$ 
uncertainty is treated on a equal footing to the PDF parameters in 
the covariance matrix. The ABM11 and HERAPDF results also include an 
uncertainty on quark masses, whereas other groups provide sets with 
a variety of masses. 

%%%%%%%%%%%%%%%%%%%%%%%%%%%%%%%%%%%%%%%%%%%%%%%%%
\begin{figure}[h]
    \begin{center}
\includegraphics[width=0.48\textwidth]{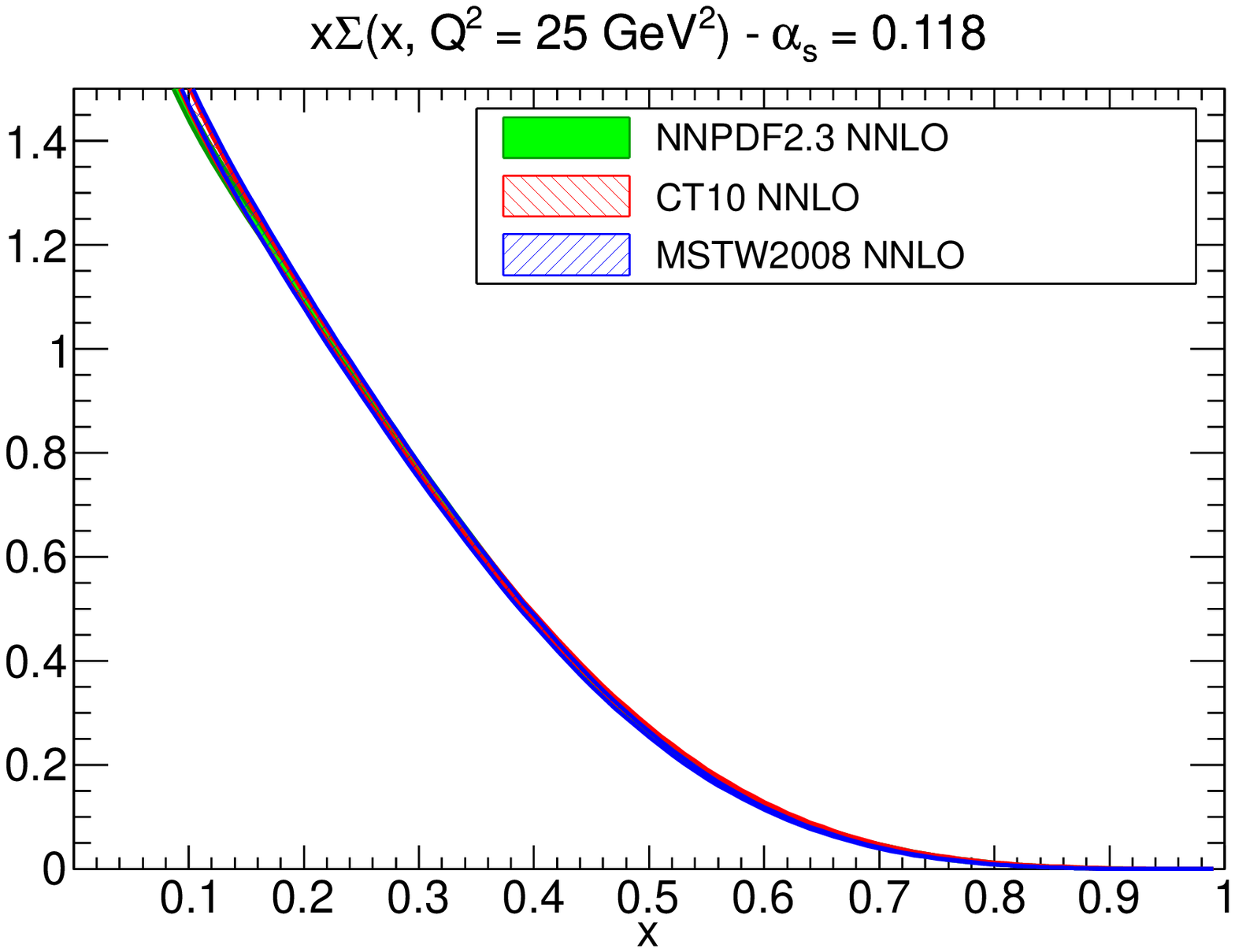}\quad
\includegraphics[width=0.48\textwidth]{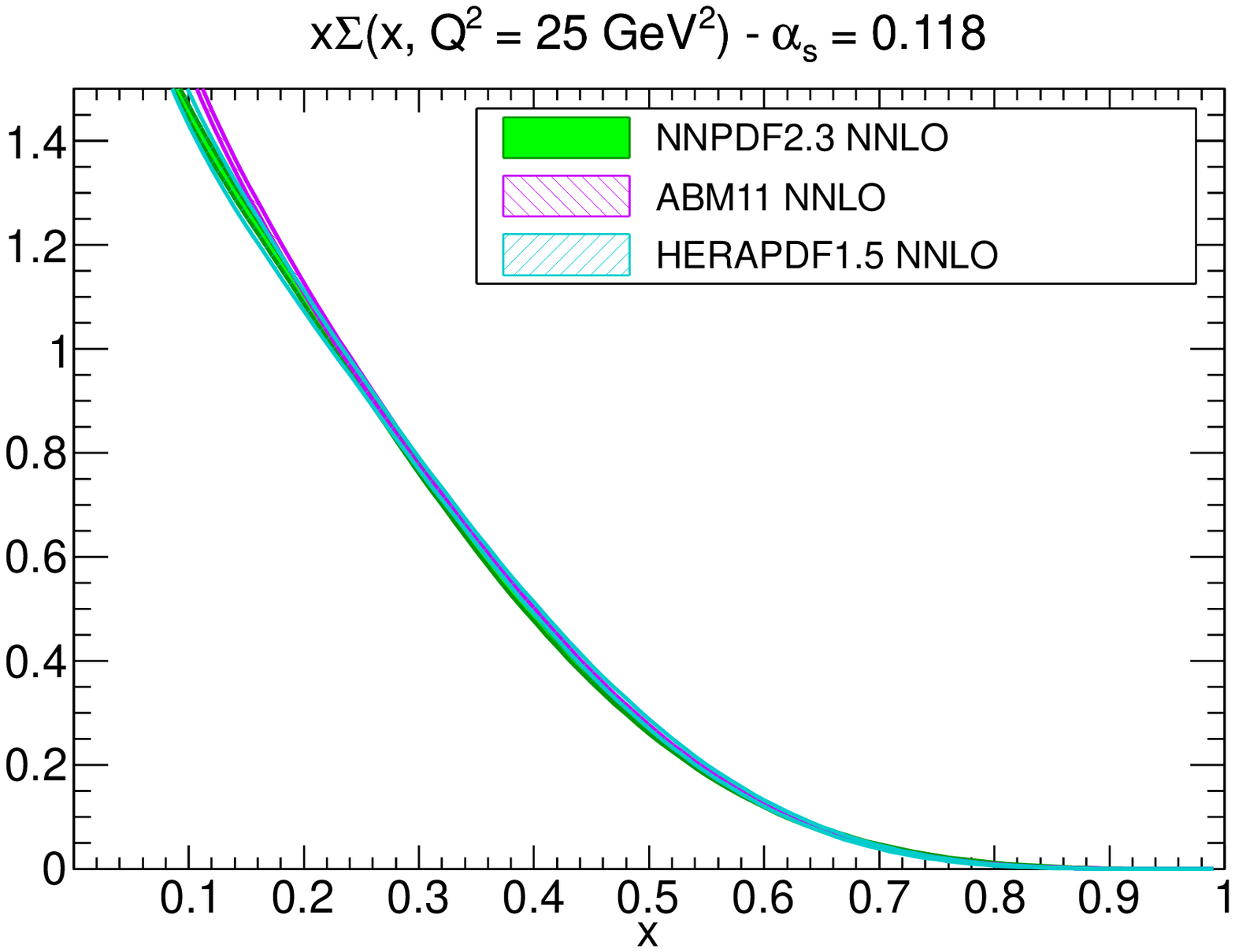}\\
 \includegraphics[width=0.48\textwidth]{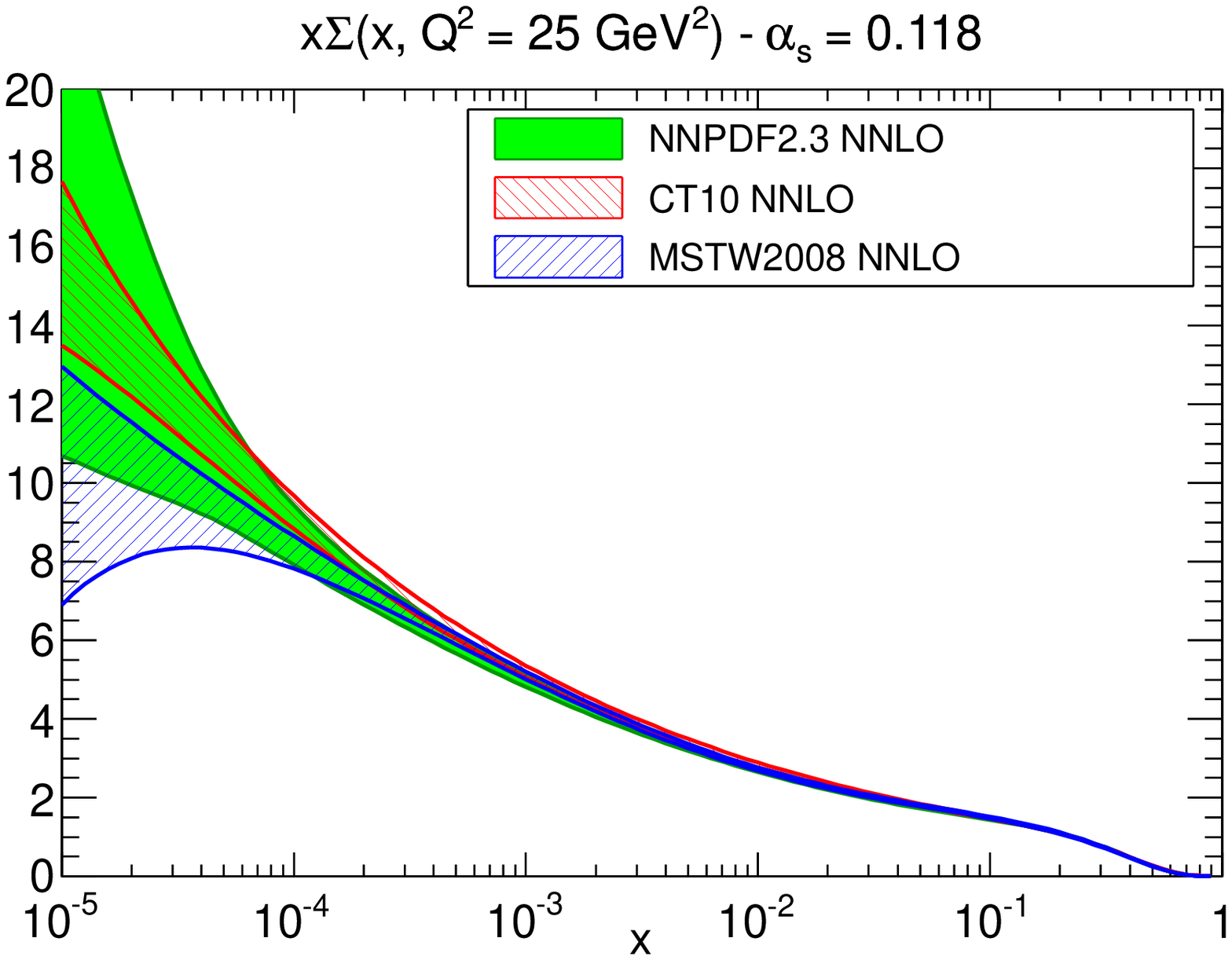}\quad
\includegraphics[width=0.48\textwidth]{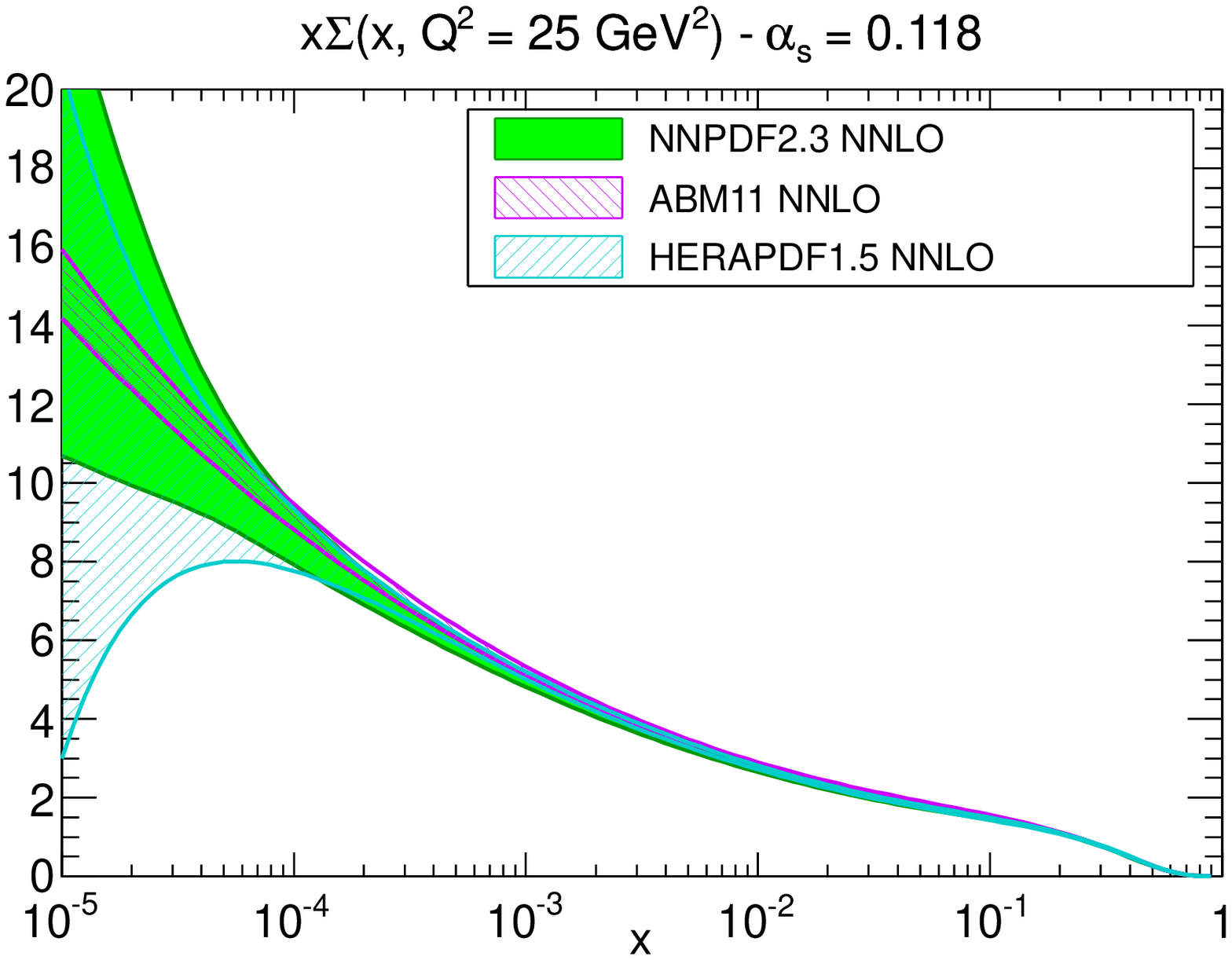}
      \end{center}
     \caption{\small 
\label{fig:PDFcomp-initscale-singlet}
The quark singlet PDFs $x\Sigma (x,Q^2)$
at $Q^2 = 25$ GeV$^2$ plotted versus $x$ on a linear scale
(upper plots) and on a logarithmic scale (lower plots). The plots on
the left
show the comparison between NNPDF2.3, CT10 and MSTW08, while
in the plots on the right we compare NNPDF2.3, HERAPDF1.5 and ABM11. All 
PDFs are shown for a common value of $\alpha_s=0.118$. }
\end{figure}
%%%%%%%%%%%%%%%%%%%%%%%%%%%%%%%%%%%%%%%%%%%%%%%%%%

%%%%%%%%%%%%%%%%%%%%%%%%%%%%%%%%%%%%%%%%%%%%%%%%%
\begin{figure}[h]
    \begin{center}
\includegraphics[width=0.48\textwidth]{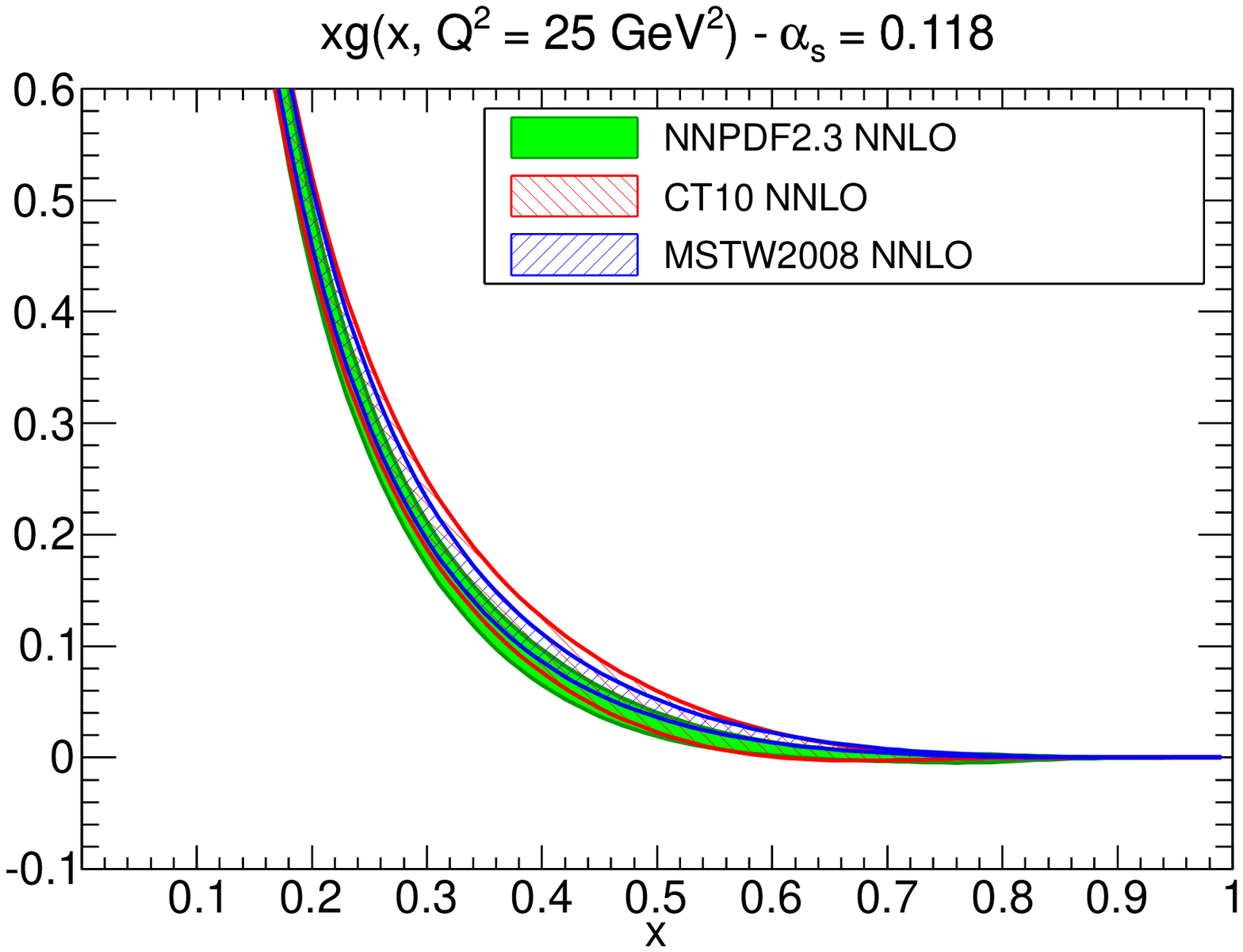}\quad
\includegraphics[width=0.48\textwidth]{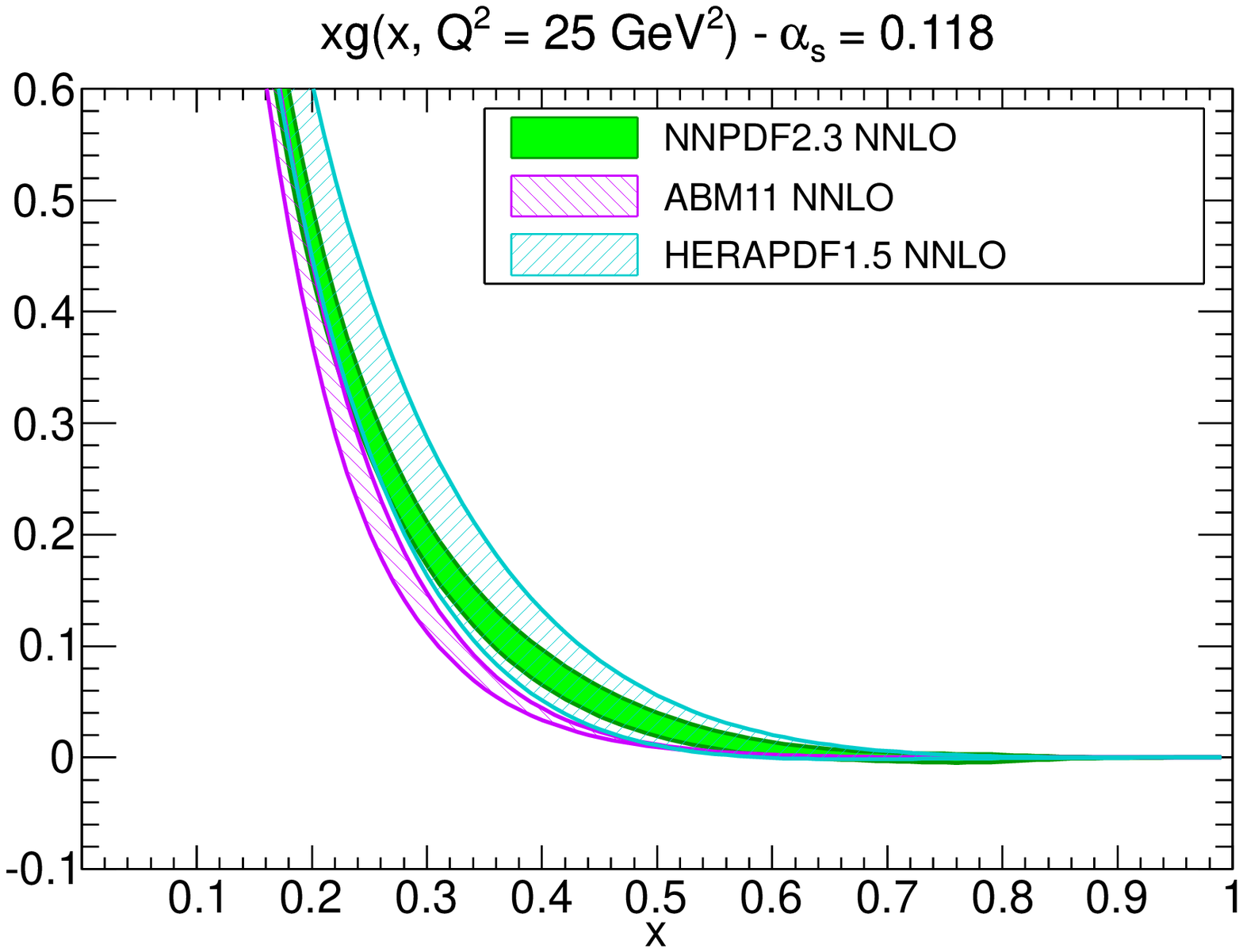}\\
 \includegraphics[width=0.48\textwidth]{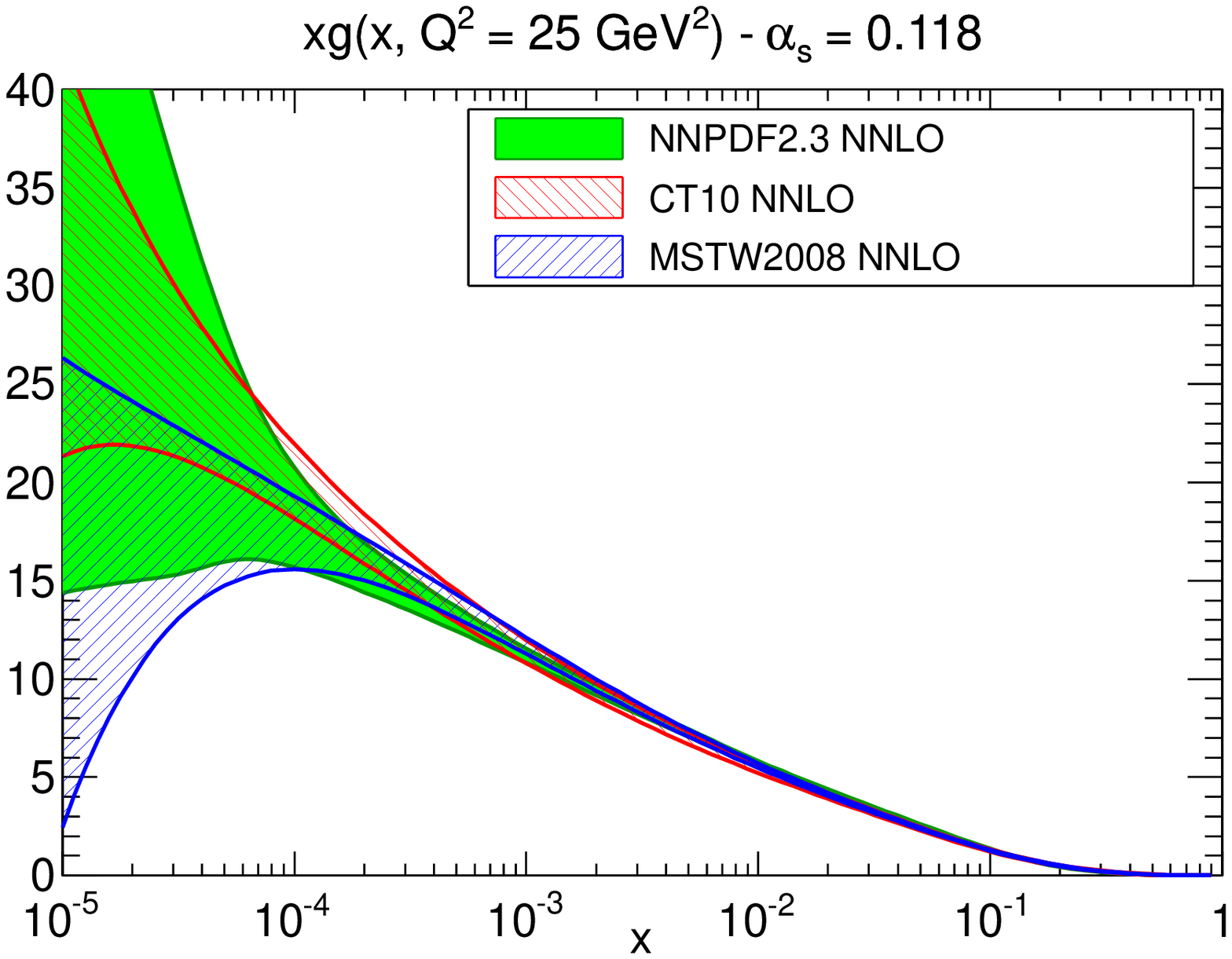}\quad
\includegraphics[width=0.48\textwidth]{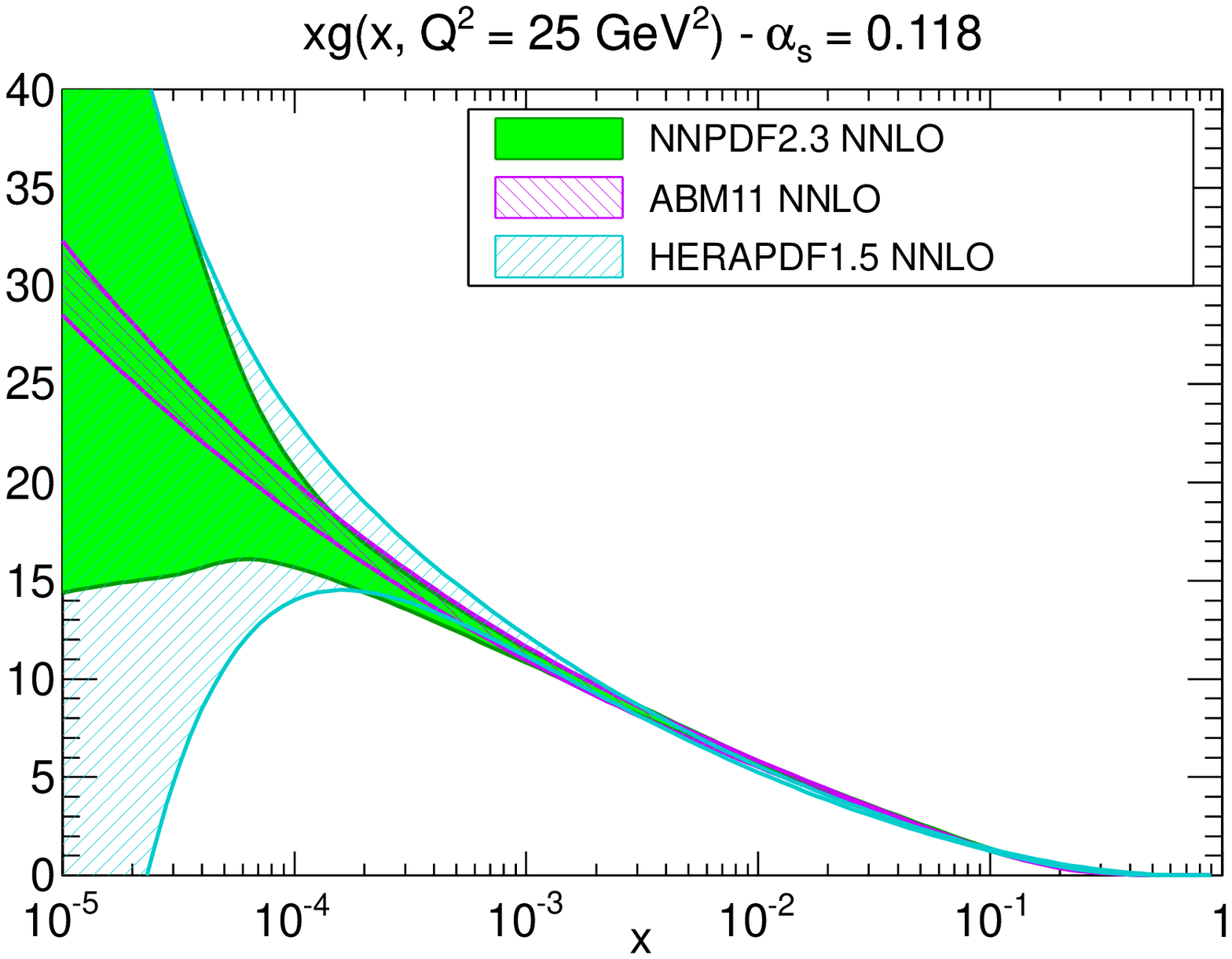}
      \end{center}
     \caption{\small 
    \label{fig:PDFcomp-initscale-g} 
Same as Fig.~\ref{fig:PDFcomp-initscale-singlet}, but for the gluon PDF.}
\end{figure}
%%%%%%%%%%%%%%%%%%%%%%%%%%%%%%%%%%%%%%%%%%%%%%%%%%

In Fig.~\ref{fig:PDFcomp-initscale-singlet}  we show the total
quark singlet PDF $\Sigma (x,Q^2) = \sum_{i=1}^5 \left[ q_i (x,Q^2) + \bar q_i(x,Q^2)\right]$, both on a linear and on a logarithmic scale, while in
Fig.~\ref{fig:PDFcomp-initscale-g} we show various gluon PDFs $g(x,Q^2)$, 
also on linear and logarithmic scales. There is a good agreement between 
all the sets for the quark singlet, though the uncertainty band at small 
$x$ is rather wider for NNPDF and HERAPDF. The gluons of CT10, MSTW and
NNPDF are also in reasonable agreement:
the PDF one-sigma uncertainty bands
overlap for all the range of $x$. Differences are larger for ABM11.
At small $x$ the ABM11 gluon has much smaller uncertainties than other groups,
even for $x$ values where there is little constraint from the data,
reflecting perhaps the more restrictive underlying PDF parametrization.
At high $x$ the ABM11 gluon is smaller than that of CT, MSTW and NNPDF, 
though the uncertainty band overlaps that of HERAPDF in most places. 
For HERAPDF1.5 the gluon at large $x$ has larger uncertainties due to 
the lack of collider data, while
at small $x$ it is close to the other PDF sets as expected, since
in this region it is only the precise HERA-I data that provides any
handle on the gluon.

%%%%%%%%%%%%%%%%%%%%%%%%%%%%%%%%%%%%%%%%%%%%%%%%%
\begin{figure}[h]
    \begin{center}
\includegraphics[width=0.48\textwidth]{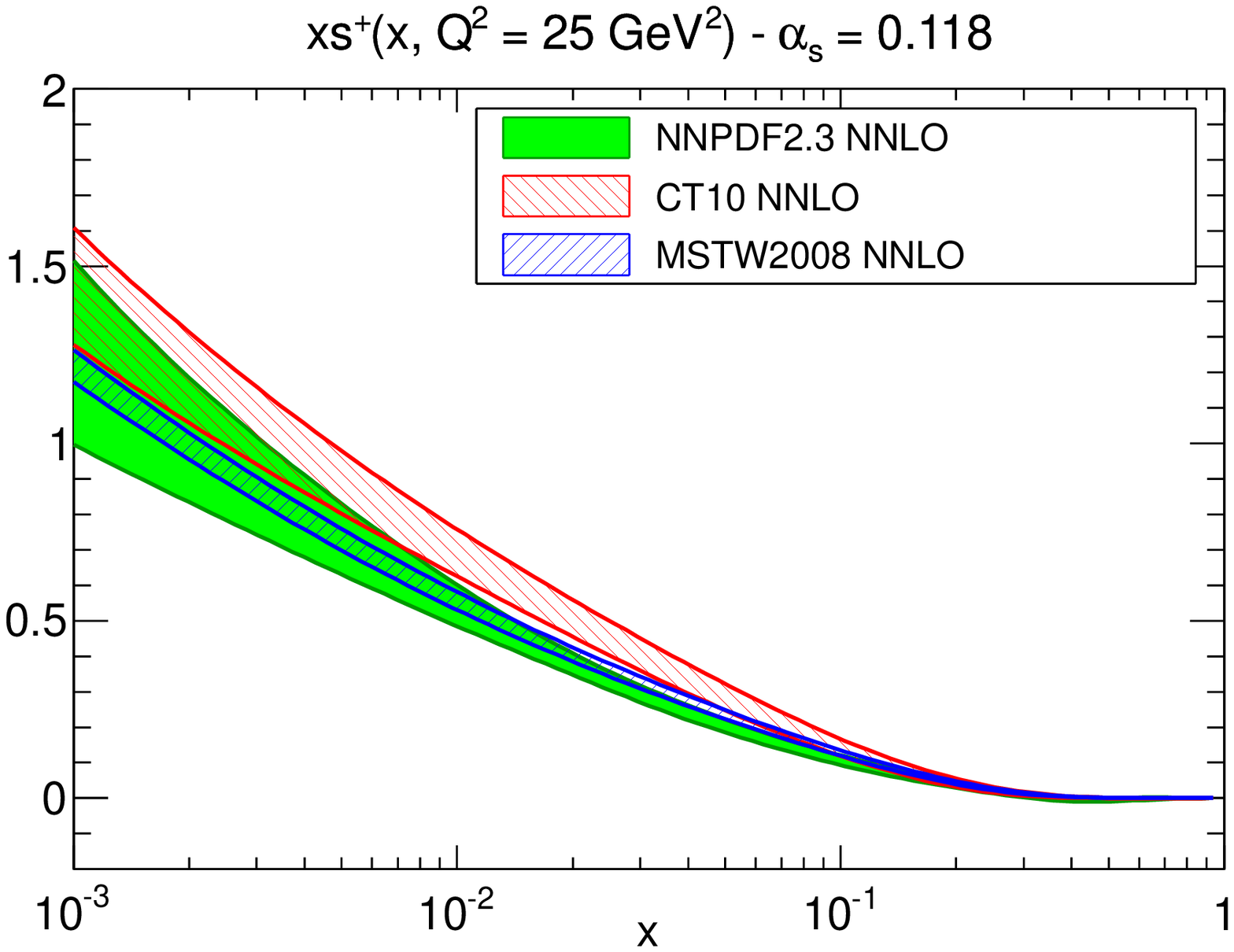}\quad
\includegraphics[width=0.48\textwidth]{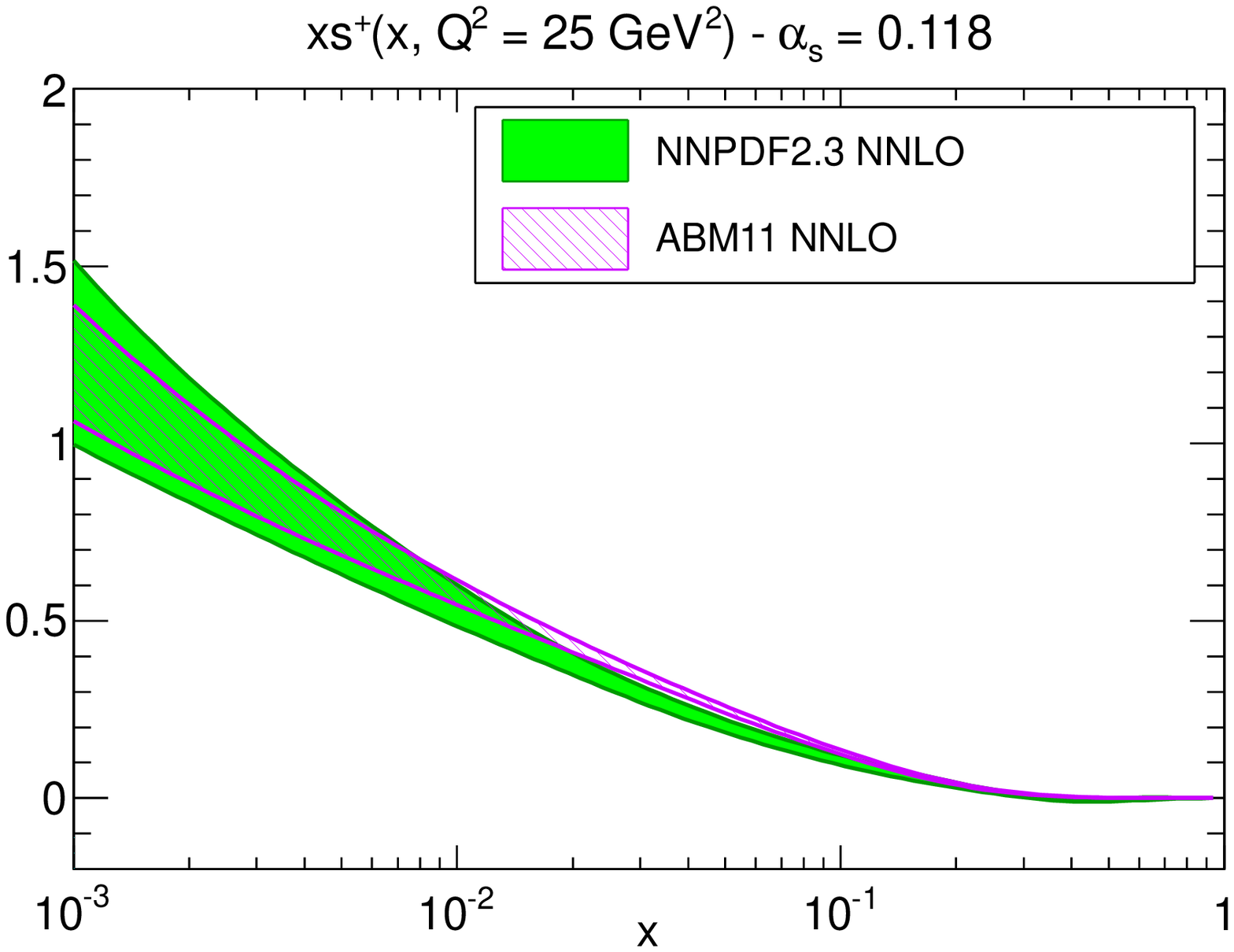}
      \end{center}
     \caption{\small The total strange PDFs
       $xs^+=x(s+\bar s)$ 
at $Q^2 = 25$ GeV$^2$. The plot on the left 
show the comparison between NNPDF2.3, CT10 and MSTW08, while
in the plot on the right we compare NNPDF2.3 and ABM11; HERAPDF1.5 is
not included as it does not have an independent parametrization of strangeness. 
    \label{fig:PDFcomp-initscale-str} }
\end{figure}
%%%%%%%%%%%%%%%%%%%%%%%%%%%%%%%%%%%%%%%%%%%%%%%%%%

The total strangeness $s^+(x,Q^2) = s(x,Q^2) + \bar s(x,Q^2)$ 
is shown on a logarithmic scale in 
Fig~\ref{fig:PDFcomp-initscale-str}; HERAPDF1.5 is not included
because it does not have an independent strangeness parametrization,
as HERA data alone do not allow disentangling of the  strange contribution.
The CT10 strange distribution is somewhat higher than that of other
groups. The origin of this difference is under study, which
is likely due to different non-perturbative parametrization of the PDFs
and differences in the heavy quark treatment of neutrino dimuon data.
Both theoretical studies and data from the LHC, both from 
electroweak vector boson production, and from the exclusive
$W+c$ data, should shed light on this issue in the future. First ATLAS
data did give some indication on strangeness ~\cite{Aad:2012sb} at
small $x$, but they are still not accurate enough~\cite{Ball:2012cx}
to lead to definite conclusions.

%%%%%%%%%%%%%%%%%%%%%%%%%%%%%%%%%%%%%%%%%%%%%%%%%
\begin{figure}[ht]
    \begin{center}
      \includegraphics[width=0.48\textwidth]{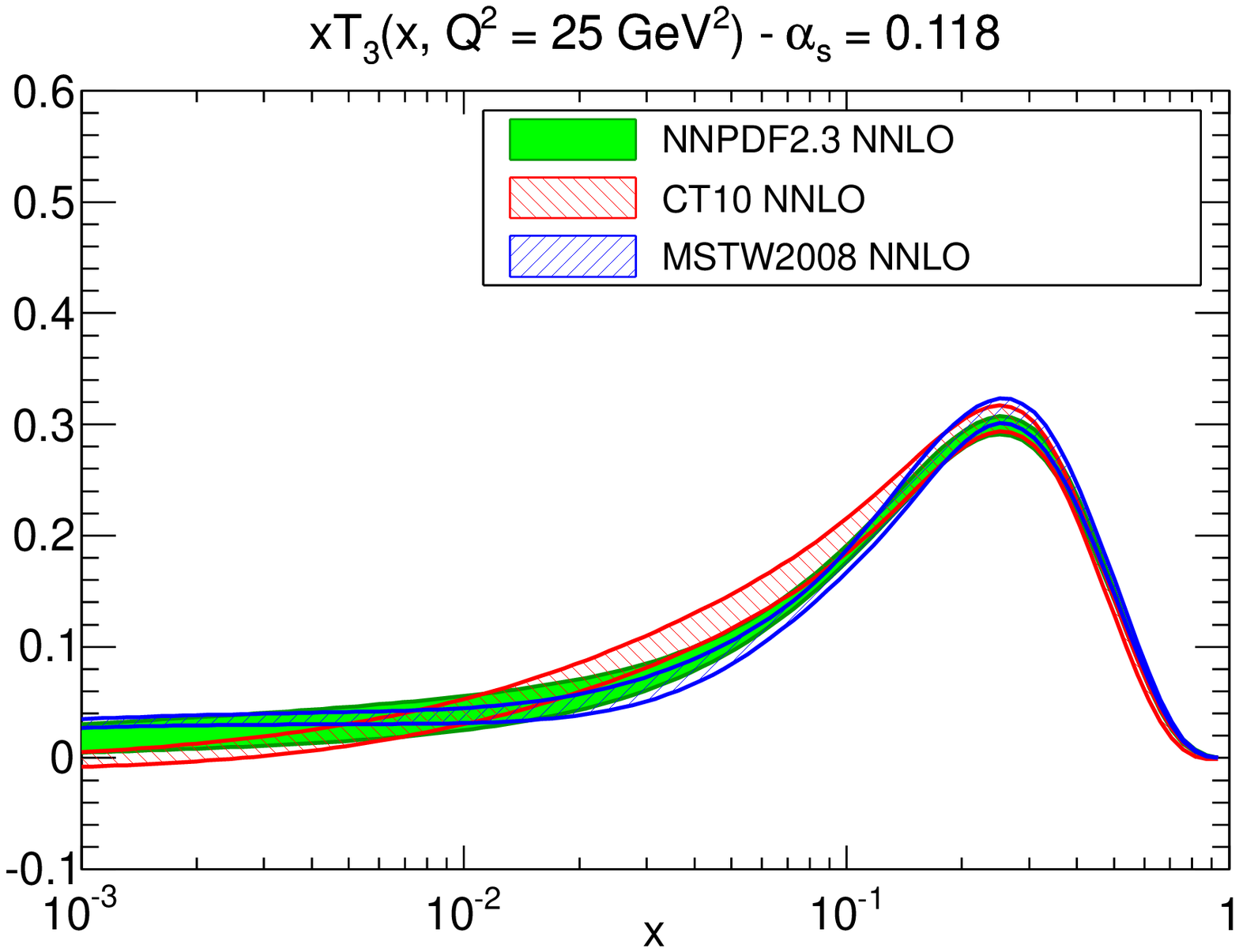}\quad
\includegraphics[width=0.48\textwidth]{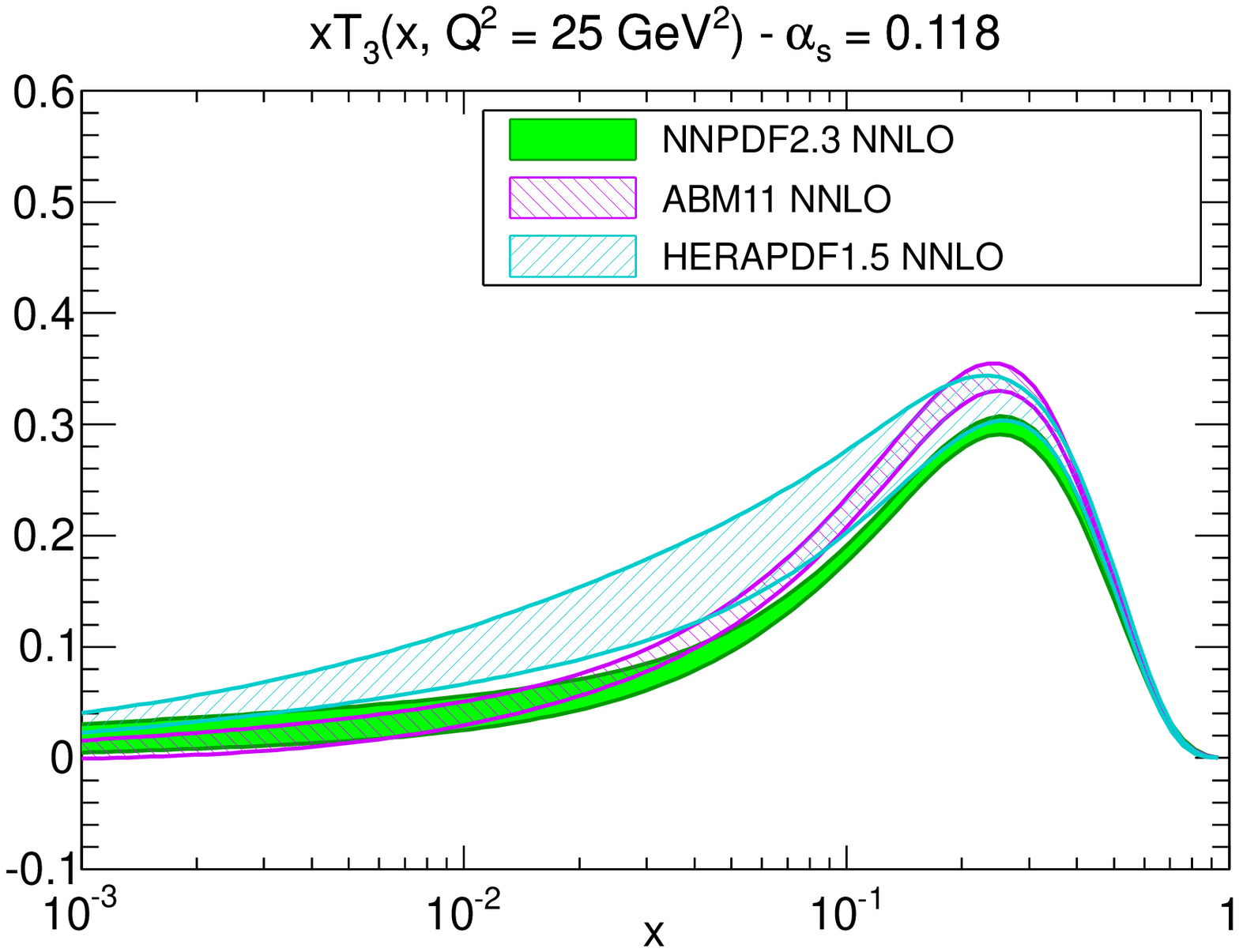}\\
\includegraphics[width=0.48\textwidth]{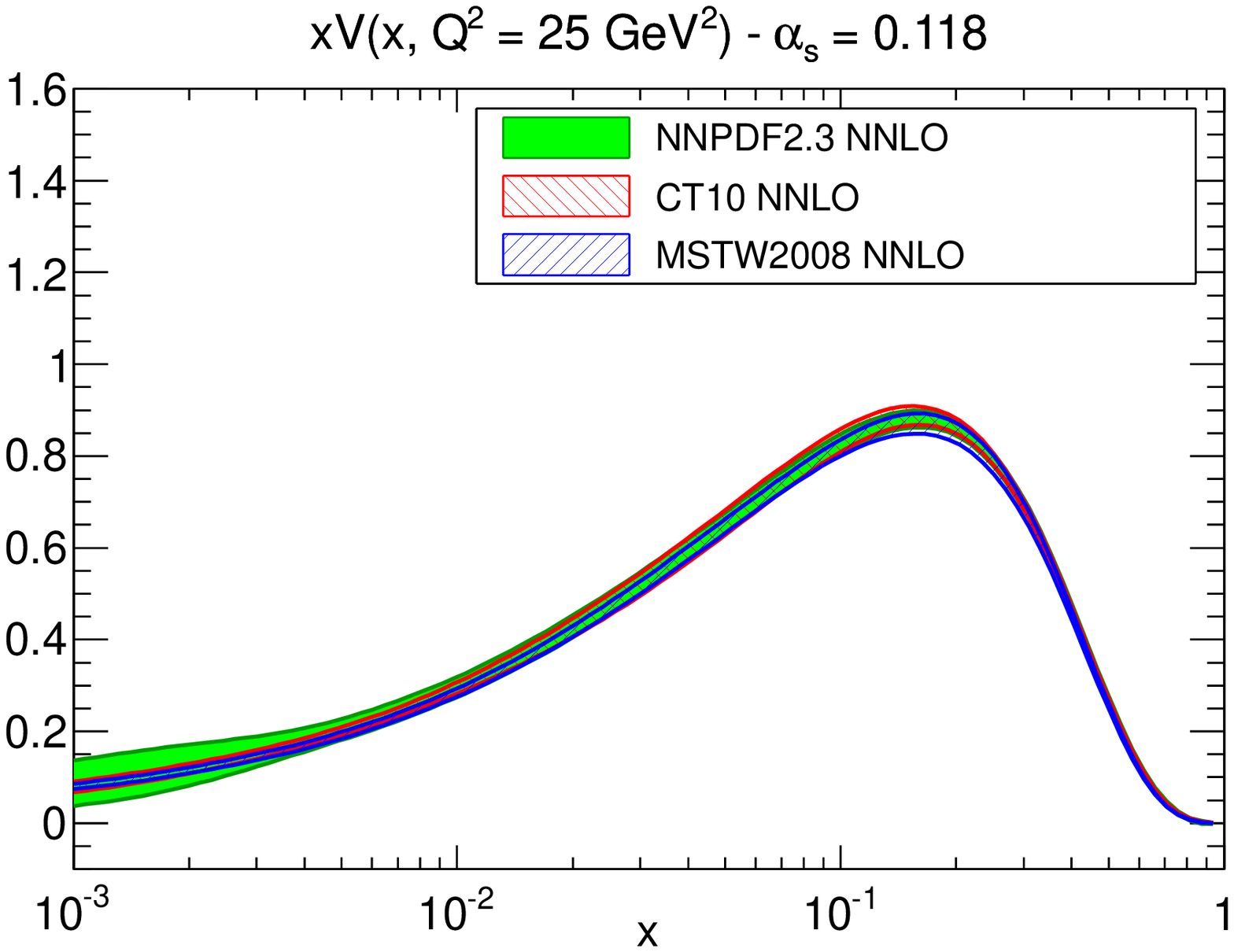}\quad
\includegraphics[width=0.48\textwidth]{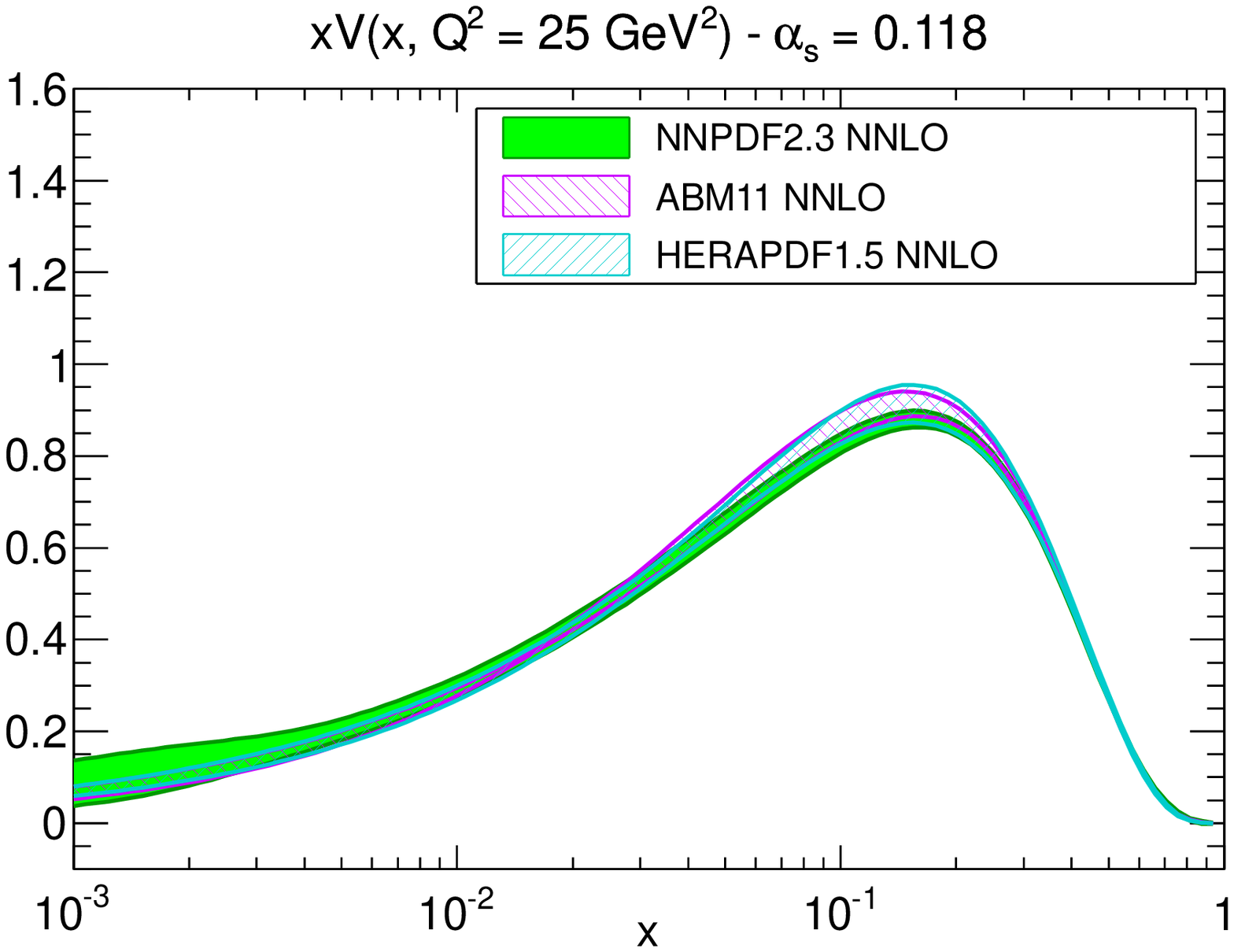}
      \end{center}
     \caption{\small 
Same as Fig.~\ref{fig:PDFcomp-initscale-singlet} for the non singlet triplet 
$xT_3(x)$ and the total valence $xV(x)$ PDFs defined in Eq.~(\ref{nonsingpdf}).
    \label{fig:PDFcomp-initscale-t3} }
\end{figure}
%%%%%%%%%%%%%%%%%%%%%%%%%%%%%%%%%%%%%%%%%%%%%%%%%%

%%%%%%%%%%%%%%%%%%%%%%%%%%%%%%%%%%%%%%%%%%%%%%%%%
\begin{figure}[ht]
    \begin{center}
      \includegraphics[width=0.48\textwidth]{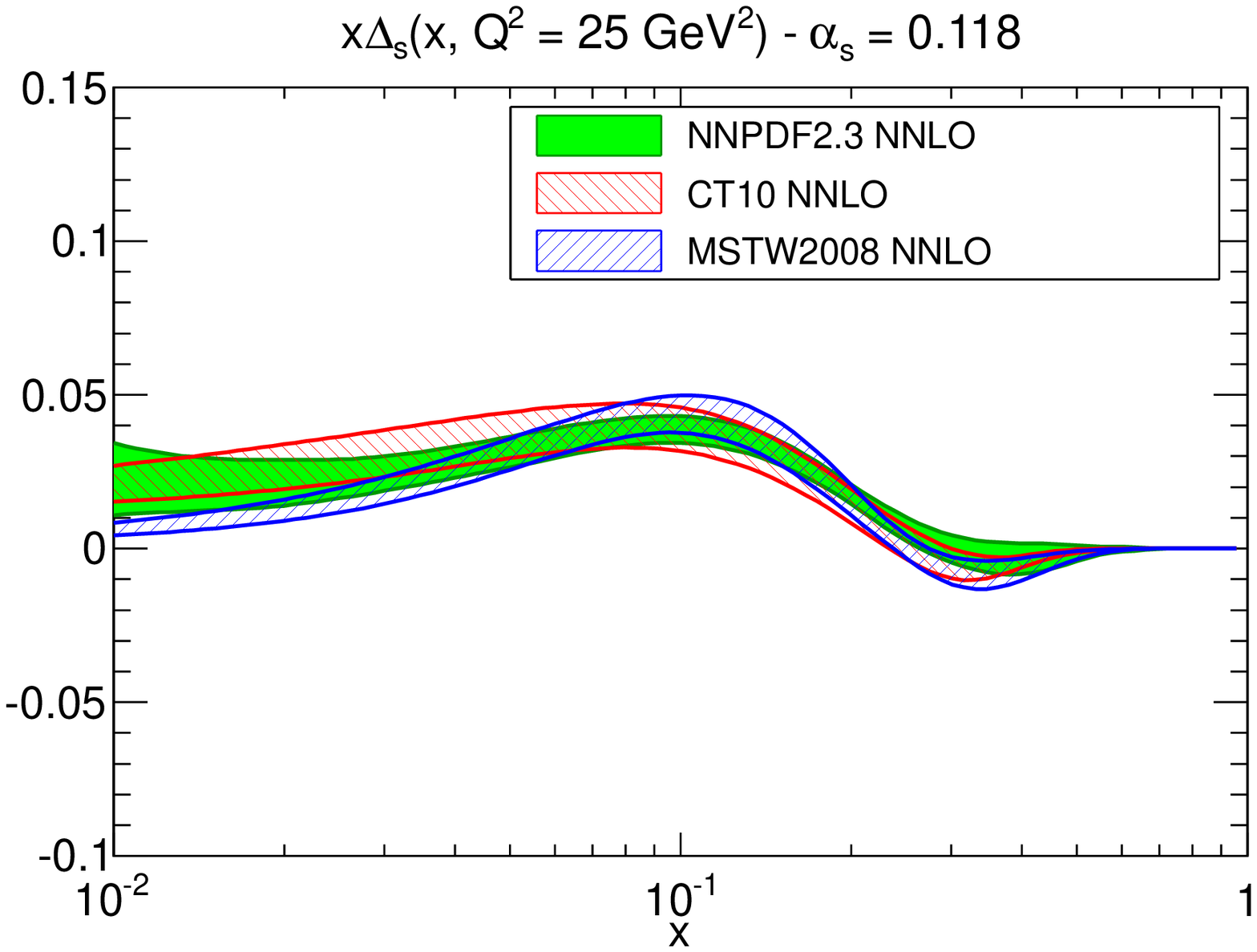}\quad
\includegraphics[width=0.48\textwidth]{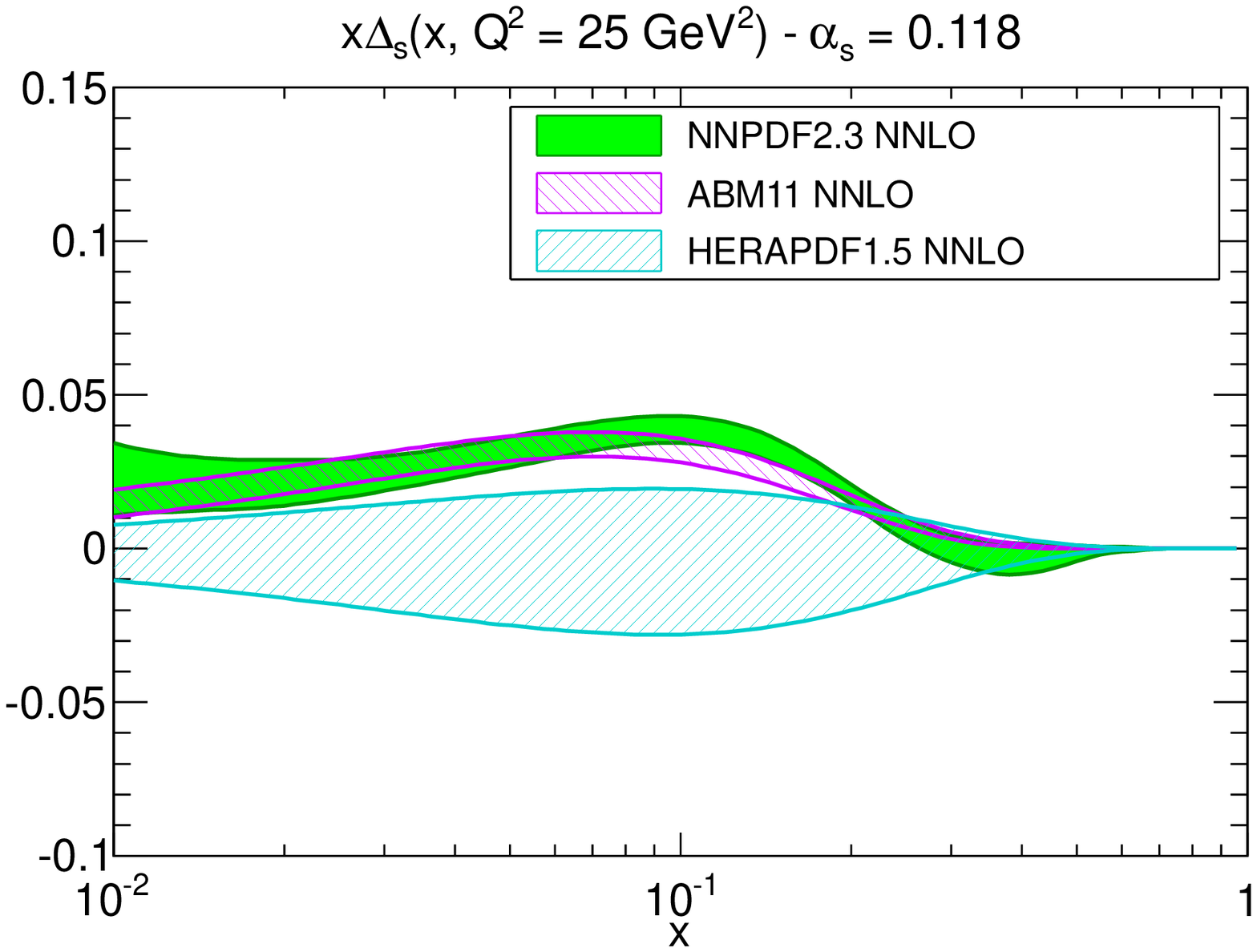}\\
\includegraphics[width=0.48\textwidth]{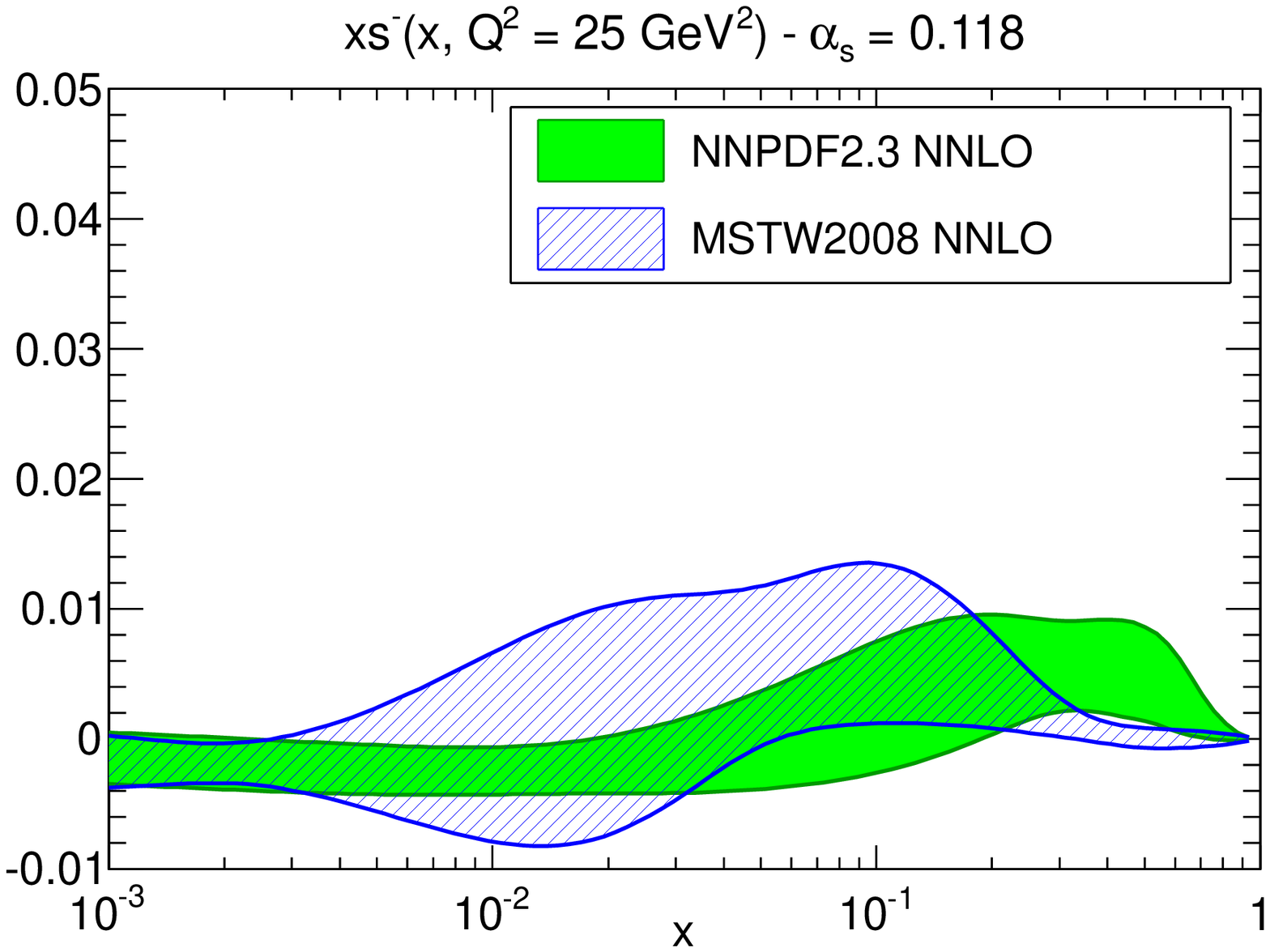}
%\quad
%\includegraphics[width=0.48\textwidth]{plots/pdf_xsminus_log_band_comparison_others_0118_b.eps}
      \end{center}
     \caption{\small
Same as Fig.~\ref{fig:PDFcomp-initscale-t3} for 
the the sea asymmetry $x\Delta_S = x(\bar{d}-\bar{u})$ and  the strange
asymmetry  $xs^- =x( s-\bar{s})$.
 In the latter case we show only the results for MSTW08 and NNPDF2.3,
the only PDF sets that introduce an independent parametrization of
the strangeness asymmetry.
    \label{fig:PDFcomp-initscale-deltas} }
\end{figure}
%%%%%%%%%%%%%%%%%%%%%%%%%%%%%%%%%%%%%%%%%%%%%%%%%%

Finally we compare
non-singlet distributions:
 the nonsinglet triplet and the total
 valence PDFs, respectively defined as 
\begin{align}
T_3&=u+\bar{u}-d-\bar{d}\nonumber\\
V&=u-\bar{u}+d-\bar{d}+s-\bar{s} \label{nonsingpdf}
\end{align}
in Fig.~\ref{fig:PDFcomp-initscale-t3},
and the quark sea asymmetry
$\Delta_S = \bar{d}-\bar{u}$ and the strangeness
asymmetry $s^- = s-\bar{s}$ in Fig.~\ref{fig:PDFcomp-initscale-deltas}.
There is reasonable agreement for $T_3$ and $V$, except
for ABM11, for which $T_3$ at large $x$ is significantly higher than in the 
other sets. This is due to a larger $u$ distribution in this region. 
The HERAPDF1.5 PDF uncertainties in $T_3$ are rather larger, 
reflecting the fact that HERA data does not provide much information
on quark flavor separation. All sets are in a broad agreement on the
light sea 
asymmetry, apart from HERAPDF1.5, which does not include 
the  Drell-Yan and electroweak boson production data and cannot 
separate $\bar u$ and $\bar d$ flavors. Only MSTW08 and NNPDF2.3
provide independent parametrizations of the strange asymmetry PDF  
and are in reasonable agreement within uncertainties.

\subsection{Parton luminosities}

Now we compare parton luminosities. 
At a hadron collider, all factorizable observables for the production 
of a final state with mass $M_X$ depend on parton
distributions through a parton luminosity, which, 
following Ref.~\cite{Campbell:2006wx}, we define 
 as
\be
\Phi_{ij}\lp M_X^2\rp = \frac{1}{s}\int_{\tau}^1
\frac{dx_1}{x_1} f_i\lp x_1,M_X^2\rp f_j\lp \tau/x_1,M_X^2\rp \ ,
\label{eq:lumdef}
\ee
where $f_i(x,M^2)$ is a PDF at a scale $M^2$, 
and $\tau \equiv M_X^2/s$. 
As the PDFs, all parton luminosities will be compared for
a common value of the strong coupling $\alpha_s=0.118$.
The parton luminosities are displayed as ratios to the 
NNPDF2.3 set. We assume a center-of-mass energy of 8 TeV.

%%%%%%%%%%%%%%%%%%%%%%%%%%%%%%%%%%%%%%%%%%%%%%%%%
\begin{figure}[ht]
    \begin{center}
      \includegraphics[width=0.48\textwidth]{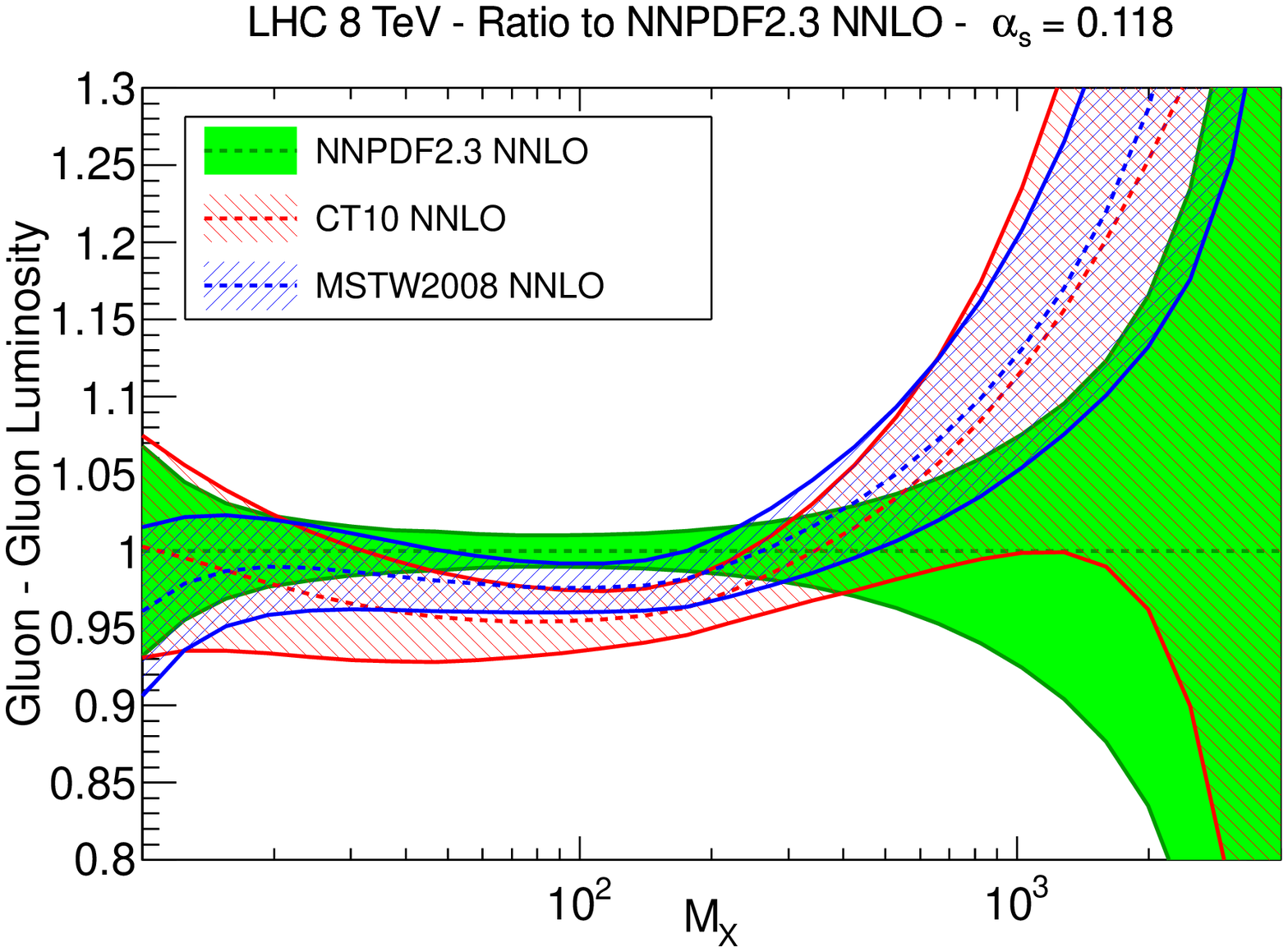}\quad
\includegraphics[width=0.48\textwidth]{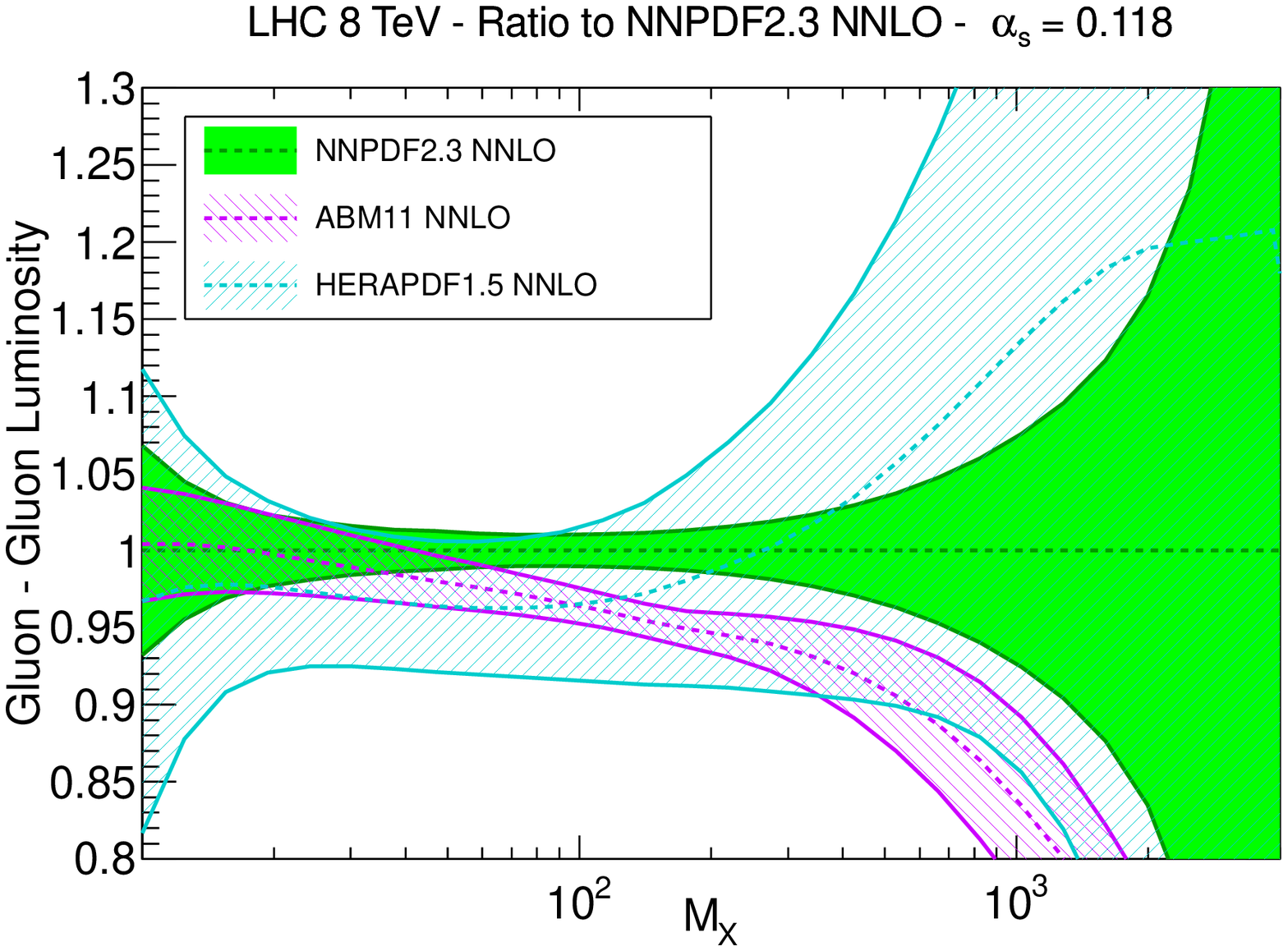}\\
  \includegraphics[width=0.48\textwidth]{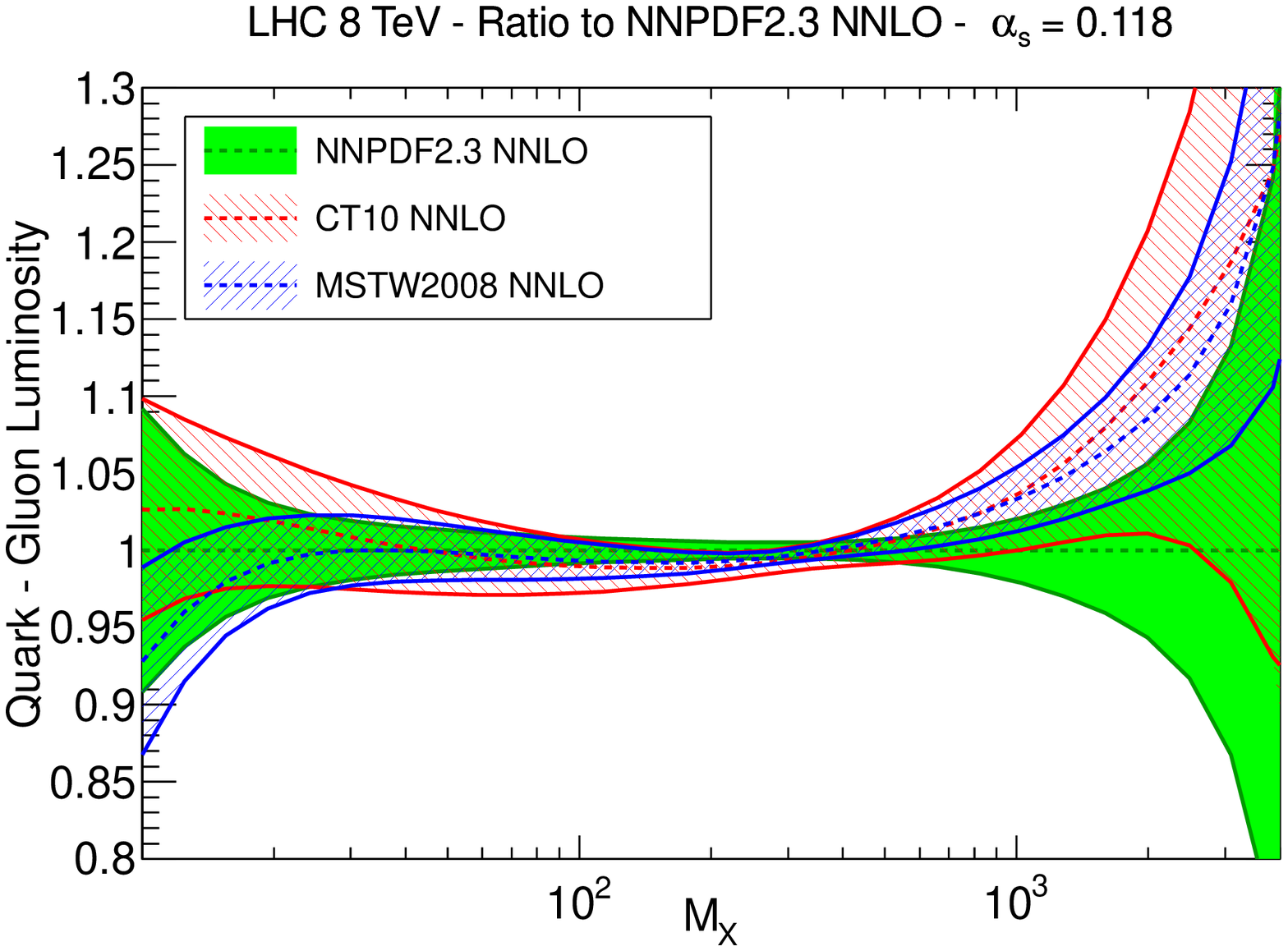}\quad
\includegraphics[width=0.48\textwidth]{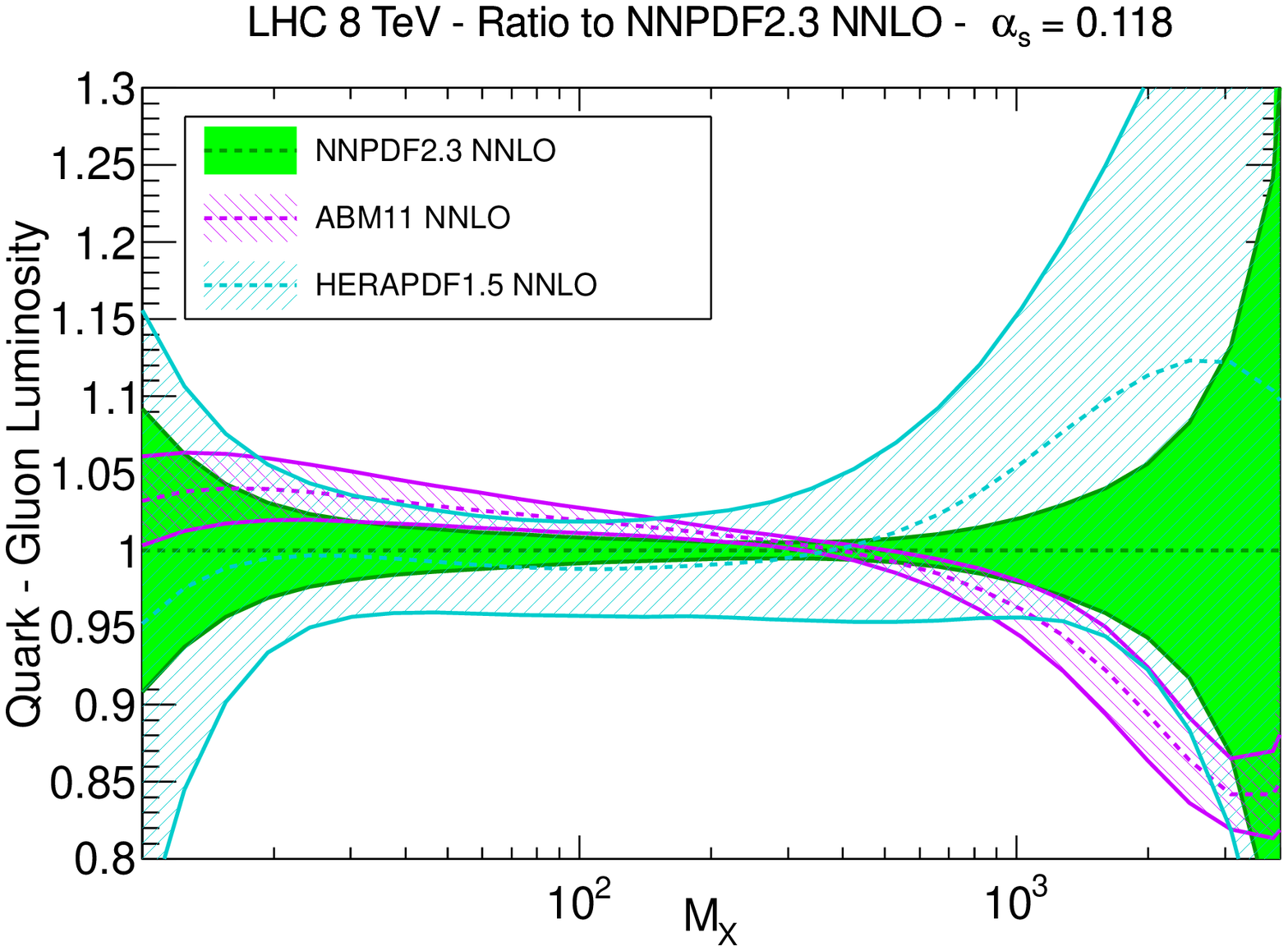}
      \end{center}
     \caption{\small
The gluon-gluon (upper plots)
and quark-gluon (lower plots) 
luminosities, Eq.~(\ref{eq:lumdef}), for the production
of a final state of invariant mass $M_X$ (in GeV) at LHC 8 TeV.  The left plots
show the comparison between NNPDF2.3, CT10 and MSTW08, while
in the right plots we compare NNPDF2.3, HERAPDF1.5 and MSTW08. 
All luminosities are computed at a common value of $\alpha_s=0.118$.
    \label{fig:PDFlumi-gg} }
\end{figure}
%%%%%%%%%%%%%%%%%%%%%%%%%%%%%%%%%%%%%%%%%%%%%%%%%%

%%%%%%%%%%%%%%%%%%%%%%%%%%%%%%%%%%%%%%%%%%%%%%%%%
\begin{figure}[ht]
    \begin{center}
      \includegraphics[width=0.48\textwidth]{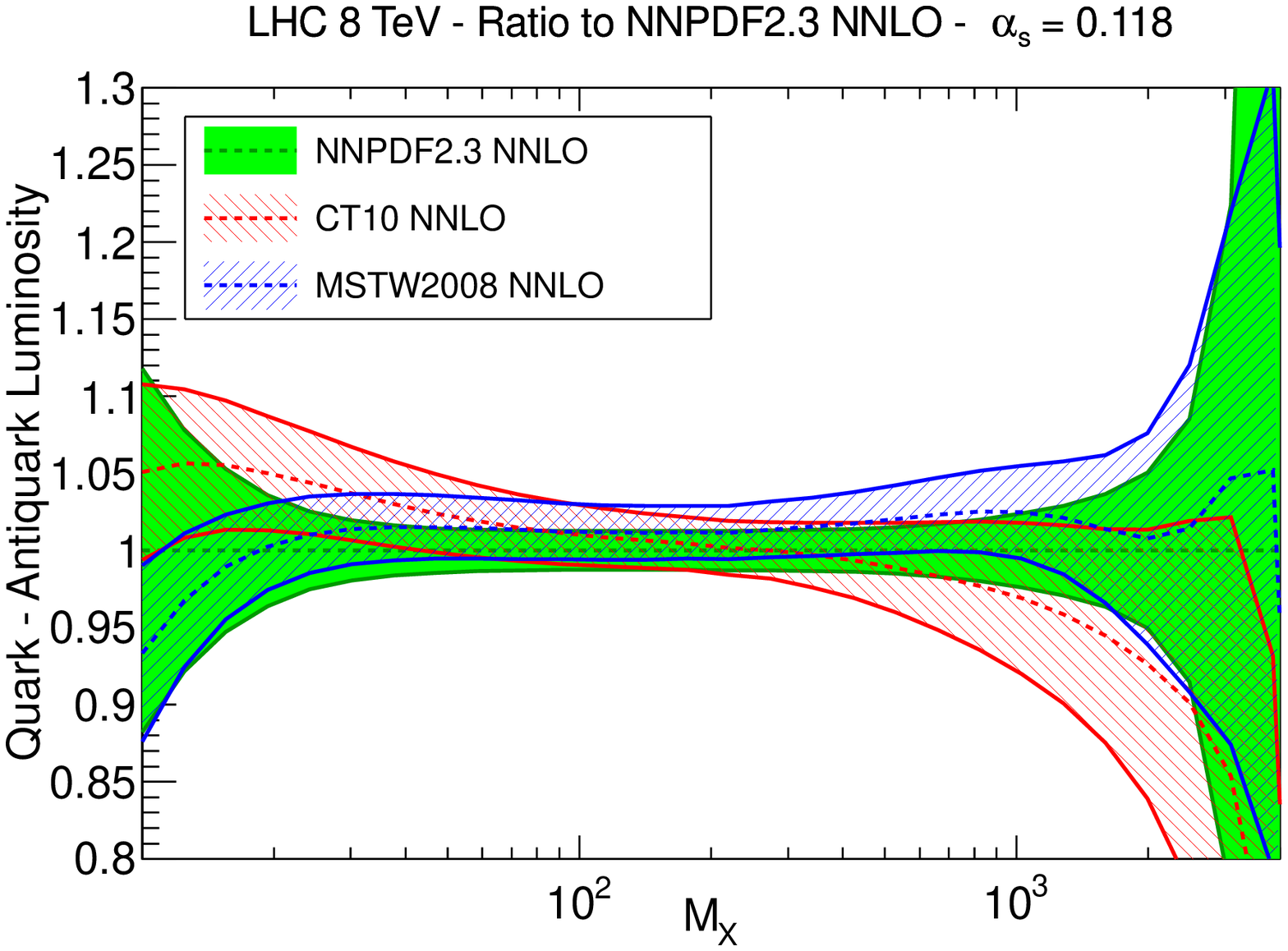}\quad
\includegraphics[width=0.48\textwidth]{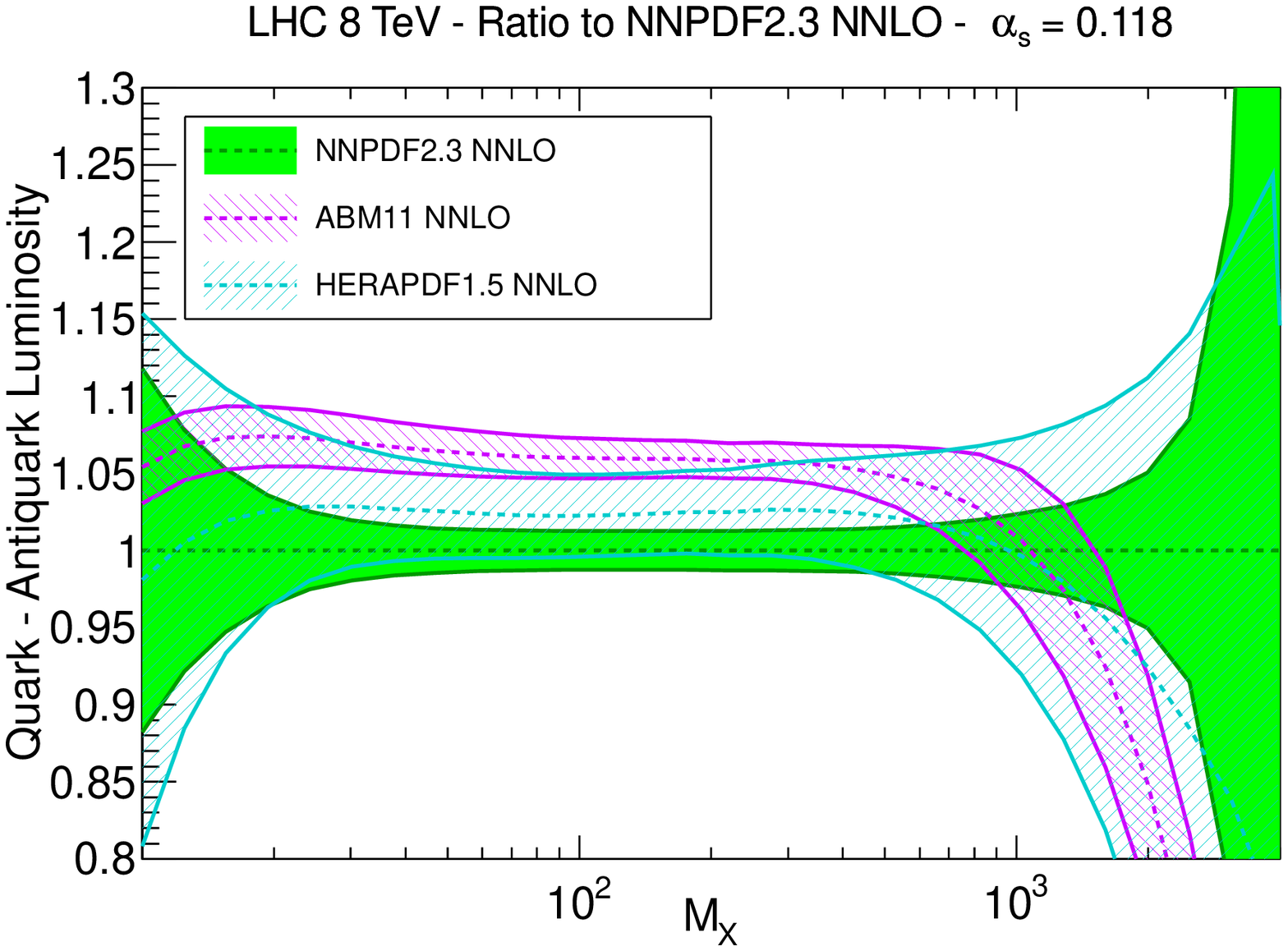}\\
  \includegraphics[width=0.48\textwidth]{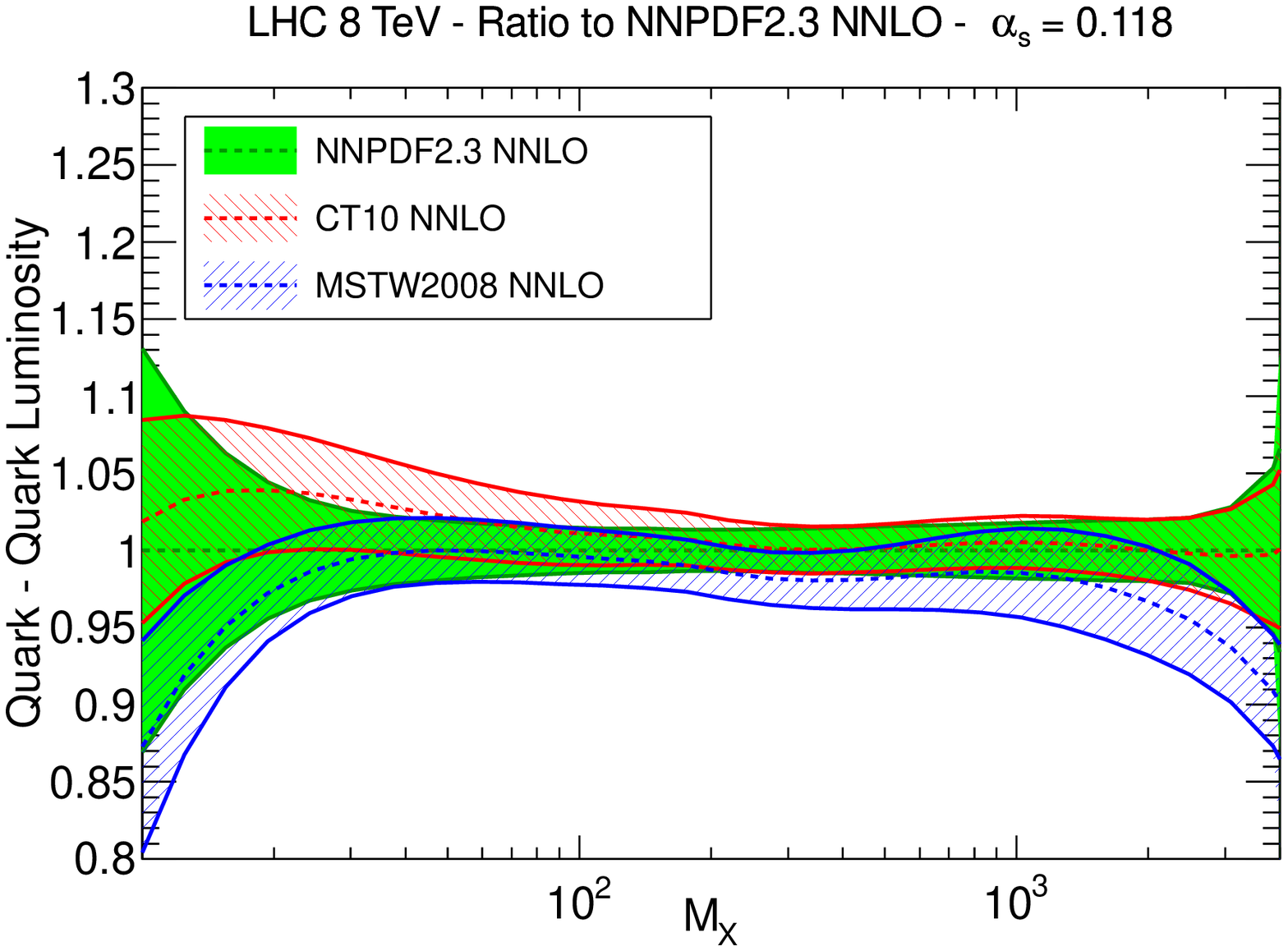}\quad
\includegraphics[width=0.48\textwidth]{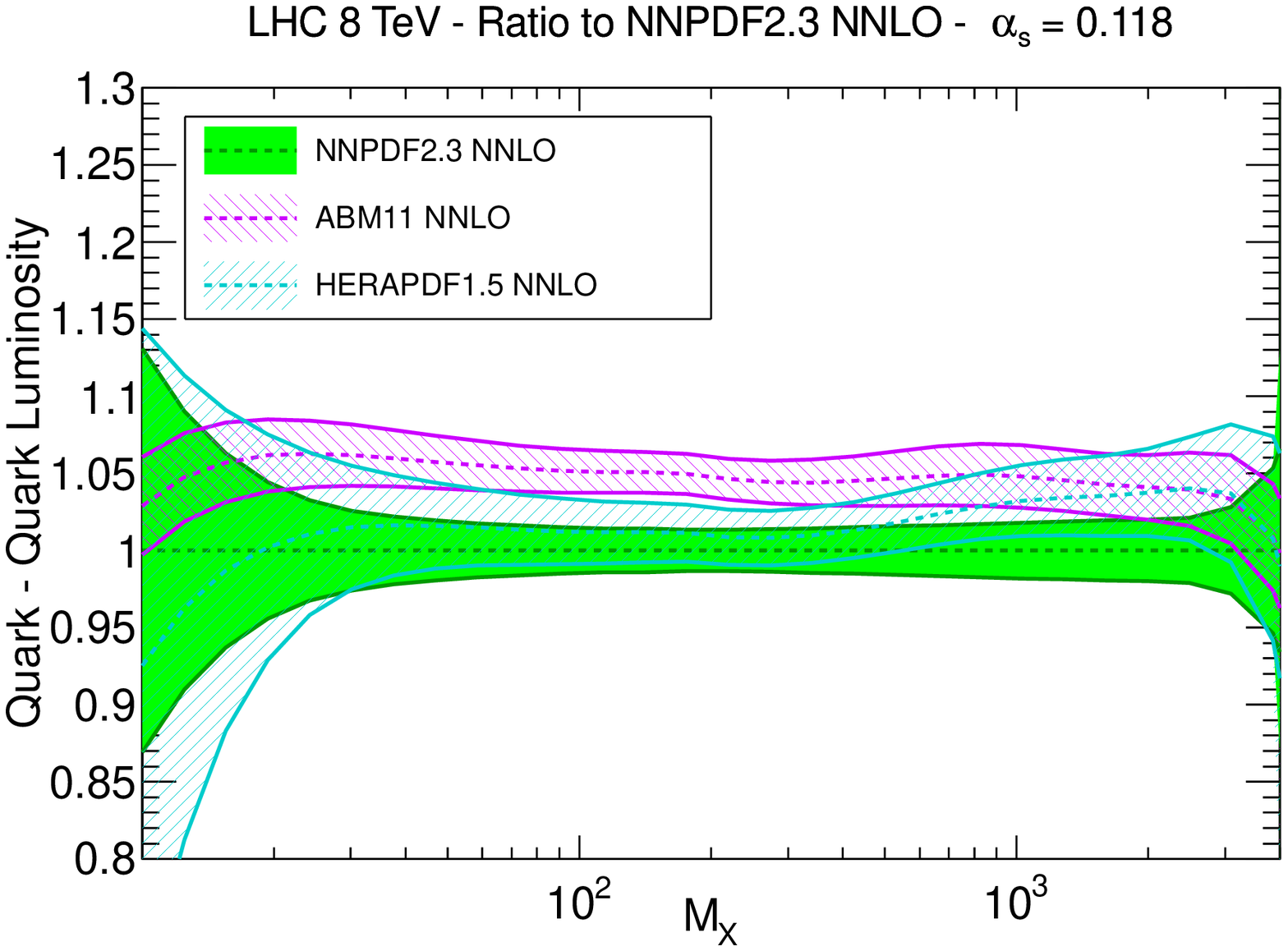}
      \end{center}
     \caption{\small
Same as  Fig.~\ref{fig:PDFlumi-gg} 
for the quark-antiquark (upper plots)
and quark-quark (lower plots) luminosities.
    \label{fig:PDFlumi-qq}} 
\end{figure}
%%%%%%%%%%%%%%%%%%%%%%%%%%%%%%%%%%%%%%%%%%%%%%%%%%
 
The gluon-gluon and quark-gluon luminosities are
shown in  Fig.~\ref{fig:PDFlumi-gg}, and the
quark-quark and quark-antiquark luminosities are
shown in  Fig.~\ref{fig:PDFlumi-qq}.
There is a reasonably good agreement between the NNPDF2.3, MSTW08 and 
CT10 PDF sets for the full range of invariant masses. 
However, the PDF uncertainties increase dramatically at $M_X > 1$~TeV, 
relevant for searches and characterization of heavy particles. 
Future data from the LHC on high-$E_T$ jet
production and high-mass Drell-Yan process should be able to provide
constraints in this  region.\footnote{When using high-mass data in PDF fits one should be careful in avoiding possible contamination from beyond the Standard Model (BSM) physics.
There are various ways to achieve this, in the particular case of jets, one could include in the fit only the data measuring high $p_T$ jet cross-section in the forward region, where the two leading jets are well separated and span a similar range of Bjorken x values of the PDFs, but with a smaller invariant mass, 
thus, being less sensitive to BSM dynamics. } Differences with  other PDFs 
are more pronounced
for the ABM11 and HERAPDF1.5 PDF sets. For HERAPDF1.5, there is generally 
an agreement in central values, but the uncertainty is rather larger
in some $x$ ranges, particularly for the gluon luminosity, 
but also to some extent for 
the quark-antiquark one. For ABM11 instead, the quark-quark and 
quark-antiquark luminosity are systematically higher by over
5\% below 1 TeV, and above this the quark-antiquark luminosity becomes 
much softer than either NNPDF2.3 or MSTW08. The gluon-gluon 
luminosity becomes smaller 
than all the other PDFs at high invariant masses, overlapping only 
with the very large HERAPDF1.5 uncertainty. 

%%%%%%%%%%%%%%%%%%%%%%%%%%%%%%%%%%%%%%%%%%%%%%%%%
\begin{figure}[ht]
    \begin{center}
      \includegraphics[width=0.48\textwidth]{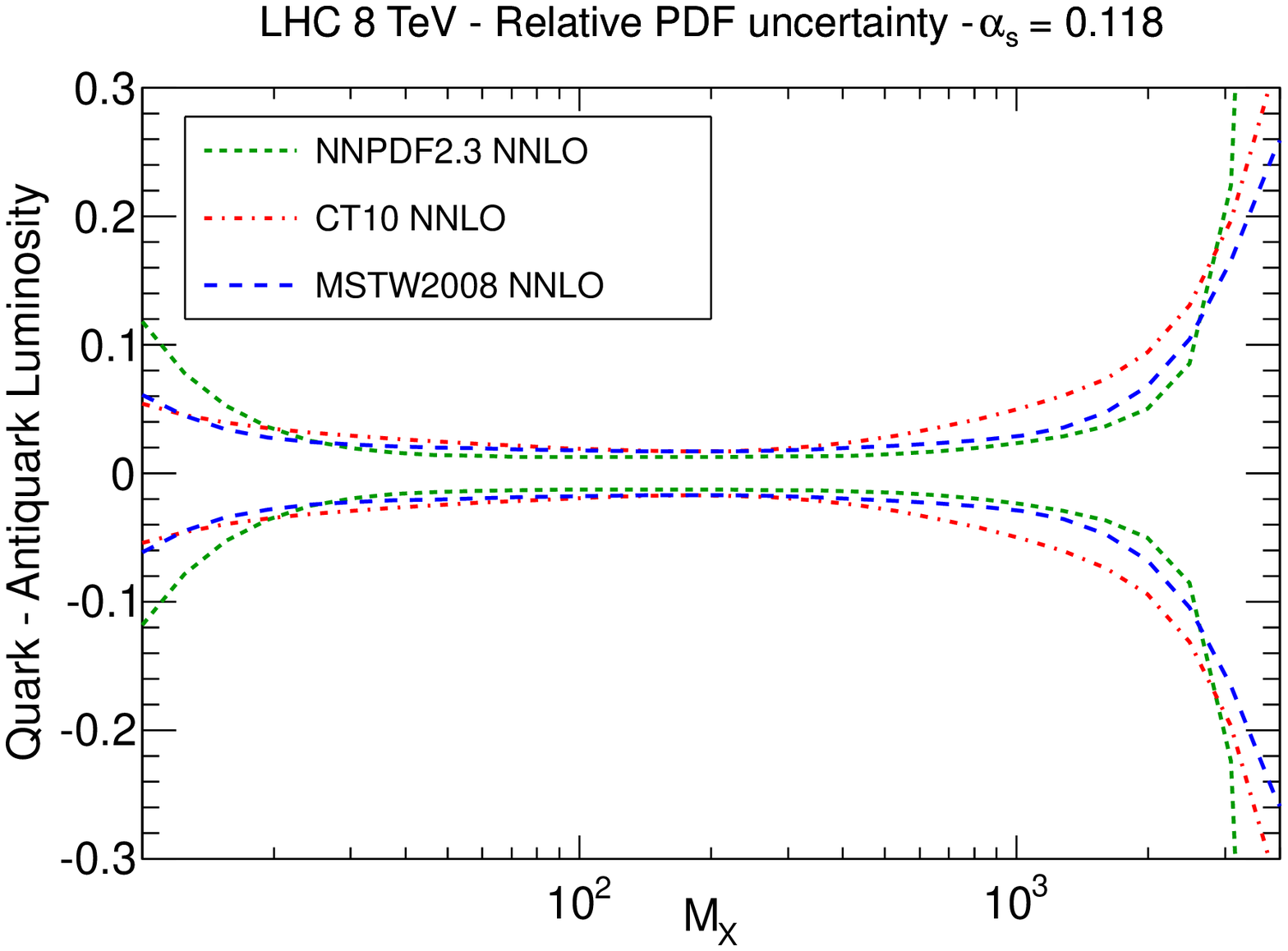}\quad
\includegraphics[width=0.48\textwidth]{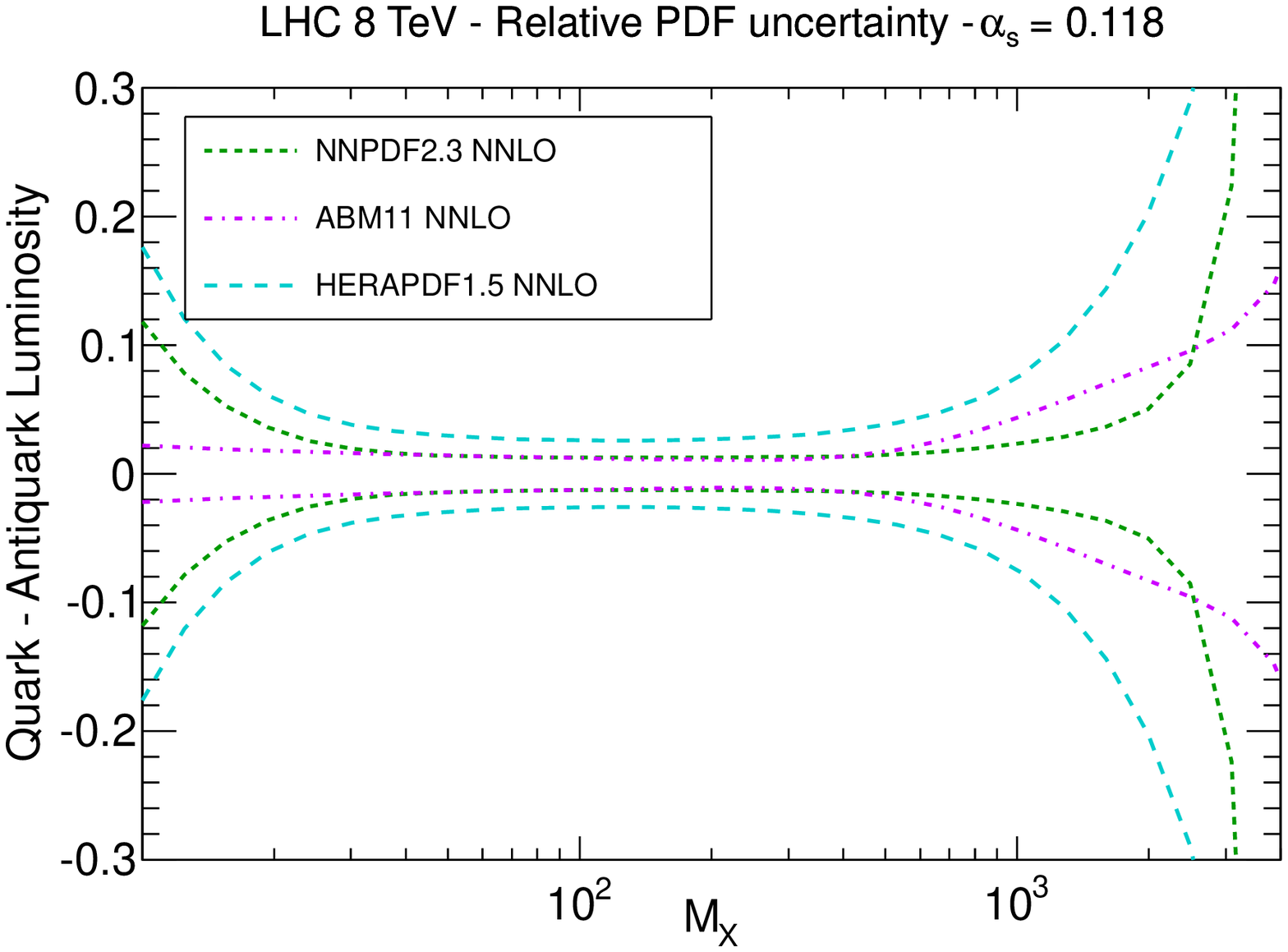}\\
  \includegraphics[width=0.48\textwidth]{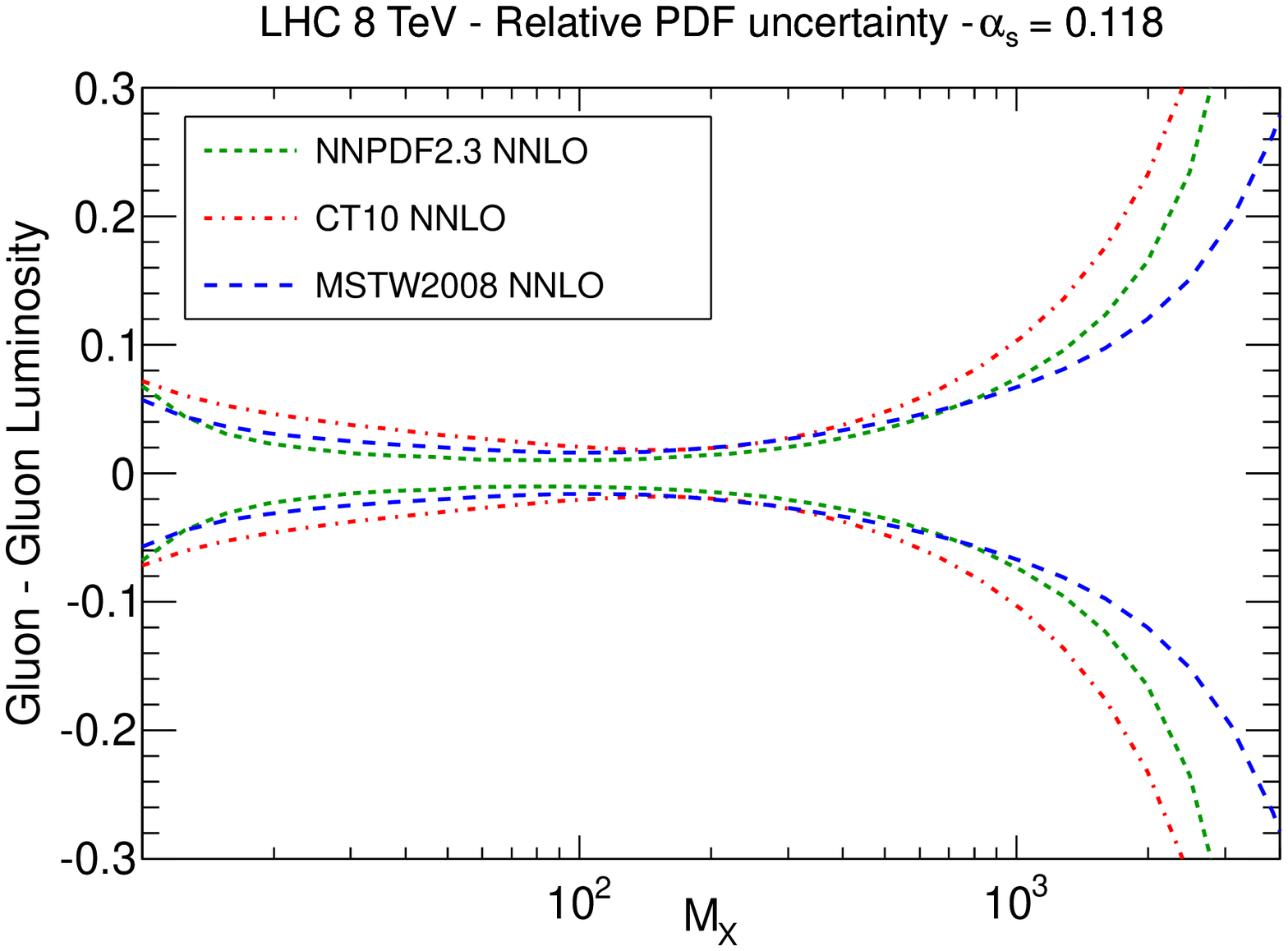}\quad
\includegraphics[width=0.48\textwidth]{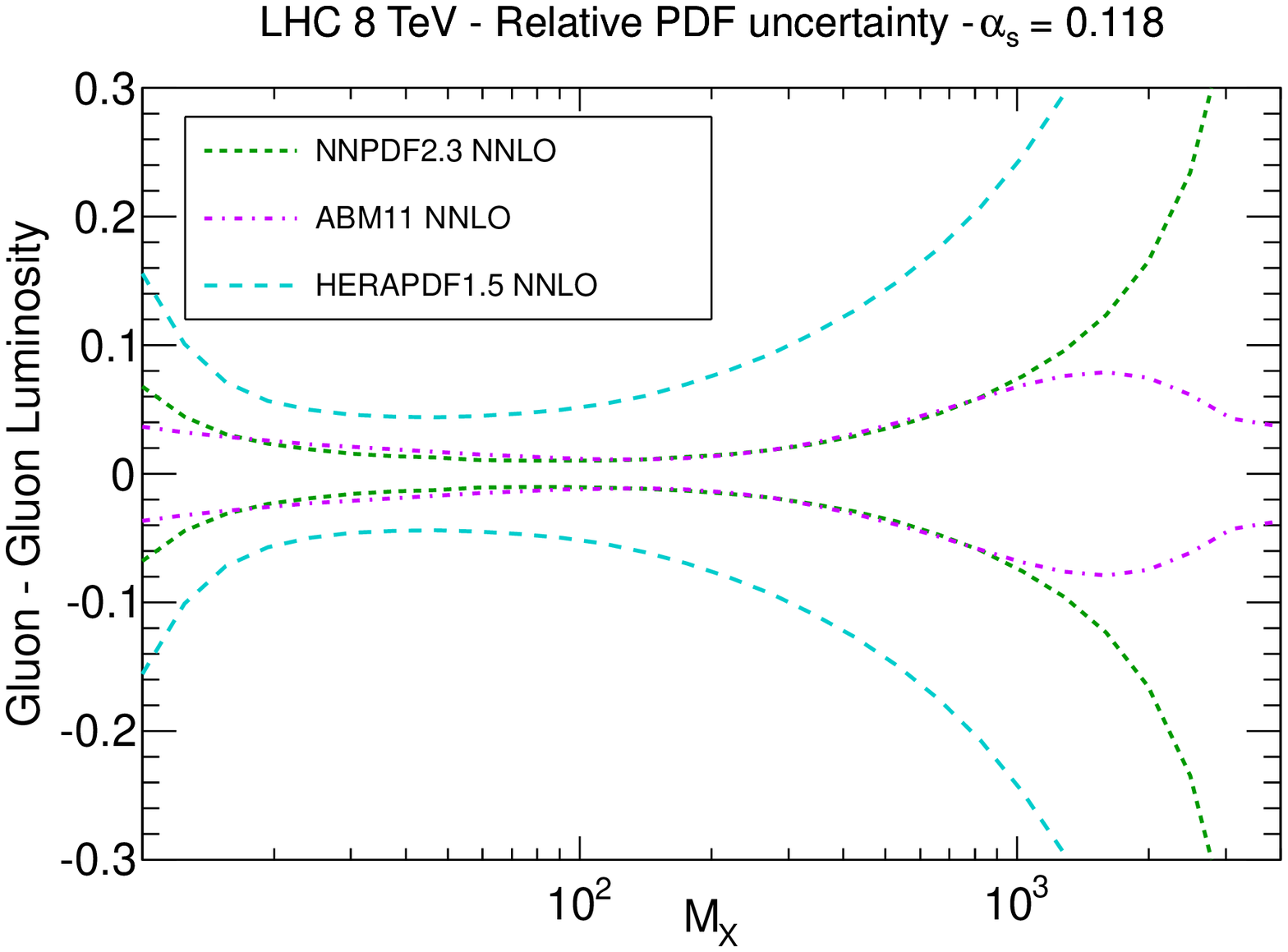}
      \end{center}
     \caption{\small
The relative PDF uncertainties in the quark-antiquark 
luminosity (upper plots)
and in the gluon-gluon luminosity (lower plots), 
for the production
of a final state of invariant mass $M_X$ (in GeV) at the LHC 8 TeV.  
All luminosities are computed at a common value of $\alpha_s=0.118$.
    \label{fig:PDFlumi-rel} }
\end{figure}
%%%%%%%%%%%%%%%%%%%%%%%%%%%%%%%%%%%%%%%%%%%%%%%%%%

It is also useful to compare the relative PDF uncertainties in the parton
luminosities. In  Fig.~\ref{fig:PDFlumi-rel}
we show this relative PDF uncertainty for the quark-antiquark
and gluon-gluon luminosities. Here we see clearly the much larger 
HERAPDF1.5 uncertainty. At high invariant mass, the uncertainty in the 
ABM11 gluon-gluon luminosity becomes smaller, despite the fact that
this is an extrapolation region due to the scarcity of experimental data.

The larger quark-antiquark luminosity from ABM11 as compared to
the other PDF sets could be inferred from the PDF comparison plots at 
lower $Q^2$: the ABM gluon is a little larger than the central value of the 
other groups below about $x=0.05$, and this drives more quark and 
antiquark evolution at small $x$ values. 
It has been recently 
suggested~\cite{Thorne:2012az}, based on 
results of a NLO fit to DIS data only, that some of these features
could be at least in part the consequence of the
ABM treatment of heavy quark contributions
(see also~\cite{CooperSarkar:2007ny}). 
Indeed, while CT, MSTW and NNPDF use  
a variable flavour number~scheme~\cite{Forte:2010ta,Thorne:2006qt,Guzzi:2011ew},
ABM11 uses a fixed flavour number scheme for heavy-quark PDFs. 
This may explain the increase in the medium-$x$ and 
small-$x$ light quarks and gluons, and the corresponding softer
large-$x$ gluon required by the momentum sum rule, found in the ABM
fits~\cite{Thorne:2012az}, though more studies would be required in
order to
conclusively establish this.

%%%%%%%%%%%%%%%%%%%%%%%%%%%%%%%%%
%%%%%%%%%%%%%%%%%%%%%%%%%%%%%%%%%
\begin{table}[h]
\centering
\begin{tabular}{|c|c|c|c|}
\hline
 & $Q_0^2$ [GeV$^2$]& $Q^2_{\rm min}$ [GeV$^2$] & $W^2_{\rm min}$ [GeV$^2$]\\
\hline
\hline
ABM11 & 9  & 2.5 & 3.24 \\
CT10 & 1.69 & 4.0 & 12.25 \\
HERAPDF1.5 & 1.9 & 3.5 & - \\
MSTW08 & 1  & 2.0 & 15.0\\
NNPDF2.3 & 2.0 & 3.0 & 12.5 \\
\hline
\end{tabular}
\caption{\small  Kinematical cuts in $Q^2$
and $W^2=Q^2\ \left(1/x-1\right)$ 
applied to DIS data in various PDF determinations. The scale $Q^2_0$
at which PDFs are parametrized is also shown.
For ABM11 there 
is also a maximum $Q^2\le  1000~{\rm GeV}^2$ cut. \label{tab:kincuts}}
\end{table}
%%%%%%%%%%%%%%%%%%%%%%%%%%%%%
%%%%%%%%%%%%%%%%%%%%%%%%%%%%%

As an alternative explanation,
a higher twist contribution has been invoked to
explain part of the differences between ABM11 and the other
PDF groups. 
While ABM fit a higher twist contribution, all groups minimize the
impact of higher twists by suitable kinematic cuts in $Q^2$ and 
$W^2=Q^2 \left(1/x-1\right)$. 
The HERAPDF fit includes 
no data at low $W^2$, so that no cut is required.
In addition,
NNPDF2.3 includes exactly kinematical target mass 
corrections~\cite{Ball:2008by},
known to be a substantial part of the higher twist corrections. 

The  kinematical cuts $Q^2_{\rm min}$
and $W^2_{\rm min}$ applied to the fitted DIS data sets
are summarized for each group in
Table~\ref{tab:kincuts} (the value of the scale $Q^2_0$ where the
PDFs are parametrized is also shown for completeness).
It should be observed that the ABM11 fit also
imposes an upper cut $Q^2_{\rm max}=10^3$ GeV$^2$ on the HERA data. 
Stability under variation of the 
default MSTW08 kinematical cuts was studied in Ref.~\cite{Thorne:2011kq}.
The inclusion of higher twists in MRST fits has previously been shown 
to lead to only a small
effect on high-$Q^2$ PDFs~\cite{Martin:1998np,Martin:2003sk}, and an ongoing 
extension of the study in~\cite{Thorne:2012az}  suggests this is 
qualitatively the same with more up-to-date PDFs. This conclusion has 
been confirmed in similar studies by NNPDF.

\clearpage
\section{LHC inclusive cross sections}
\label{sec:LHCincl}

In this section we compute inclusive cross sections at 8 TeV
for various benchmark processes and compare the results
for all NNLO PDF sets.
 We consider electroweak gauge boson production,
top quark pair production and Higgs boson production 
in various channels. We will provide results
for $\alpha_s=0.117$ and $\alpha_s=0.119$. 
The Higgs case is discussed in more detail in Sect.~\ref{sec:higgs}, 
together with the interplay between the PDF and $\alpha_s$ 
uncertainties. The comparisons of data and theory predictions for 7 TeV inclusive cross sections has been discussed
in detail in previous benchmark studies~\cite{Watt:2011kp,Watt:2012np}, 
Similar comparisons, but regarding various differential distributions, 
will be discussed in the next section.

For these inclusive benchmark cross sections, we use
the following codes and settings:
\begin{itemize}
\item Higgs boson production cross sections in the gluon fusion
  channel have been computed at NNLO with the {\tt iHixs} code~\cite{Anastasiou:2011pi}.  The central renormalization and
factorization scales have been taken to be
  $\mu_F=\mu_R=m_H$. This is the same choice used for the default
predictions for Higgs production in gluon fusion adopted by
the Higgs Cross Section Working Groups~\cite{Dittmaier:2011ti}. 
In all the Higgs production
cross sections, we take $m_H=125$ GeV.
\item Higgs production in the Vector Boson Fusion (VBF) channel
has been computed at NNLO 
with the {\tt VBF@NNLO} code~\cite{Bolzoni:2010xr}, with the
scale choice  $\mu_F=\mu_R=m_H$.
\item Higgs production in association with $W$ and $Z$ bosons
has been computed at NNLO with the {\tt VH@NNLO} program~\cite{Brein:2003wg,Brein:2012ne}. Also in this case the scale choice is
 $\mu_F=\mu_R=m_H$.
\item Higgs production in association with
a top quark pair, $t\bar{t}H$, has been computed at LO with the
{\tt MCFM} program~\cite{Campbell:2002tg}. Here the scale
choice is  $\mu_F=\mu_R=2m_t+m_H$.
\item Electroweak gauge boson production has been computed at NNLO
  using the {\tt Vrap} code~\cite{Anastasiou:2003ds}. The central
  scale choice is $\mu_R=\mu_F=M_V$.
\item Top quark pair production has been computed at NNLO$_{\rm approx}$+NNLL with the
  {\tt top++} code~\cite{Czakon:2011xx}, including
    the latest development of the calculation of the complete NNLO
    corrections to the $q\bar{q}\to t\bar{t}$ production, documented
  in~\cite{Baernreuther:2012ws}, as implemented
in {\tt v1.3}. The factorization and renormalization scales
have been set to  $\mu_R=\mu_F=m_t$.  The settings of the theoretical
    calculations are the default ones in Ref.~\cite{Cacciari:2011hy}.
    In all calculations we use $m_t=173.2$ GeV.
\end{itemize}

Let us emphasize that in this work we consider only
PDF uncertainties, and it is beyond the scope of this paper
to provide a careful assessment of all
relevant theoretical uncertainties into consideration for
each of the studied processes.Before any strong 
statements can be made
 about the constraining 
power of various experimental data to discriminate between PDF sets, 
 relevant theoretical uncertainties should be properly included.

\begin{figure}[ht!]
\centering
\epsfig{width=0.47\textwidth,figure=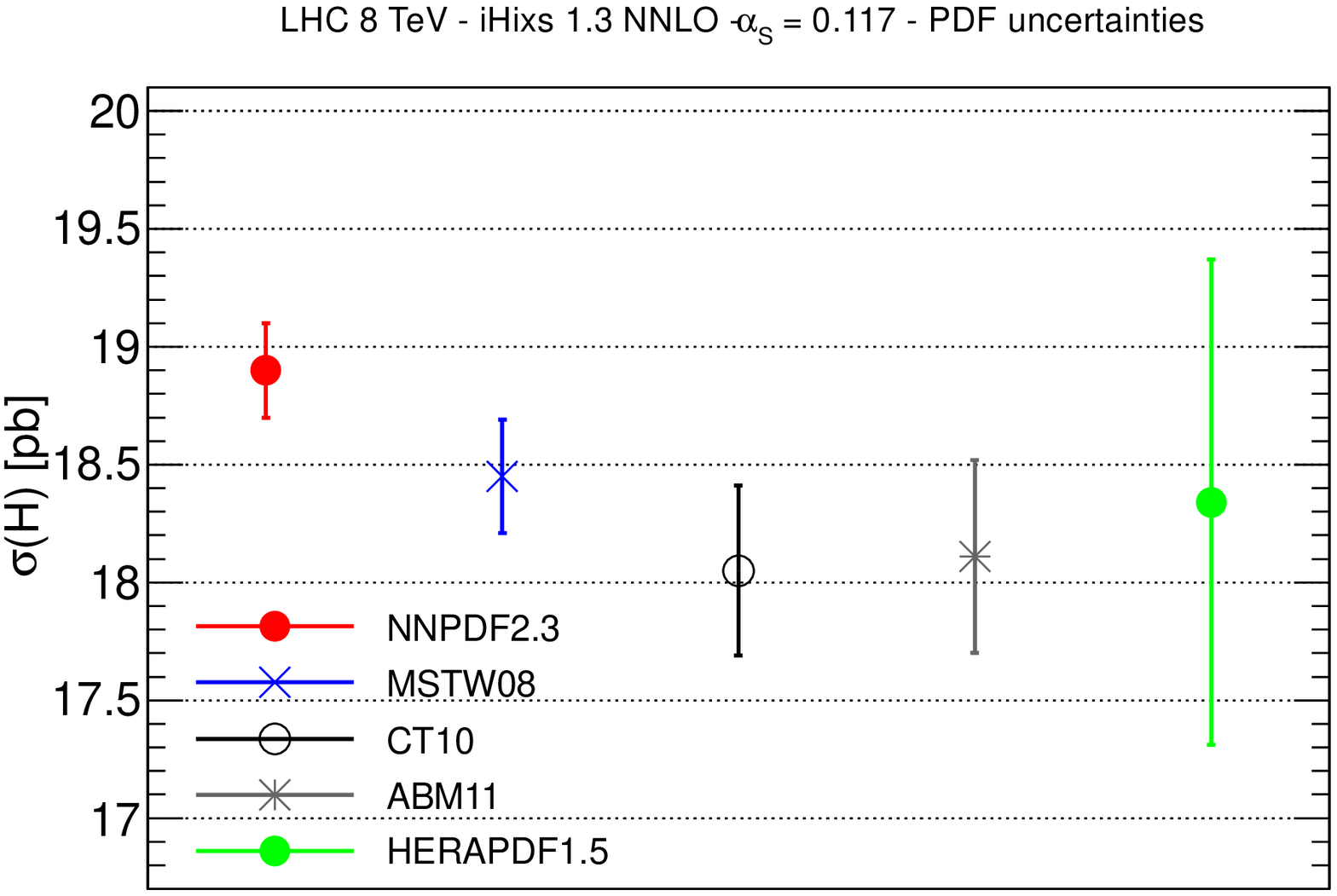}\quad
\epsfig{width=0.47\textwidth,figure=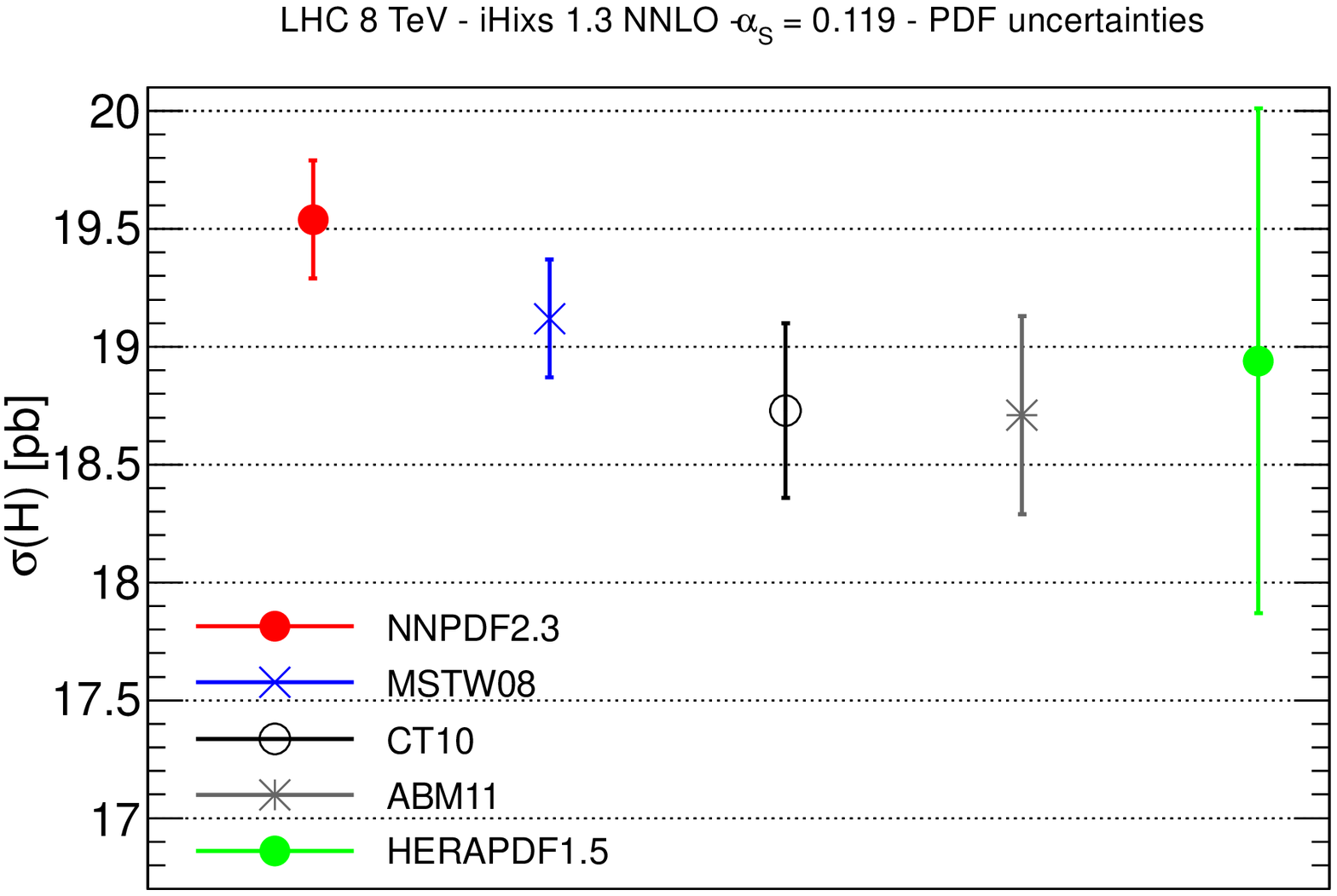}
\epsfig{width=0.47\textwidth,figure=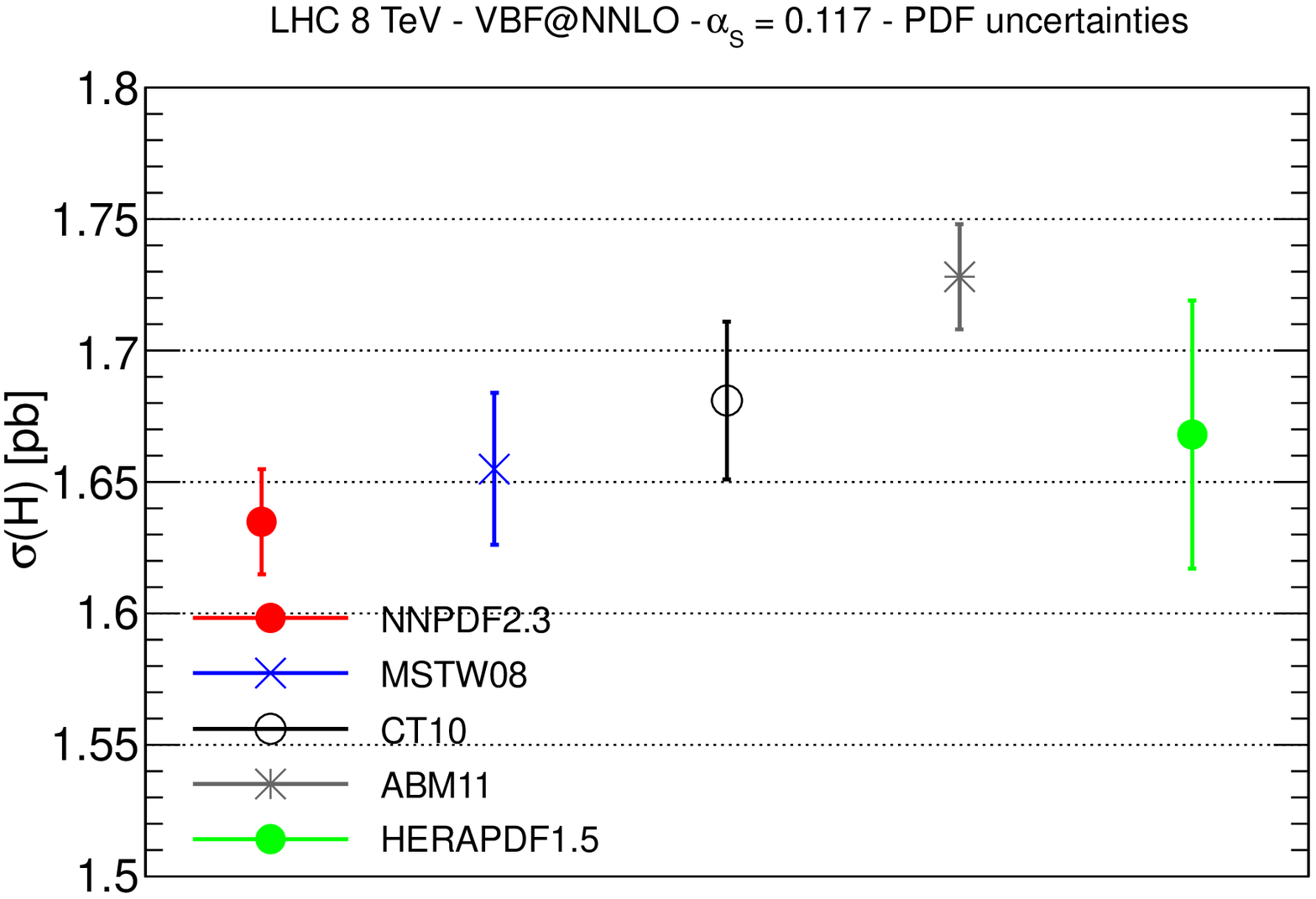}\quad
\epsfig{width=0.47\textwidth,figure=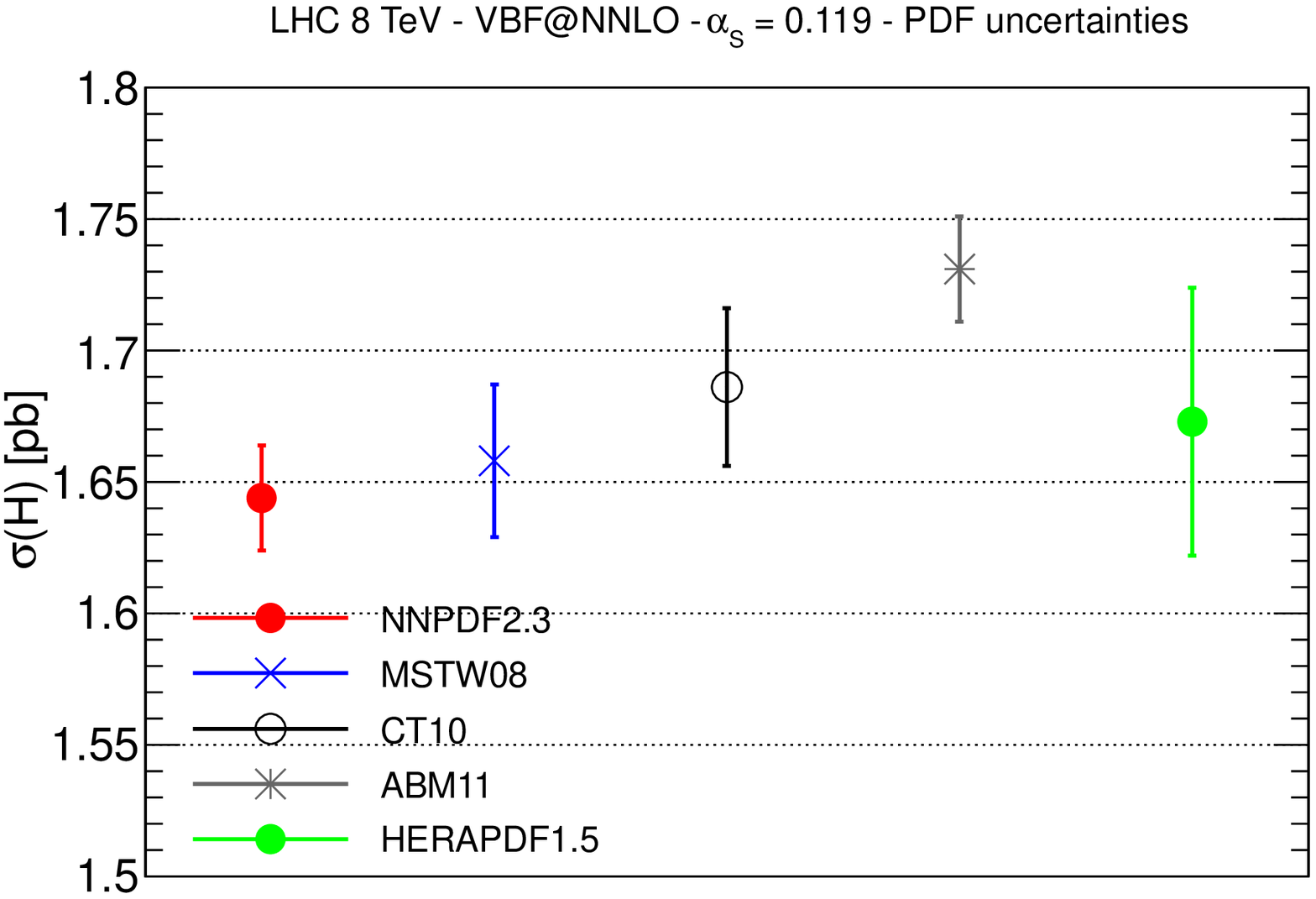}
\epsfig{width=0.47\textwidth,figure=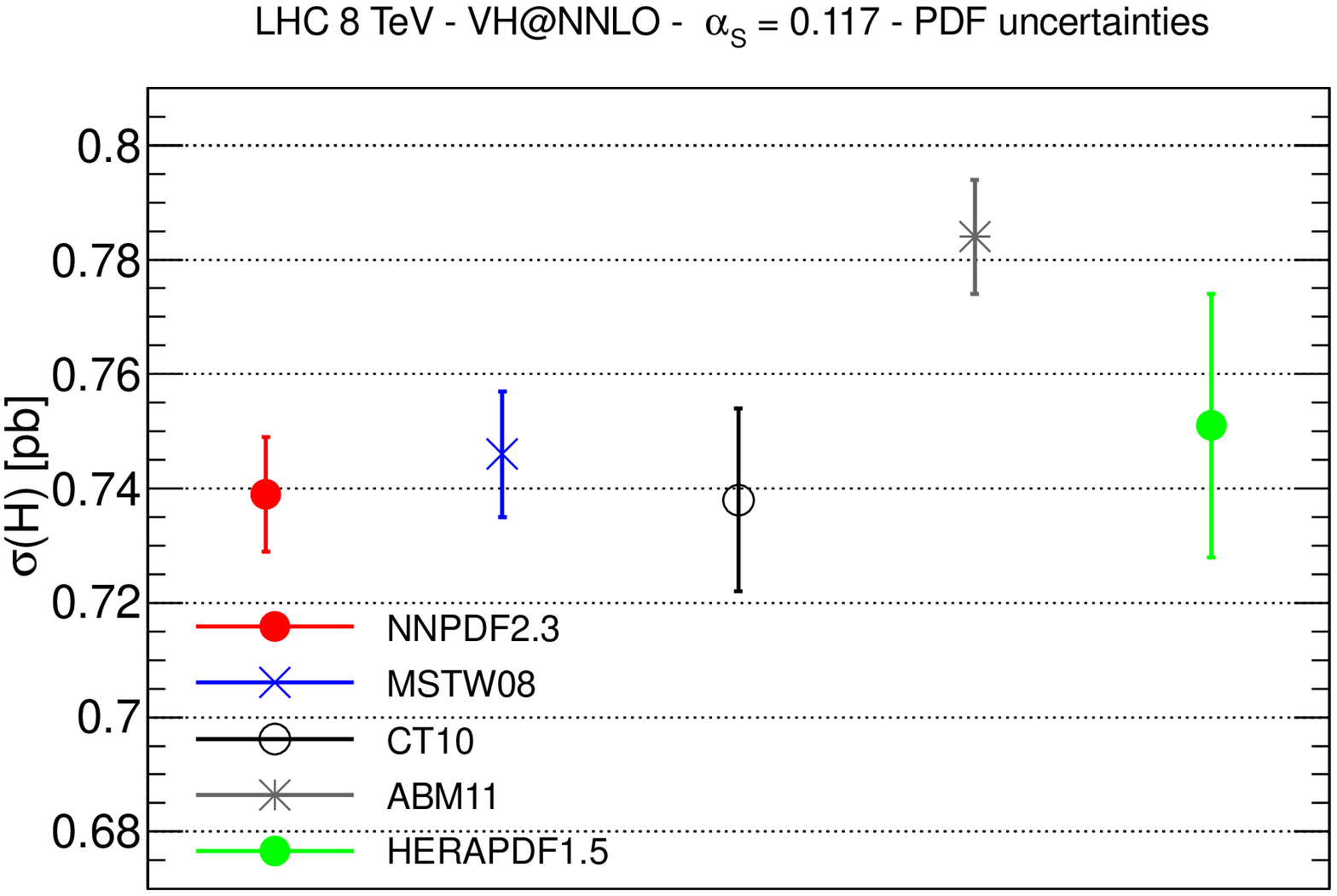}\quad
\epsfig{width=0.47\textwidth,figure=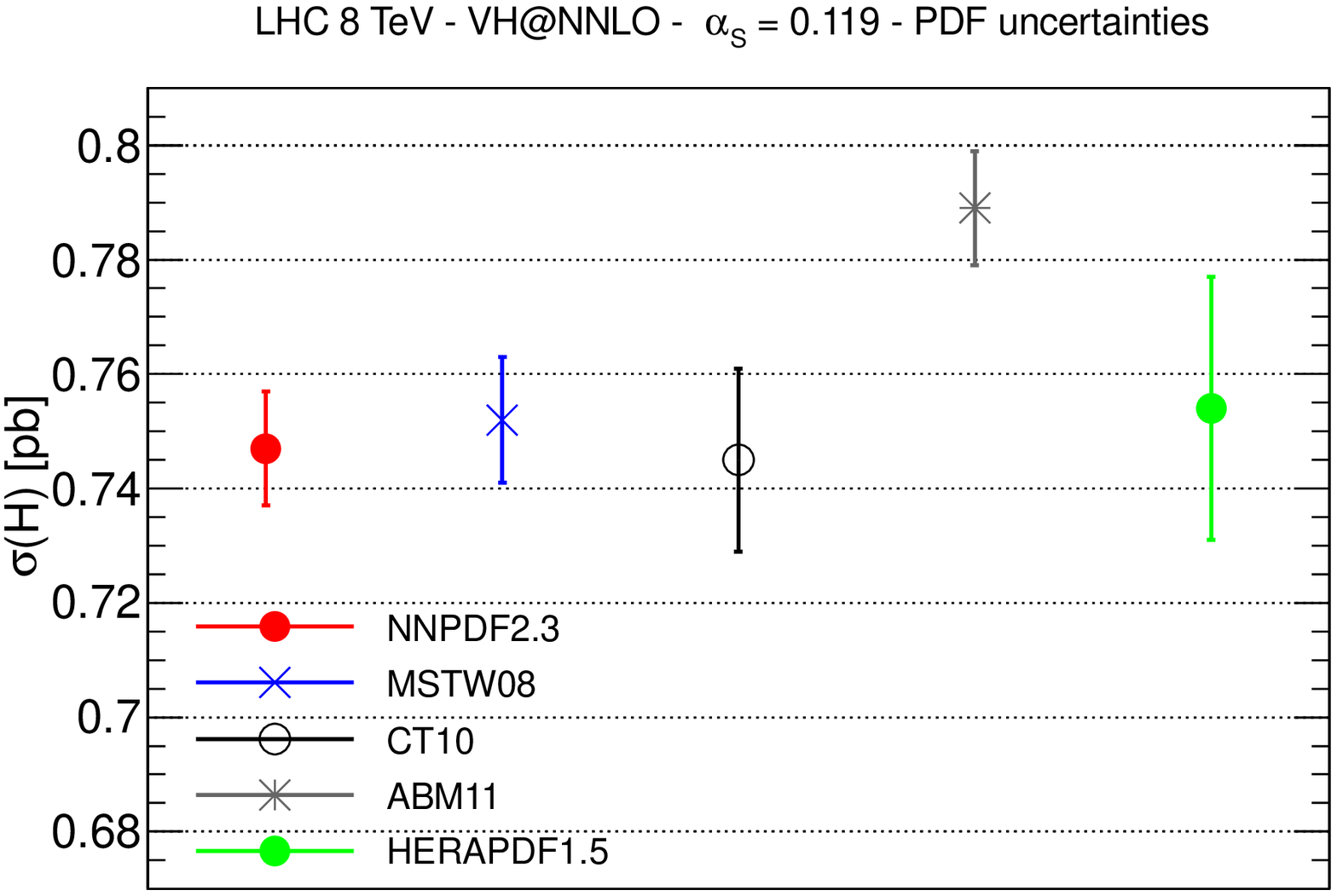}
\epsfig{width=0.47\textwidth,figure=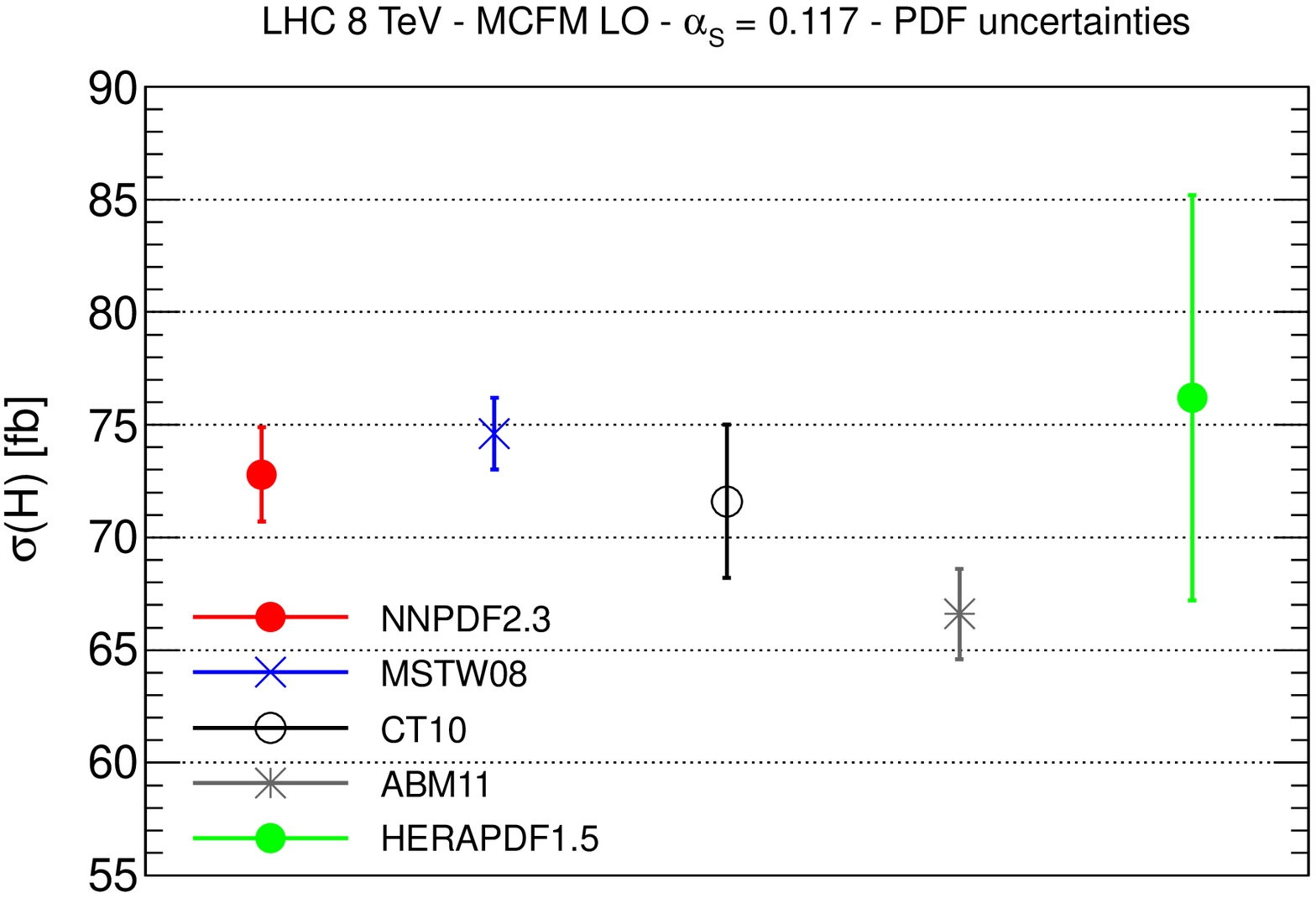}\quad
\epsfig{width=0.47\textwidth,figure=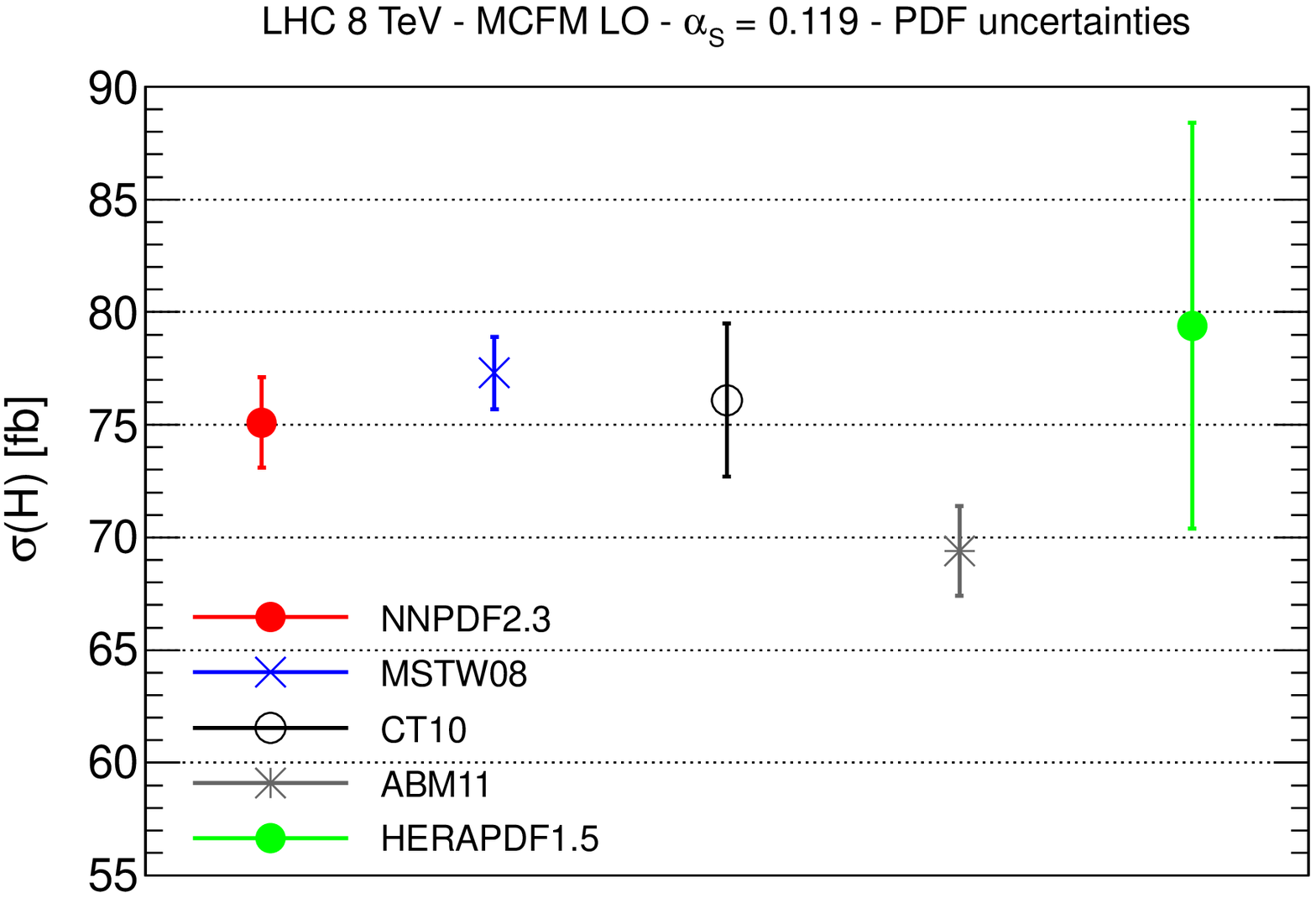}
\caption{\small Comparison of the 
predictions for the LHC Standard Model Higgs boson
cross sections at 8 TeV obtained using
various NNLO PDF sets. From top to bottom we show gluon fusion, vector 
boson fusion, associated production (with $W$), and associated 
production with a $t\bar{t}$ pair.
The left hand plots show 
results for $\alpha_S(M_Z)=0.117$, while on the right we have 
$\alpha_S(M_Z)=0.119$.
\label{fig:8tev-higgs}}
\end{figure}

We begin with the Higgs production cross sections. 
Results at 8 TeV for all relevant production channels and 
different PDF sets and $\alpha_S(M_Z)$ values have been collected 
in Table~\ref{tab:higgs}.
In all cases the same value of $\alpha_S$ is used consistently in
both the PDFs and in the matrix element calculation. Results
are also represented graphically in 
Fig.~\ref{fig:8tev-higgs}.  Note that the error
bands shown correspond to the PDF uncertainty only, with the exception 
for ABM11 and, to a lesser extent, HERAPDF.

%%%%%%%%%%%%%%%
%%%%%%%%%%%%%%%%%%%%%%%%%%%%%%%%%%%%%%%%%%%%%%%%%
\begin{table}[ht!]
  \centering
  \footnotesize
 \begin{tabular}{c||c|c|c|c|c}
 \hline
 \multicolumn{6}{c}{Gluon Fusion (pb)}\\
 \hline
 $\alpha_S(M_Z)$ & NNPDF2.3 & MSTW08& CT10& ABM11 & HERAPDF1.5 \\
 \hline
 \hline
0.117         &  18.90 $\pm$ 0.20  &  18.45  $\pm$ 0.24    &  18.05 $\pm$ 0.36  &  18.11 $\pm$ 0.41    &  18.34  $\pm$ 1.03 \\ 
0.119         &  19.54 $\pm$ 0.25   &  19.12 $\pm$ 0.25   &   18.73 $\pm$ 0.37 &  18.71 $\pm$ 0.42  & 18.94 $\pm$ 1.07   \\
 \hline
\multicolumn{6}{c}{}\\
 \end{tabular}  
 \begin{tabular}{c||c|c|c|c|c}
 \hline
\multicolumn{6}{c}{Vector Boson Fusion (pb)}\\
 \hline
 $\alpha_S(M_Z)$ & NNPDF2.3 & MSTW08& CT10& ABM11 & HERAPDF1.5 \\
 \hline
 \hline
0.117         &   1.635 $\pm$ 0.020   &  1.655   $\pm$ 0.029     &  1.681 $\pm$ 0.030   & 1.728 $\pm$ 0.020     &  1.668 $\pm$ 0.051  \\ 
0.119         &  1.644 $\pm$ 0.020    &  1.658 $\pm$ 0.029    &     1.686$\pm$ 0.030 &  1.731  $\pm$ 0.020  &  1.673 $\pm$  0.051  \\
 \hline
\multicolumn{6}{c}{}\\
 \end{tabular}  
\begin{tabular}{c||c|c|c|c|c}
 \hline
\multicolumn{6}{c}{$WH$ production (pb)}\\
 \hline
 $\alpha_S(M_Z)$ & NNPDF2.3 & MSTW08& CT10& ABM11 & HERAPDF1.5 \\
 \hline
 \hline
0.117         & 0.739  $\pm$ 0.010  &  0.746   $\pm$ 0.011     &   0.738 $\pm$  0.016 &  0.784  $\pm$ 0.010     & 0.751  $\pm$ 0.023  \\ 
0.119         &  0.747 $\pm$ 0.010   & 0.752  $\pm$ 0.011   &  0.745  $\pm$ 0.016 &   0.789 $\pm$  0.010 &  0.754 $\pm$ 0.023   \\
 \hline
\multicolumn{6}{c}{}\\
 \end{tabular}  
 \begin{tabular}{c||c|c|c|c|c}
 \hline
\multicolumn{6}{c}{$t\bar{t}H$ associated production (fb)}\\
 \hline
 $\alpha_S(M_Z)$ & NNPDF2.3 & MSTW08& CT10& ABM11 & HERAPDF1.5 \\
 \hline
 \hline
0.117         & 72.8  $\pm$ 2.1  &   74.6  $\pm$ 1.6     &  71.6 $\pm$ 3.4    &  66.6 $\pm$ 2.0  &  76.2 $\pm$ 9.0    \\ 
0.119         &  75.1 $\pm$ 2.0    &  77.3 $\pm$ 1.6    &  76.1  $\pm$   3.4  & 69.4 $\pm$ 2.0   & 79.4 $\pm$ 9.0     \\
 \hline
 \end{tabular} 
  \caption{\label{tab:higgs}  
\small The cross sections for Higgs production
at 8 TeV in various channels
using the settings described in the text. From top to
bottom: gluon fusion, vector boson fusion,
$WH$ production and $t\bar{t}H$ production. 
We have assumed a Standard Model Higgs boson with mass $m_H=125$ GeV.
We show the results for two different values of $\alpha_S(M_Z)$, 
0.117 and 0.119.
}
\end{table}
%%%%%%%%%%%%%%%%%%%%%%%%%%%%%%%%%%%%%%%%%%%

%%%%%%%%%%%%%%%%%%%%%%%%%%%%%%%%%%%%%%%%%%
%%%%%%%%%%%%%%%%%%%%%%%%%%%%%%%%%%%%%%%%%

The main features which emerge from the plots are the following:
\begin{itemize}
\item The relative sizes of the cross sections obtained using different
  PDF sets are almost independent of $\alpha_s$:  when $\alpha_s$ is
  varied all cross sections get rescaled by a comparable amount.
\item The ABM11 and HERAPDF1.5 central predictions for
gluon fusion are contained within the
envelope of the NNPDF2.3, CT10 and MSTW results.
However, the HERAPDF1.5 uncertainty is bigger than this envelope. 
The agreement with ABM11 would be 
spoiled if their default  
value of $\alpha_s(M_Z)=0.1134$ were used.  
\item For VBF, $WH$ and $t\bar{t}H$ production, there is
a reasonable agreement between CT10, MSTW and NNPDF2.3 both
in central values and in the size of PDF uncertainties. 
 ABM11 instead leads to rather different results, even when
a common value of $\alpha_s$ is used. 
For quark-initiated processes, like VBF and $WH$, the ABM11 cross section
is higher than that of the other sets, especially for $WH$ production.
For $t\bar{t}H$, which  receives the largest contribution from 
gluon-initiated diagrams, the
ABM11 cross section is smaller.
\item The HERAPDF1.5 PDF uncertainties are distinctly larger, 
especially for $ggH$ and $t \bar tH$, mostly due  to fact that HERA data do
not 
constrain well the large-$x$ gluon.
\end{itemize}
A more detailed discussion of the interplay of PDF and $\alpha_s$
uncertainties for Higgs production, focused on the
gluon fusion channel, will be presented in Sect.~\ref{sec:higgs} below.

An interesting result from  Table~\ref{tab:higgs} is that 
the CT10 and NNPDF2.3 prediction for Higgs production
via gluon fusion do not agree within the respective
1--sigma errors, with MSTW lying in between.
It is not clear to the authors which is the origin
of this discrepancy. 
It could be related to differences in the gluon
parametrization, different datasets, or differences in the
statistical methodology.
On purely statistically grounds, some discrepancy at
the one or two sigma level is not surprising, given different
 data sets and methodologies, and 
in spite of the unfortunate location of the discrepancy at the phenomenologically important mass of $m_H$=125 GeV.

%%%%%%%%%%%%%%%
%%%%%%%%%%%%%%%%%%%%%%%%%%%%%%%%%%%%%%%%%%%%%%%%%
\begin{table}
  \centering
  \footnotesize
 \begin{tabular}{c||c|c|c|c|c}
 \hline
\multicolumn{6}{c}{$t\bar{t}$ production (pb)}\\
 \hline
 $\alpha_S(M_Z)$ & NNPDF2.3 & MSTW08& CT10& ABM11 & HERAPDF1.5 \\
 \hline
 \hline
0.117         & 217.9 $\pm$ 4.8  &   222.5 $\pm$ 5.5 & 218.0 $\pm$ 7.8  & 199.7 $\pm$ 5.5  & 225.1 $\pm$ 26.1  \\ 
0.119         & 227.8 $\pm$ 5.0  & 232.1 $\pm$ 5.8   & 227.6  $\pm$ 8.2    &   211.2 $\pm$ 5.8  & 237.5 $\pm$ 27.5  \\ 
 \hline
 \end{tabular}  
  \caption{\label{tab:top}  
\small Same as Tab.~\ref{tab:higgs}  for the cross sections for top quark
pair production
at 8 TeV at NNLO$_{\rm approx}$+NNLL, using {\tt top++}
with the settings described in the text.
We have assumed a top quark mass of $m_t=173.2$ GeV.
}
\end{table}
%%%%%%%%%%%%%%%%%%%%%%%%%%%%%%%%%%%%%%%%%%%

%%%%%%%%%%%%%%%%%%%%%%%%%%%%%%%%%%%%%%%%%%
\begin{figure}[ht!]
\centering
\epsfig{width=0.47\textwidth,figure=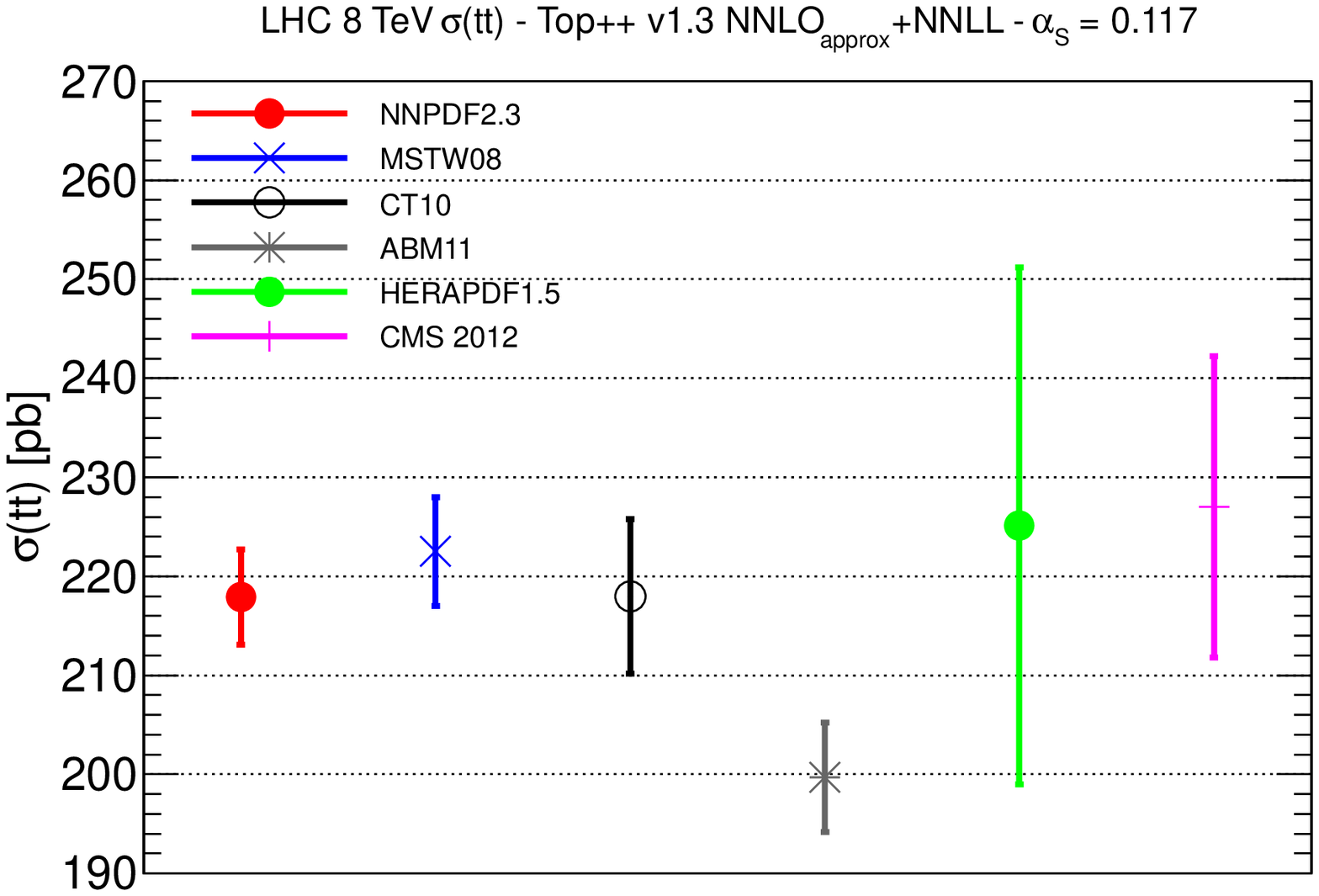}\quad
\epsfig{width=0.47\textwidth,figure=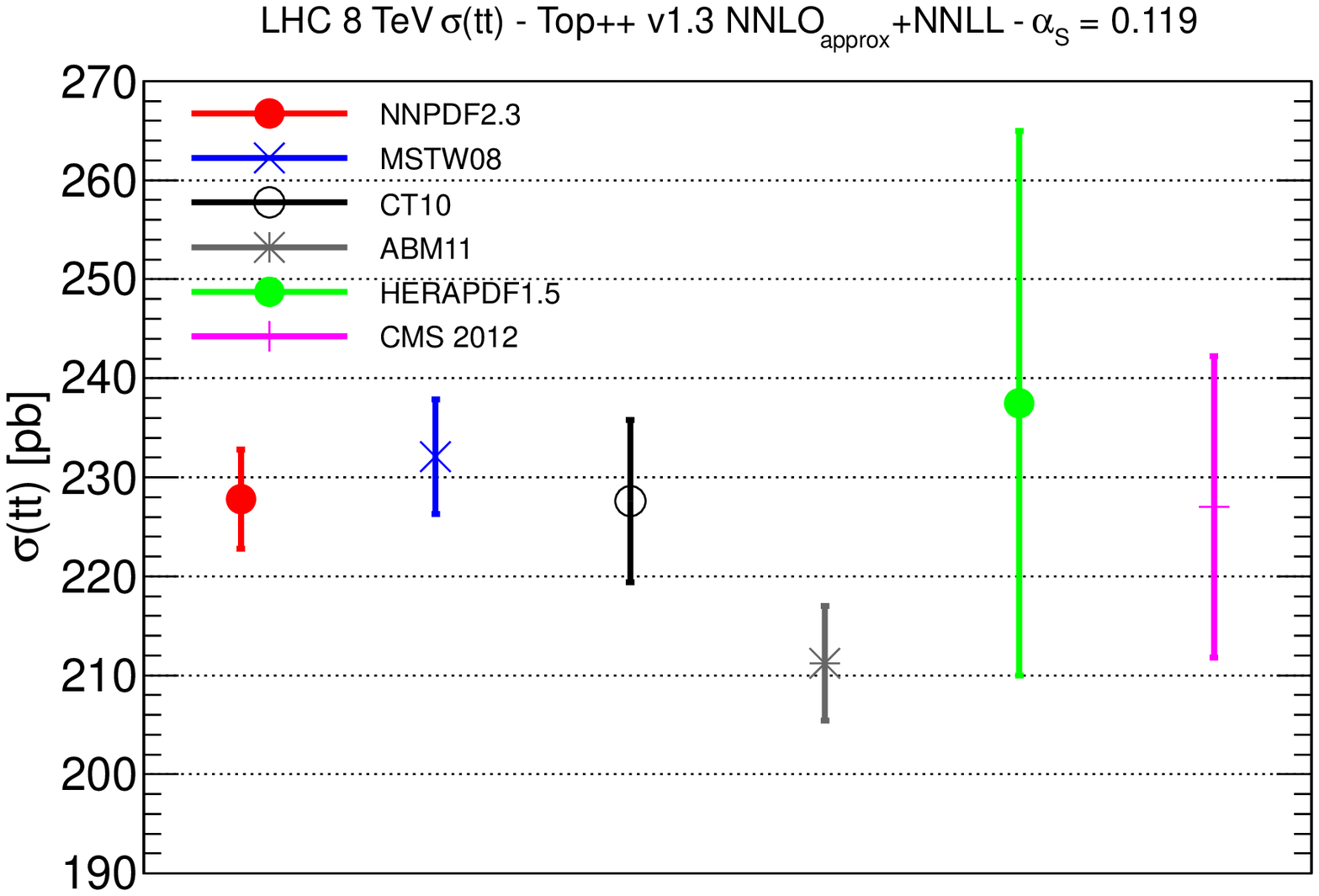}
\caption{\small Comparison of the 
predictions for the top quark
pair production at LHC 8 TeV obtained using
various NNLO PDF sets. Left plot: results for $\alpha_S(M_Z)=0.117$.
Right plot: results for $\alpha_S(M_Z)=0.119$. 
We also
show the recent CMS 8 TeV measurements.
\label{fig:8tev-ttbar}}
\end{figure}
%%%%%%%%%%%%%%%%%%%%%%%%%%%%%%%%%%%%%%%%%

Next we consider 
inclusive top quark pair production. 
Theoretical progress towards the full NNLO result has
been made recently~\cite{Cacciari:2011hy,Baernreuther:2012ws,Czakon:2012pz,Czakon:2012zr,Aliev:2010zk,Moch:2012mk}, 
including the recent calculation of the full NNLO $qg$ 
initiated contribution~\cite{Czakon:2012pz} (which amounts to a 
small ${\cal O}(1\%)$ correction, contrary to previous approximate estimates~\cite{Moch:2012mk}).
The approximate NNLO top quark pair production cross sections
at 8 TeV for different PDF sets and for different values
of $\alpha_S(M_Z)$ are been collected in Table~\ref{tab:top}.
In all cases the same value of $\alpha_S$ is used consistently in
the PDFs and in the matrix element calculation. Results
are also shown in Fig.~\ref{fig:8tev-ttbar}, and compared
to the recent CMS measurements~\cite{cmstop}\footnote{We take the average
of the cross section in the di-lepton and lepton+jets final states.}. 
The variation in the cross sections 
with $\alpha_s$ shows that the $t\bar{t}$ total
cross section has some sensitivity to the value of $\alpha_s$. 
This sensitivity has been recently used by CMS to provide the first ever
determination of  $\alpha_s$ from top
cross sections~\cite{cmstopas}.
For the $t\bar{t}$ cross section, we see a reasonable agreement
between NNPDF2.3, CT10 and MSTW,
while ABM11 is somewhat lower. Using the
default value of $\alpha_s=0.1134$ in ABM11 would make
the difference even more marked. The HERAPDF1.5 central 
value is in good agreement with the global fits but, as usual, the
PDF uncertainties are larger.

%%%%%%%%%%%%%%%
%%%%%%%%%%%%%%%%%%%%%%%%%%%%%%%%%%%%%%%%%%%%%%%%%
\begin{table}
  \centering
  \footnotesize
 \begin{tabular}{c||c|c|c|c|c}
 \hline
& \multicolumn{5}{|c}{$\sigma(W^+)$ (nb)} \\
 \hline
 $\alpha_S(M_Z)$ & NNPDF2.3 & MSTW08& CT10& ABM11 & HERAPDF1.5 \\
 \hline
 \hline
0.117         &  6.937 $\pm$ 0.097 & 6.967 $\pm$ 0.118     & 6.990 $\pm$ 0.150  & 7.419 $\pm$ 0.107  & 7.088 $\pm$ 0.189 \\ 
0.119         &  7.045 $\pm$ 0.094  & 7.072  $\pm$ 0.118   & 7.107  $\pm$  0.151  &  7.509  $\pm$  0.105 & 7.140  $\pm$ 0.191 \\ 
 \hline
\multicolumn{5}{c}{}\\
 \end{tabular}  
 \begin{tabular}{c||c|c|c|c|c}
 \hline
& \multicolumn{5}{|c}{$\sigma(W^-)$ (nb)} \\
 \hline
 $\alpha_S(M_Z)$ & NNPDF2.3 & MSTW08& CT10& ABM11 & HERAPDF1.5 \\
 \hline
 \hline
0.117         & 4.855 $\pm$ 0.058 &  4.945 $\pm$ 0.083     & 4.857 $\pm$ 0.111  & 5.073 $\pm$ 0.079  & 4.987 $\pm$ 0.117 \\ 
0.119         & 4.906 $\pm$ 0.061  & 5.004 $\pm$ 0.083  &  4.940 $\pm$ 0.112  &  5.136  $\pm$  0.078  & 5.027 $\pm$ 0.118 \\ 
 \hline
\multicolumn{5}{c}{}\\
 \end{tabular}  
 \begin{tabular}{c||c|c|c|c|c}
 \hline
& \multicolumn{5}{|c}{$\sigma(Z)$ (nb)} \\
 \hline
 $\alpha_S(M_Z)$ & NNPDF2.3 & MSTW08& CT10& ABM11 & HERAPDF1.5 \\
 \hline
 \hline
0.117         &  1.120 $\pm$ 0.013 &  1.128 $\pm$ 0.019     & 1.126 $\pm$ 0.024  & 1.179 $\pm$ 0.016  & 1.135 $\pm$ 0.033 \\ 
0.119         & 1.127 $\pm$ 0.013 &  1.141 $\pm$ 0.019   &   1.144 $\pm$ 0.019   & 1.192  $\pm$ 0.017  & 1.145 $\pm$ 0.033 \\ 
 \hline
\multicolumn{5}{c}{}\\
 \end{tabular}  
 \begin{tabular}{c||c|c|c|c|c}
 \hline
& \multicolumn{5}{|c}{$\sigma(W^+)/\sigma(W^-)$} \\
 \hline
 $\alpha_S(M_Z)$ & NNPDF2.3 & MSTW08& CT10& ABM11 & HERAPDF1.5 \\
 \hline
 \hline
0.117         &  1.429 $\pm$ 0.013 & 1.409 $\pm$ 0.011     & 1.439 $\pm$ 0.013  & 1.462 $\pm$ 0.015  & 1.421 $\pm$ 0.013 \\ 
0.119         & 1.436 $\pm$ 0.012  &  1.413 $\pm$ 0.011   & 1.439   $\pm$ 0.013   & 1.462   $\pm$ 0.015  & 1.420  $\pm$ 0.013 \\ 
 \hline
\multicolumn{5}{c}{}\\
 \end{tabular}  
 \begin{tabular}{c||c|c|c|c|c}
 \hline
& \multicolumn{5}{|c}{$\sigma(W)/\sigma(Z)$} \\
 \hline
 $\alpha_S(M_Z)$ & NNPDF2.3 & MSTW08& CT10& ABM11 & HERAPDF1.5 \\
 \hline
 \hline
0.117         &  10.523 $\pm$ 0.035 & 10.560 $\pm$ 0.018     & 10.521 $\pm$ 0.068  & 10.595 $\pm$ 0.024  & 10.639 $\pm$ 0.057 \\ 
0.119         & 10.604 $\pm$ 0.035  &  10.583 $\pm$ 0.018   & 10.532   $\pm$ 0.068   & 10.608  $\pm$ 0.024  & 10.626  $\pm$ 0.057 \\ 
 \hline
 \end{tabular}  
  \caption{\label{tab:ewk}  
\small The inclusive cross sections for electroweak
gauge boson production
at 8 TeV at NNLO using the {\tt Vrap} code, obtained using
various NNLO PDF sets, for different values of
$\alpha_s$.
 From top to bottom we show the results
for the $W^+$, $W^-$ and $Z$ total cross sections
and then for the $W^+/W^-$ and $W/Z$ cross section ratios.
}
\end{table}
%%%%%%%%%%%%%%%%%%%%%%%%%%%%%%%%%%%%%%%%%%%%%%%%%

%%%%%%%%%%%%%%%%%%%%%%%%%%%%%%%%%%%%%%%%%%
\begin{figure}[ht!]
\centering
\epsfig{width=0.47\textwidth,figure=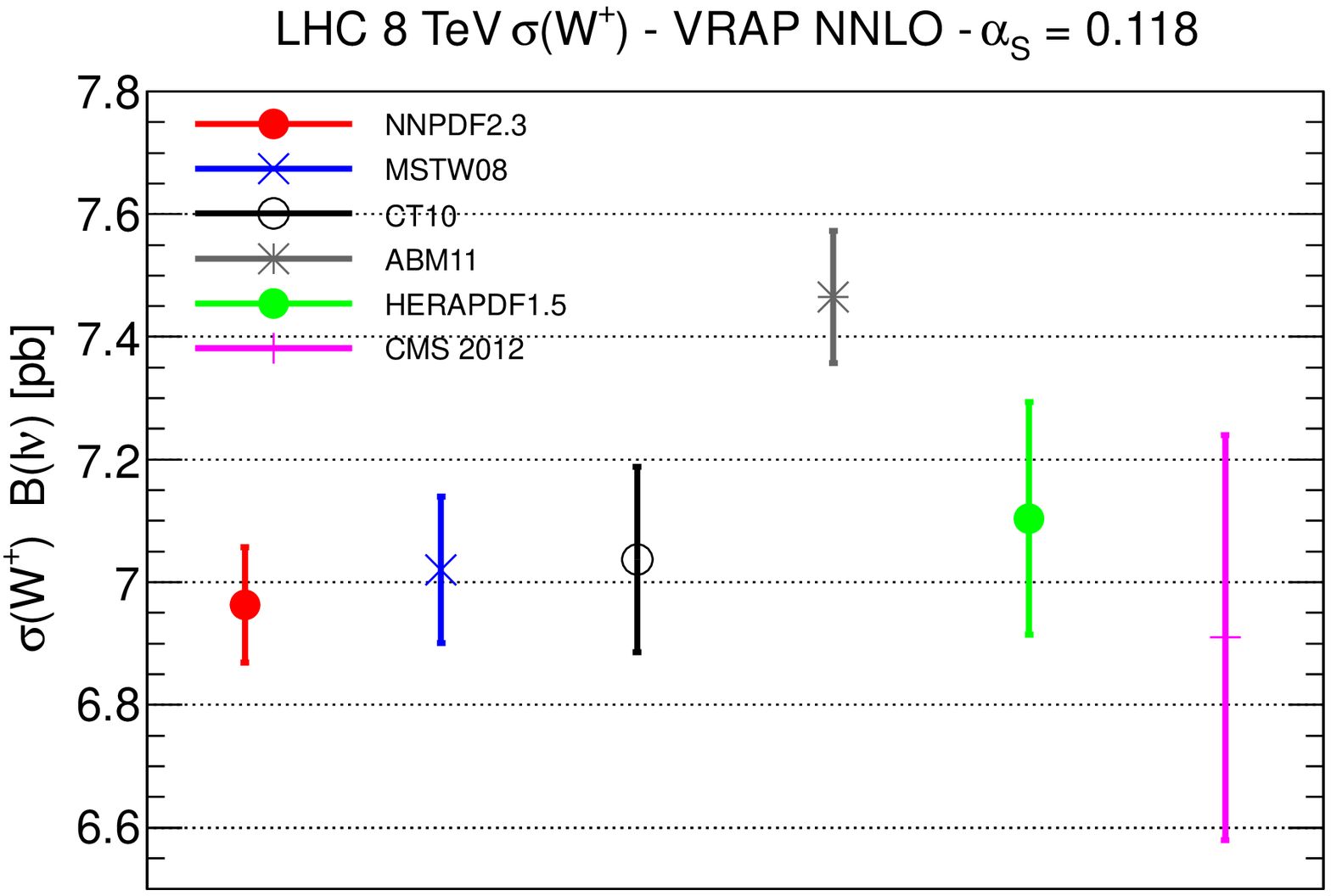}\quad
\epsfig{width=0.47\textwidth,figure=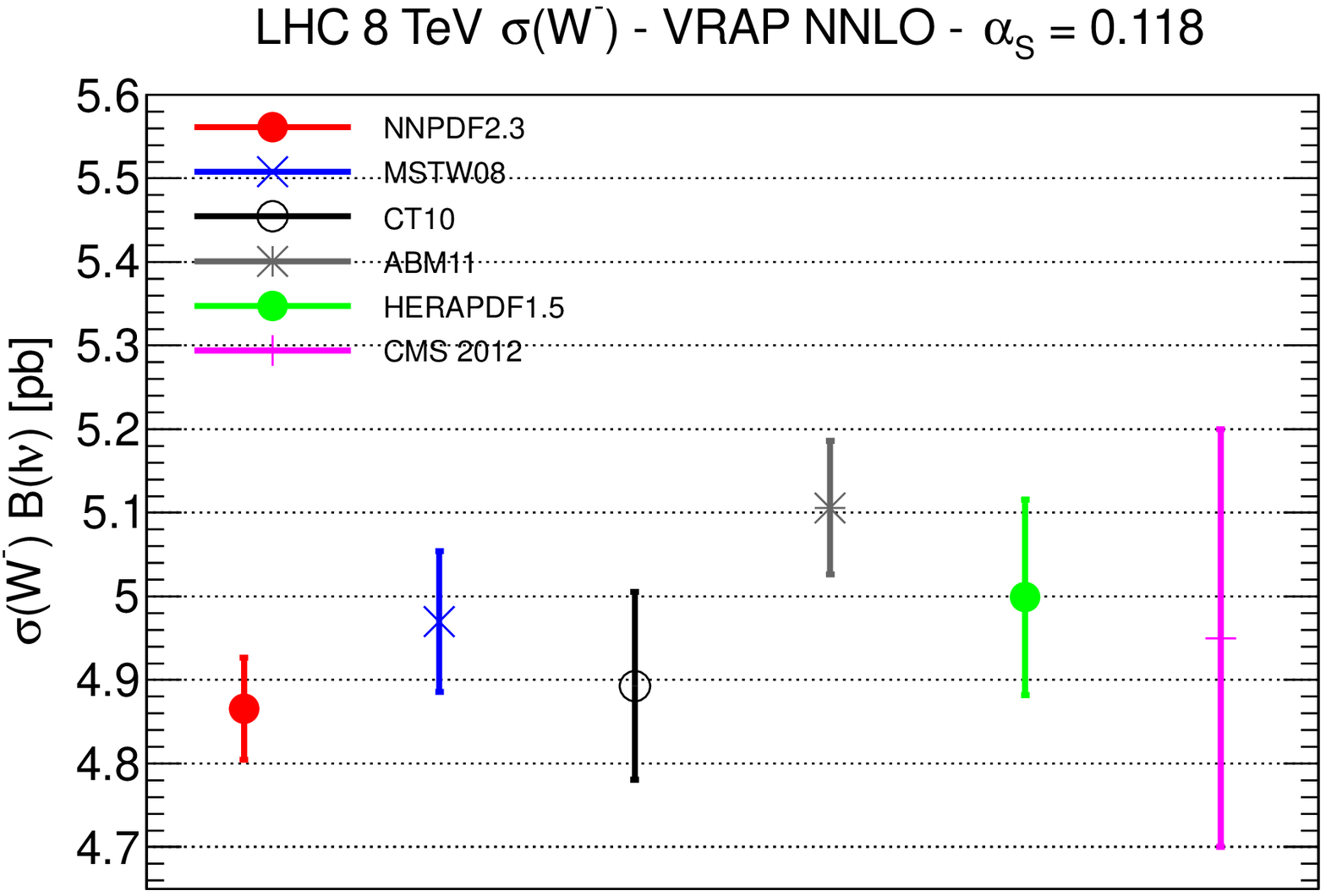}
\epsfig{width=0.47\textwidth,figure=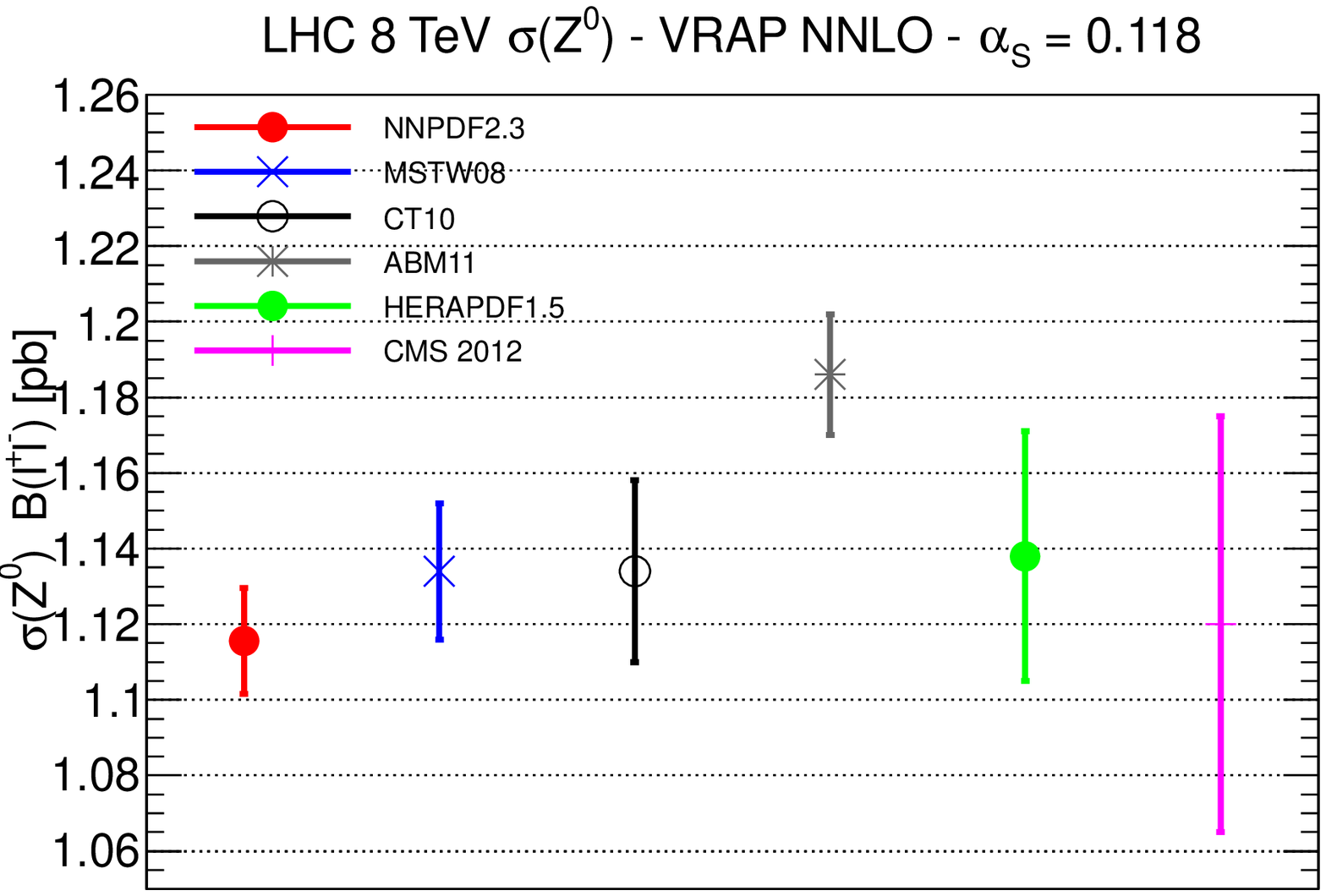}\\
\epsfig{width=0.47\textwidth,figure=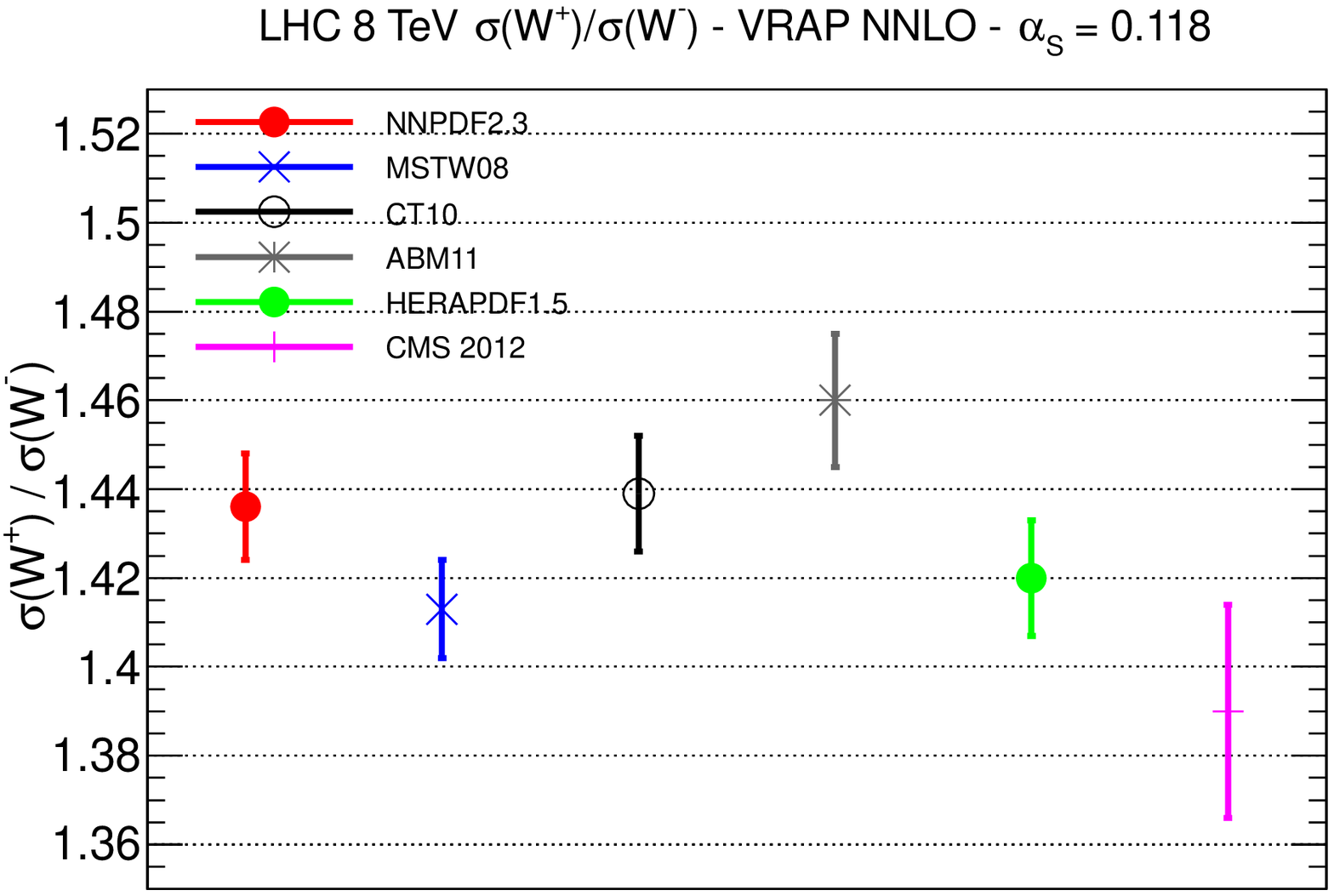}\quad
\epsfig{width=0.47\textwidth,figure=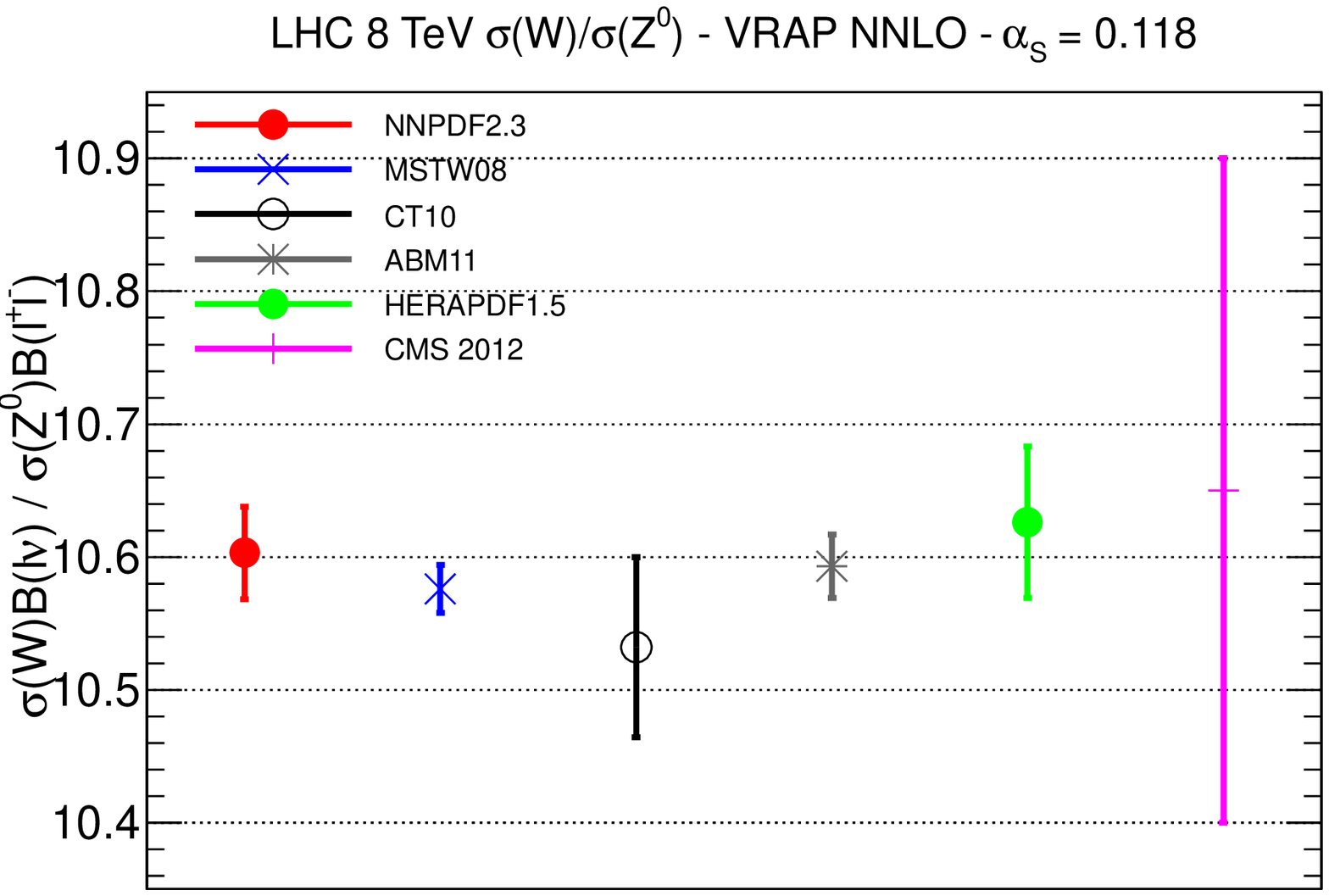}
\caption{\small Comparison of the 
predictions for inclusive cross sections for
electroweak gauge boson production
between different PDF sets at LHC 8 TeV. In all cases
the branching ratios to leptons have  been included. From top to
bottom and from left to right we show the 
$W^+$, $W^-$, and $Z$ inclusive cross sections,
and then the $W^+/W^-$ and 
$W/Z$ ratios. All cross sections are compared
at a common value of $\alpha_S(M_Z)=0.118$. We also
show the recent CMS 8 TeV measurements.
\label{fig:8tev-ewk}}
\end{figure}
%%%%%%%%%%%%%%%%%%%%%%%%%%%%%%%%%%%%%%%%%

Finally, we discuss the inclusive electroweak
gauge boson production at 8 TeV. Here we can also
compare with the recent CMS measurements~\cite{cmsewk}.
The cross section results for
$\alpha_s=0.117$ and $0.119$ 
are collected in Table~\ref{tab:ewk}, where
from top to bottom we show the results
for the $W^+$, $W^-$ and $Z$ total cross sections
and then for the $W^+/W^-$ and $W/Z$ cross section ratios.
Results are collected graphically and compared to the recent
CMS data in Fig.~\ref{fig:8tev-ewk}. In the figure we 
show results only for $\alpha_S(M_Z)=0.118$, since the strong coupling
dependence of these cross sections is rather mild, particularly for 
the cross section ratios.

We find good agreement between MSTW, CT10 and NNPDF2.3 and 
HERAPDF1.5: this is to be expected, since from Fig.~\ref{fig:PDFlumi-qq}  
we know that the respective $q\bar{q}$ parton luminosities are
similar in the relevant regions. On the other hand, ABM11 leads to 
systematically higher cross sections (particularly for the $u$-quark
dominated cross sections), consistent with 
the larger luminosities seen in Fig.~\ref{fig:PDFlumi-qq}. 
The available LHC 8 TeV data
is in good agreement with the theory predictions, perhaps
disfavoring the harder ABM11 cross sections, although the accuracy
is not enough for full discrimination. Future data for
lepton differential distributions at 8 TeV will be an
important ingredient for the next generation of PDF determinations.

%%%%%%%%%%%%%%%
%%%%%%%%%%%%%%%%%%%%%%%%%%%%%%%%%%%%%%%%%%%%%%%%%
\begin{table}[h]
  \centering
  \footnotesize
 \begin{tabular}{c||c}
 \hline
 \multicolumn{2}{c}{$\alpha_s(M_Z)=0.1134$}\\
 \hline
 Process &  ABM11  \\
 \hline
 \hline
$\sigma(gg\to H)$       &  17.01 $\pm$ 0.41 pb  \\
$\sigma(t\bar{t})$        &  181.4 $\pm$ 5.0 pb  \\
$\sigma(W^+)$       &  7.240 $\pm$ 0.104 nb  \\
$\sigma(W^-)$         &  4.944 $\pm$ 0.077 pb  \\
$\sigma(Z)$        &  1.151 $\pm$ 0.016 nb  \\
\hline
\end{tabular}
  \caption{\label{tab:xsecabm11}  
\small Benchmark cross sections 
at 8 TeV 
using the settings described as the cross
sections in in Tables~\ref{tab:higgs},~\ref{tab:top}  and~\ref{tab:ewk},  
but now for the ABM11 NNLO PDF set with the default value
of  $\alpha_S(M_Z)=0.1134$.
}
\end{table}
%%%%%%%%%%%%%%%%%%%%%%%%%%%%%%%%%%%%%%%%%%%

While in the previous discussion we have compared predictions
for $\alpha_s(M_Z)$ values close to the PDG average, often
PDF sets are used together with their default $\alpha_s(M_Z)$ values.
The only case where this difference is significant is for ABM11,
since the default value $\alpha_s(M_Z)=0.1134$ is not close
to the values explored above. Therefore, in Table~\ref{tab:xsecabm11} 
we collect some of the ABM11 NNLO benchmark cross sections,
but this time with the default $\alpha_s(M_Z)$ value. 
As is clear by comparing with the results in Tables~\ref{tab:higgs},~\ref{tab:top}  and~\ref{tab:ewk}, using this default value increases the difference
between ABM11 and the other PDF sets for Higgs production via gluon
fusion and for top quark production (predominantly via gluon
fusion at the LHC), whose cross sections are also sensitive
to the value of $\alpha_s$, while it brings ABM11 closer to the other PDF sets
(and to the CMS data) for the electroweak boson production
cross sections.

\clearpage
\section{PDF dependence of LHC differential distributions}
\label{sec:LHCdist}

We now study the PDF dependence of LHC differential distributions.
Since we want to quantify the agreement between data and theory, 
we consider only the LHC data sets for which the the full experimental
covariance matrix is available. These  were all taken at 7 TeV centre 
of mass energy: the 8 TeV data on differential
distributions have yet to be released. We 
will provide a comparison of theory and data for electroweak
vector boson and inclusive jet production, and examine whether these
data can discriminate between the PDFs.\footnote{In addition 
to these sets, ATLAS data on differential top quark pair
production have been recently presented~\cite{top:2012hg}.
They include the experimental covariance matrix, hence they could
be included in global PDF fits to constrain the gluon PDF.
We do not consider inclusive photon production, since the covariance matrix
is not available. The impact of the photon data on the 
PDF analysis was studied in Ref.~\cite{d'Enterria:2012yj}.
}
In the next section we  will present a more detailed  study of
jet production, including comparison between different codes, a
discussion of scale dependence, and a study of systematic shifts for
each PDF set in the description of ATLAS data. We will also
provide comparisons for the Tevatron Run II jet production
experiments, updating hence the analysis of Ref.~\cite{Thorne:2011kq},
based on previous PDF sets.

Specifically, the experimental data that we consider in this 
section is:
\begin{itemize}
\item The ATLAS measurement of the $W$ lepton and $Z$ rapidity 
distributions from the 2010 dataset ($36~{\rm pb}^{-1}$)~\cite{Aad:2011dm}.
\item The CMS measurement of the electron asymmetry
with the 2011 dataset ($840~{\rm pb}^{-1}$)~\cite{Chatrchyan:2012xt}.
\item The LHCb measurements of the $W^+$ and $W^-$ lepton
level 
rapidity distributions in the forward region from the 2010 data set~\cite{Aaij:2012vn}.
\item The ATLAS  measurement of the inclusive jet
production from the 2010 dataset 
($36~{\rm pb}^{-1}$)~\cite{Aad:2011fc}. We consider
the $R=0.4$ dataset only, very similar results are obtained
if the $R=0.6$ radius is also used.\footnote{Recently, the ratio
of these jet cross sections to the 2.76 TeV ones where also presented~\cite{atlasratio},
although in preliminary form. These cross section ratios~\cite{Mangano:2012mh}
have the potential to improve the PDF constraints as compared
to the 7 TeV data alone, thanks to the cancellation of
systematic uncertainties. }
\item The Tevatron Run II inclusive jet production from the CDF
and D0 collaborations, based on the $k_t$ and code jet reconstruction
algorithms respectively~\cite{Abulencia:2007ez,D0:2008hua}.
\end{itemize}

Theoretical
predictions have been obtained as follows:
\begin{itemize}
\item For electroweak vector boson production, we have computed
differential distributions at NLO with the 
{\tt MCFM} code~\cite{Campbell:2004ch} 
interfaced to the {\tt APPLgrid} software~\cite{Carli:2010rw} that
allows  a fast computation of the observable when PDFs are varied,
and  cross checked against  the {\tt DYNNLO} code~\cite{Catani:2010en}.
For ATLAS $W,Z$ data we have also cross-checked against 
the 
{\tt APPLgrid} implementation used in the ATLAS 
strangeness determination~\cite{Aad:2012sb}. NNLO predictions have
been obtained using local K-factors determined 
with {\tt DYNNLO}. 
\item For inclusive jet production at the LHC, we have used the {\tt NLOjet++}
program interfaced to the {\tt APPLgrid} software. The scale is chosen to be 
the $p_T$ of the hardest jet in the event
within each rapidity bin. Comparisons with {\tt FastNLO}~\cite{Kluge:2006xs} 
and {\tt MEKS}~\cite{Gao:2012he} are presented in the next section. 
Note that,  even though  NNLO PDFs are used, the accuracy of the
calculation is NLO, as  NNLO partonic cross sections are not yet available. 
\item For inclusive jet production at the Tevatron, we have used the 
{\tt FastNLO}~\cite{Kluge:2006xs} computation with the default
scale choice.
\end{itemize}

For inclusive jet production, the approximate NNLO coefficient
functions, derived from threshold resummation in {\tt FastNLO},
are used for the Tevatron predictions but not for the LHC.
In the latter case they are found to be unnaturally high, 
much larger than NLO corrections in a region far from
kinematical threshold.
An improved understanding of threshold corrections at the
LHC would be required before they can be used reliably for phenomenology.

In order to provide quantitative comparisons we 
compute the $\chi^2$ using 
different PDF sets. Note that, unlike other sets, 
NNPDF2.3 already includes these data in their fit, so it necessarily
provides a  good description of all of them.
For consistency of comparison,  
we use  the  same definition Eq.~(\ref{eq:chi2}) of the $\chi^2$ 
with the experimental covariance matrix Eq.~(\ref{eq:covmat}), 
even though this is not in general the quantity which has been
minimized when determining PDFs.
Results at NLO and at NNLO are summarized 
in Tables~\ref{tab:chi2-nnlo-as0117} and \ref{tab:chi2-nnlo-as0119}, 
where common values of $\alpha_s\lp M_Z\rp=0.117$ 
and $\alpha_s\lp M_Z\rp=0.119$ respectively have been used. 

As in the previous section, it is useful
to provide as well the $\chi^2$ values for ABM11 NNLO
with the default value $\alpha_s(M_Z)=0.1134$, since this value
is far from the range explored in this paper and is used in many
phenomenological comparisons. Therefore, in Table~\ref{tab:chi2-nnlo-as0114} 
we collect the $\chi^2$ for the ABM11 NNLO LHC and
Tevatron distributions,
but this time with the default $\alpha_s(M_Z)$ value. By comparing with the
results in Tables~\ref{tab:chi2-nnlo-as0117} and~\ref{tab:chi2-nnlo-as0119},
we see that the use of $\alpha_s(M_Z)=0.1134$ somewhat improves the
description of the LHC electroweak production data, but at the price
of worsening the description of jet production, specially of the
precise CDF Run II $k_T$ inclusive jet distributions.

%%%%%%%%%%%%%%%
%%%%%%%%%%%%%%%%%%%%%%%%%%%%%%%%%%%%%%%%%%%%%%%%%
\begin{table}[t]
  \centering
  \footnotesize
%%%%%%%%%%%%%%%%%%%%%%%%%%%%%%%%%%%%%%%%%%%%%%%%%%%%%%%%%%%%%%%%%%%
 \begin{tabular}{c||c|c|c|c|c}
 \hline
 & \multicolumn{4}{c}{NLO $\alpha_s=0.117$} \\
\hline
 Dataset & NNPDF2.3 & MSTW08& CT10& ABM11 & HERAPDF1.5 \\
 \hline
 \hline
ATLAS $W,Z$           &     1.234  &   1.993  &   1.047  &   1.472  &   1.719 \\
CMS $W$ el asy        &      0.884  &   4.694 &    1.458  &   1.961 &    0.671 \\
LHCb $W$            &     0.658   &  0.869  &   0.994   &  2.272   &  2.885 \\
ATLAS jets            &     0.916   &  0.893  &   1.212   &  1.409   &  0.968 \\
 \hline
CDF RII $k_T$ jets &     0.619    &  0.635      &  1.108       &  1.961      &  1.528        \\
D0 cone jets &   0.797      & 0.819       &  0.972       & 1.149      &  1.296        \\
 \hline
 \end{tabular}
%%%%%%%%%%%%%%%%%%%%%%%%%%%%%%%%%%%%%%%%%%%%%%%%%%%%%%%%%%%%%%%%%
   \\
$\qquad$~ \\
$\qquad$~ \\
 \begin{tabular}{c||c|c|c|c|c}
 \hline
& \multicolumn{4}{c}{NNLO $\alpha_s=0.117$} \\
\hline
 Dataset & NNPDF2.3 & MSTW08& CT10& ABM11 & HERAPDF1.5 \\
 \hline
 \hline
ATLAS $W,Z$           &       1.382  &   3.194 &    1.125  &   1.923    & 1.845 \\
CMS $W$ el asy        &       0.828  &   4.140 &    1.778   &  1.602    & 0.817 \\
LHCb $W$            &       0.741  &   0.956 &    0.892  &   1.873    & 0.744 \\
ATLAS jets            &       0.862  &   0.828 &    0.940  &   0.963    & 0.848 \\
 \hline
CDF RII $k_T$ jets &   0.667      & 0.587       &  0.629      &  1.179     &  0.676        \\
D0 cone jets &    0.878     &   0.875     &  0.943      &  0.917     & 0.981        \\
 \hline
 \end{tabular}
  
  \caption{\label{tab:chi2-nnlo-as0117}  
\small The $\chi^2/N_{pt}$ values  for
the available LHC data with published correlated uncertainties,
computed using different PDF sets. 
We also include in this comparison the Tevatron Run II
inclusive jet production data.
The theoretical predictions have been computed
at NLO (upper table) and at NNLO
(lower table) using {\tt APPLgrid} 
for a common value of the strong coupling
$\alpha_s\lp M_Z\rp=0.117$. The experimental definition of 
the covariance matrix $({\rm cov})_{ij}$ is used, see
Eq.~(\ref{eq:chi2}).
}
\end{table}
%%%%%%%%%%%%%%%%%%%%%%%%%%%%%%%%%%%%%%%%%%%

%%%%%%%%%%%%%%%
%%%%%%%%%%%%%%%%%%%%%%%%%%%%%%%%%%%%%%%%%%%%%%%%%
\begin{table}[t]
  \centering
  \footnotesize
%%%%%%%%%%%%%%%%%%%%%%%%%%%%%%%%%%%%%%%%%%%%%%%%%%%
 \begin{tabular}{c||c|c|c|c|c}
 \hline
& \multicolumn{4}{c}{NLO $\alpha_s=0.119$} \\
\hline
 Dataset & NNPDF2.3 & MSTW08& CT10& ABM11 & HERAPDF1.5 \\
 \hline
 \hline
ATLAS $W,Z$           &      1.271  &      2.003  &      1.061  &      1.561  &      1.757 \\
CMS $W$ el asy        &      0.822  &      4.698  &      1.421  &      1.929  &      0.693 \\
LHCb $W$              &      0.673  &      0.919  &      1.063  &      2.332  &      4.124 \\
ATLAS jets            &      1.004  &      0.972  &      1.352  &      1.345  &      1.111 \\
 \hline
CDF RII $k_T$ jets &  0.599  &   0.642 &    1.088   &  1.662  &   1.494 \\
D0 cone jets &   0.842    & 0.865    & 1.058  &   1.062  &   1.324 \\
\hline
 \end{tabular}
%%%%%%%%%%%%%%%%%%%%%%%%%%%%%%%%%%%%%%%%%%%%%%%%%%%
  \\
$\qquad$~ \\
$\qquad$~ \\
 \begin{tabular}{c||c|c|c|c|c}
 \hline
& \multicolumn{4}{c}{NNLO $\alpha_s=0.119$} \\
\hline
 Dataset & NNPDF2.3 & MSTW08& CT10& ABM11 & HERAPDF1.5 \\
 \hline
 \hline
ATLAS $W,Z$           &     1.435   &  3.201 &    1.160  &   2.061   &  1.872 \\ 
CMS $W$ el asy        &     0.813  &   3.862  &   1.772 &    1.614  &   0.814 \\
LHCb $W$            &       0.831 &    1.050  &   0.966  &   1.970  &   0.784 \\
ATLAS jets            &    0.937   &  0.935  &   1.016  &   0.959  &   1.011 \\   
 \hline
CDF RII $k_T$ jets &        0.679    &  0.642   &  0.666   &  0.926   &  0.769       \\
D0 cone jets &         0.939  &   0.954  &   1.026   &  0.915   &  1.110        \\
\hline
 \end{tabular}
  
  \caption{\label{tab:chi2-nnlo-as0119}  
\small Same as Table~\ref{tab:chi2-nnlo-as0117}, but for 
$\alpha_s\lp M_Z\rp=0.119$.
}
\end{table}
%%%%%%%%%%%%%%%%%%%%%%%%%%%%%%%%%%%%%%%%%%%

%%%%%%%%%%%%%%%
%%%%%%%%%%%%%%%%%%%%%%%%%%%%%%%%%%%%%%%%%%%%%%%%%
\begin{table}[t]
  \centering
  \footnotesize
 \begin{tabular}{c||c}
 \hline
& \multicolumn{1}{c}{NNLO $\alpha_s=0.1135$} \\
\hline
 Dataset & ABM11   \\
 \hline
 \hline
ATLAS $W,Z$           &      1.739 \\
CMS $W$ el asy        &      1.650 \\
LHCb $W$            &        1.821 \\
ATLAS jets            &      1.195 \\
 \hline 
CDF RII $k_T$ jets &       1.932 \\
D0 cone jets &             1.195 \\
 \hline
 \end{tabular}
  
  \caption{\label{tab:chi2-nnlo-as0114}  
\small Same as Table~\ref{tab:chi2-nnlo-as0117}, but for 
$\alpha_s\lp M_Z\rp=0.1134$, for the ABM11 predictions at
NNLO.
}
\end{table}
%%%%%%%%%%%%%%%%%%%%%%%%%%%%%%%%%%%%%%%%%%%

The main conclusions which can be drawn from these comparisons 
are the following:
\begin{itemize}
\item All PDF sets lead to predictions in reasonable agreement with ATLAS
jet data. In general, the description improves when
NNLO PDFs are used as compared to NLO PDFs. While the ATLAS
jet data appear to have  only moderate constraining power, larger impact
is expected when the full 7 TeV 5 fb$^{-1}$ data from
CMS and ATLAS will become available.
\item The ATLAS and CMS electroweak data appear to have considerable
  discriminating power, and thus are likely to constrain significantly
quarks and anti-quarks at medium and small-$x$, and specifically strangeness~\cite{Aad:2012sb}. The worst
description of the electroweak data is provided by MSTW08:  
this will be discussed in more detail below.
\item The LHCb data also appears to have discriminating power.
This data is sensitive to flavor separation
at the smallest values of $x$, and to fairly high-$x$ quarks, 
thanks to the forward
coverage of the LHCb detector. Predictions obtained using all PDF sets
describe the data quite well, with the exception of ABM11. It should
be noticed that while at NNLO 
HERAPDF1.5 agrees with the data, at NLO instead it provides  a
poor description, due to the large
antiquark PDF at high $x$. 
\end{itemize}

The main reason why MSTW08 provides a 
rather poor description of the ATLAS $W,Z$, and especially of the
CMS $W$ data
is understood~\cite{Watt:2012tq, Martin:2012xx} as a consequence of
the behavior of  the
$u_v-d_v$ distribution around $x\sim 0.03$. Indeed, in
Ref.~\cite{Watt:2012tq} it is shown  that once the LHC $W$ asymmetry
data is included in MSTW08 using PDF 
reweighting~\cite{Ball:2010gb,Ball:2011gg}, the fit quality improves 
substantially. In~\cite{Martin:2012xx} it is shown that an extended 
parameterisation for quarks (and to a lesser extent a consideration 
of deuteron corrections) automatically alters the form of $u_v-d_v$
for the standard MSTW08 fit in the relevant region without including new data, 
and the predictions for the asymmetry improve
enormously --- the $\chi^2$ for the prediction for the asymmetry data 
decreases to about one per point. It is also demonstrated explicitly that 
this is a very local
discrepancy which has a very small effect on more inclusive  
cross sections, much less than PDF uncertainties. 

We can also compare the agreement of the different PDF sets with the 
data by examining plots, although of course this will be
less quantitative than the $\chi^2$ comparison. Note, in particular
that the correlated systematical error (shown as a band in the bottom of each
plot) is quite large, and typically dominates over the uncorrelated
statistical uncertainty.  As a consequence, it is  difficult to judge the fit
quality by simple inspection of the plots.
The main motivation to show the plots is to provide a link
between the quantitative $\chi^2$ numbers and the visual data versus theory
comparisons, that are frequently used, and to make clear
that the quantitative information can be provided only by the quantitative
estimator. 
So this  plots only serve the purpose of giving a rough 
indication of the trend of
the data versus theory comparison, for example, 
one see from plots if there are systematic 
differences between predictions and data or just fine details in shape.

\begin{figure}[ht]
    \begin{center}
      \includegraphics[width=0.48\textwidth]{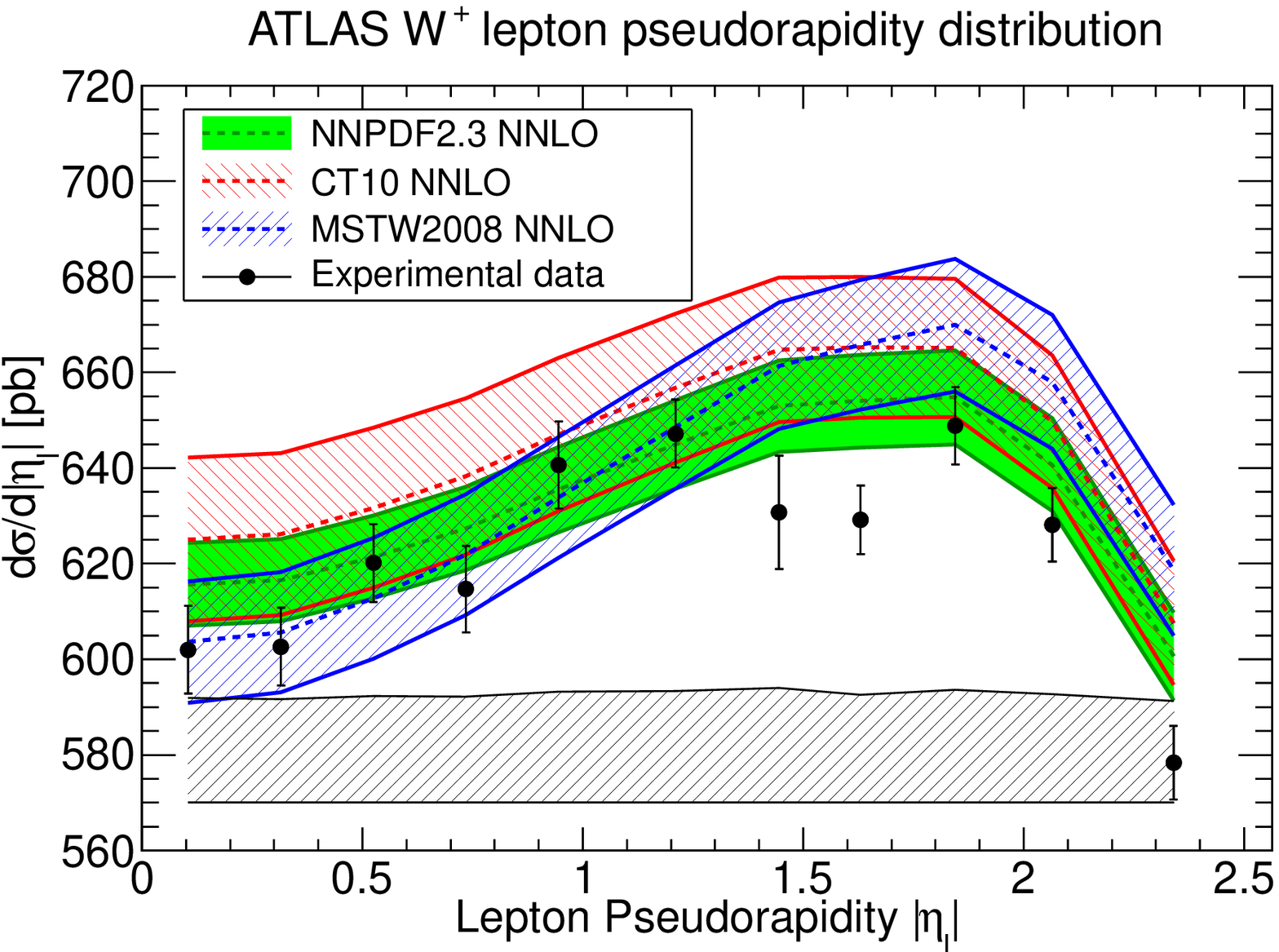}\quad
    \includegraphics[width=0.48\textwidth]{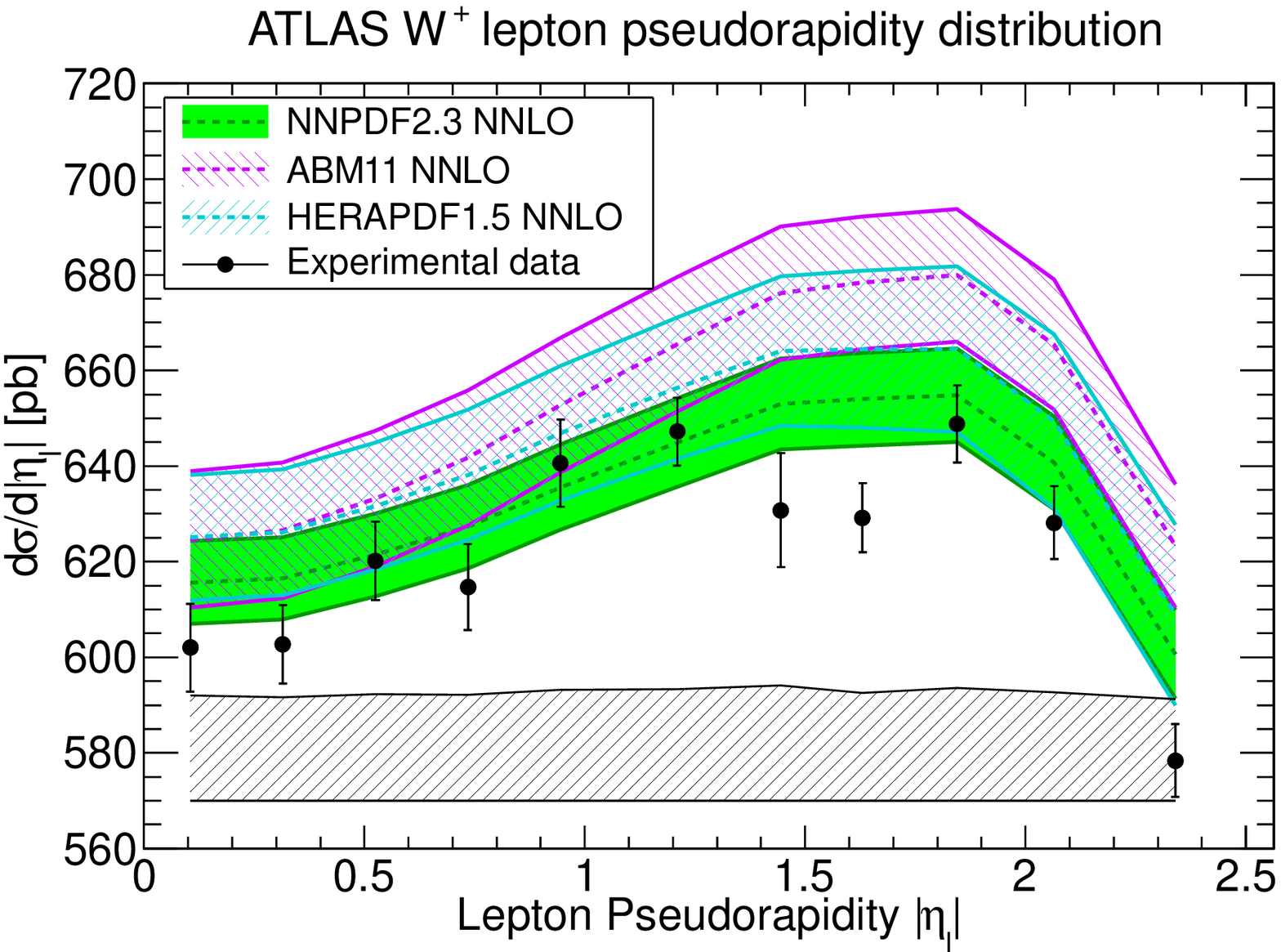}\\
   \includegraphics[width=0.48\textwidth]{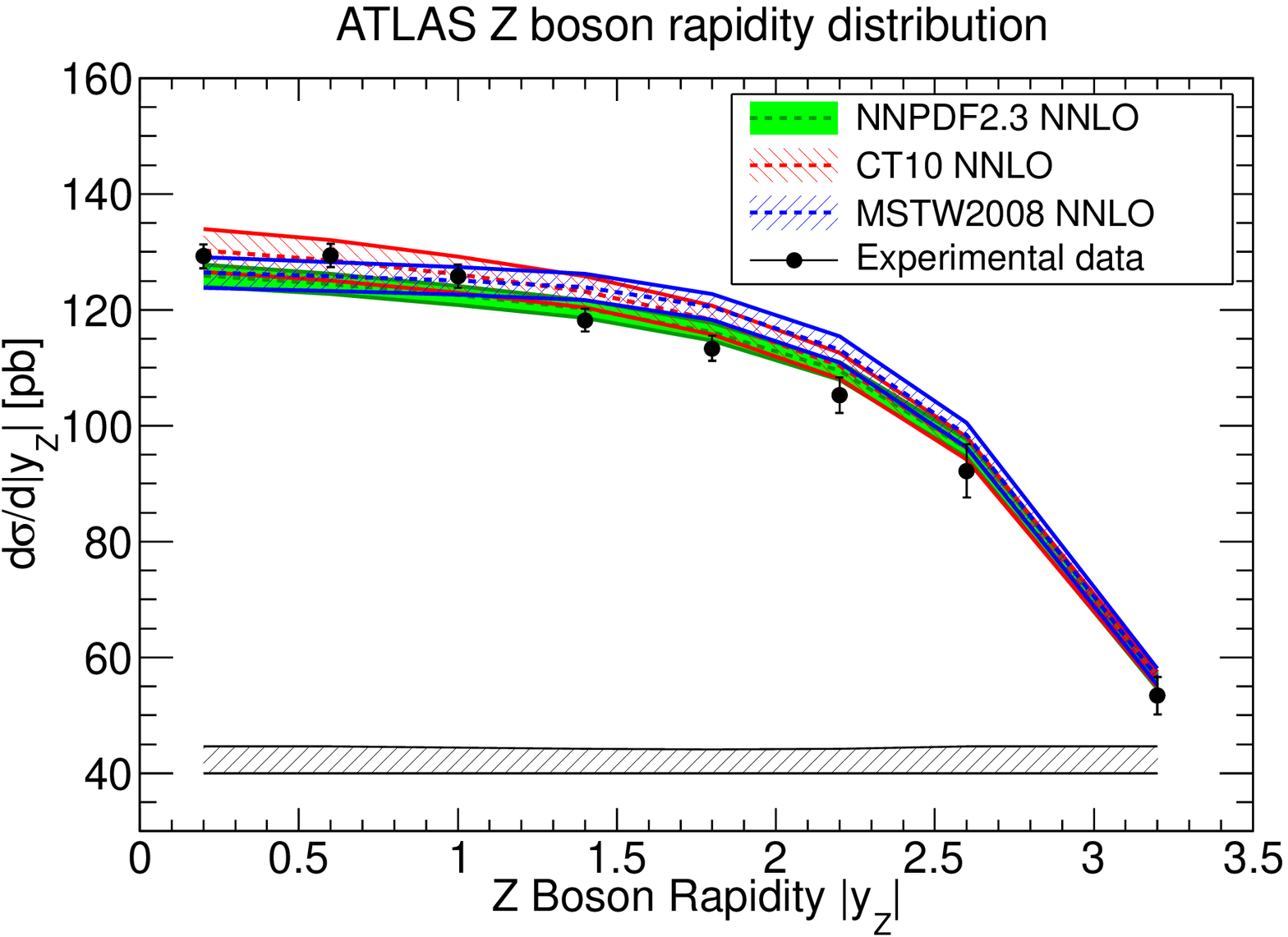}\quad
  \includegraphics[width=0.48\textwidth]{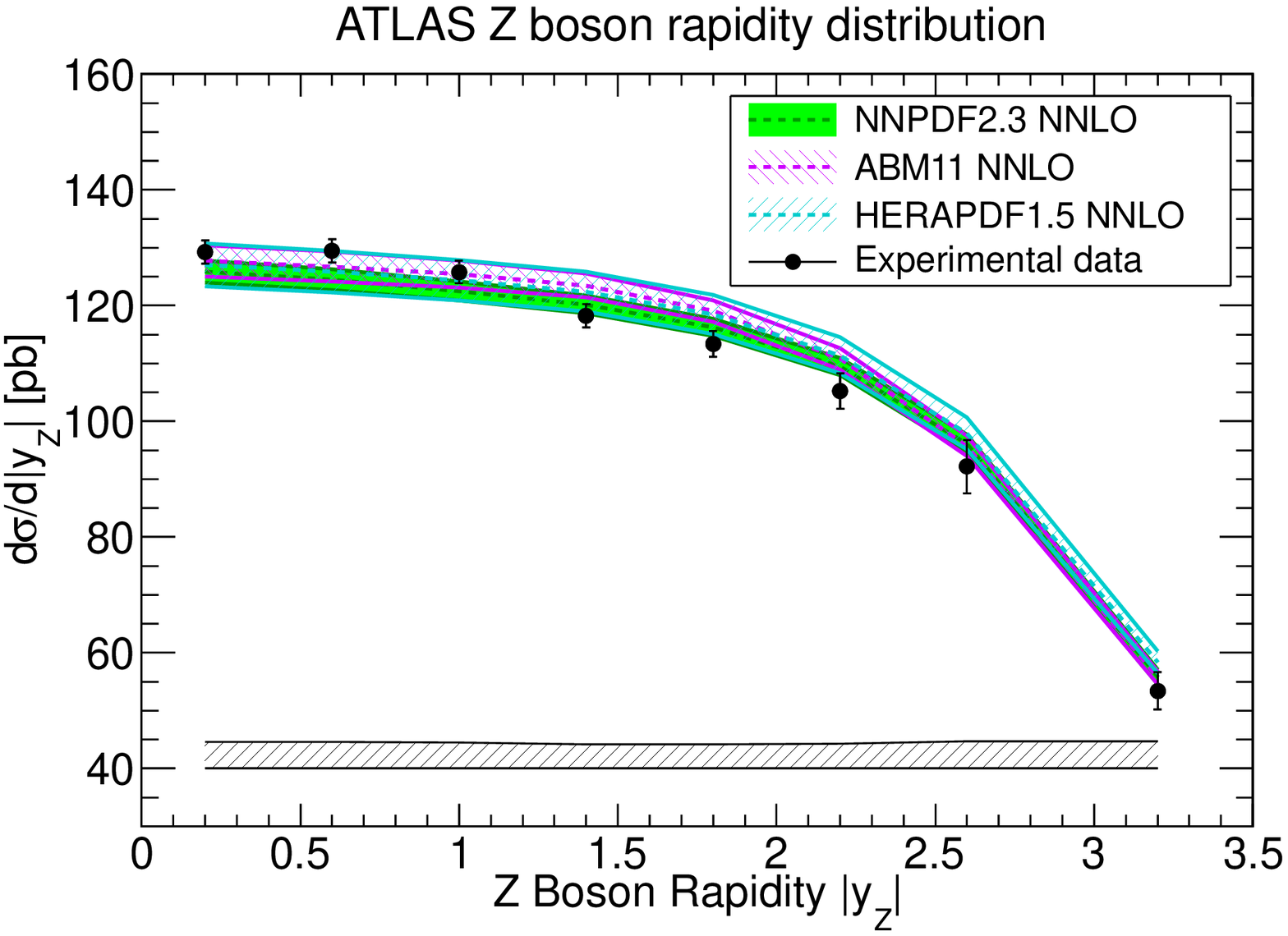}
    \end{center}
    \vskip-0.5cm
    \caption{\small Comparison of the ATLAS electroweak
vector boson production data
with the NNPDF2.3, CT10 and MSTW2008  predictions with 
$\alpha_{s} = 0.118$. The error bars correspond to statistical
uncertainties, while the band in the bottom of the plot indicates the
correlated systematics (including normalization errors).
}
    \label{fig:LHCdataplots-ewk1}
\end{figure}

\begin{figure}[ht]
    \begin{center}
  \includegraphics[width=0.48\textwidth]{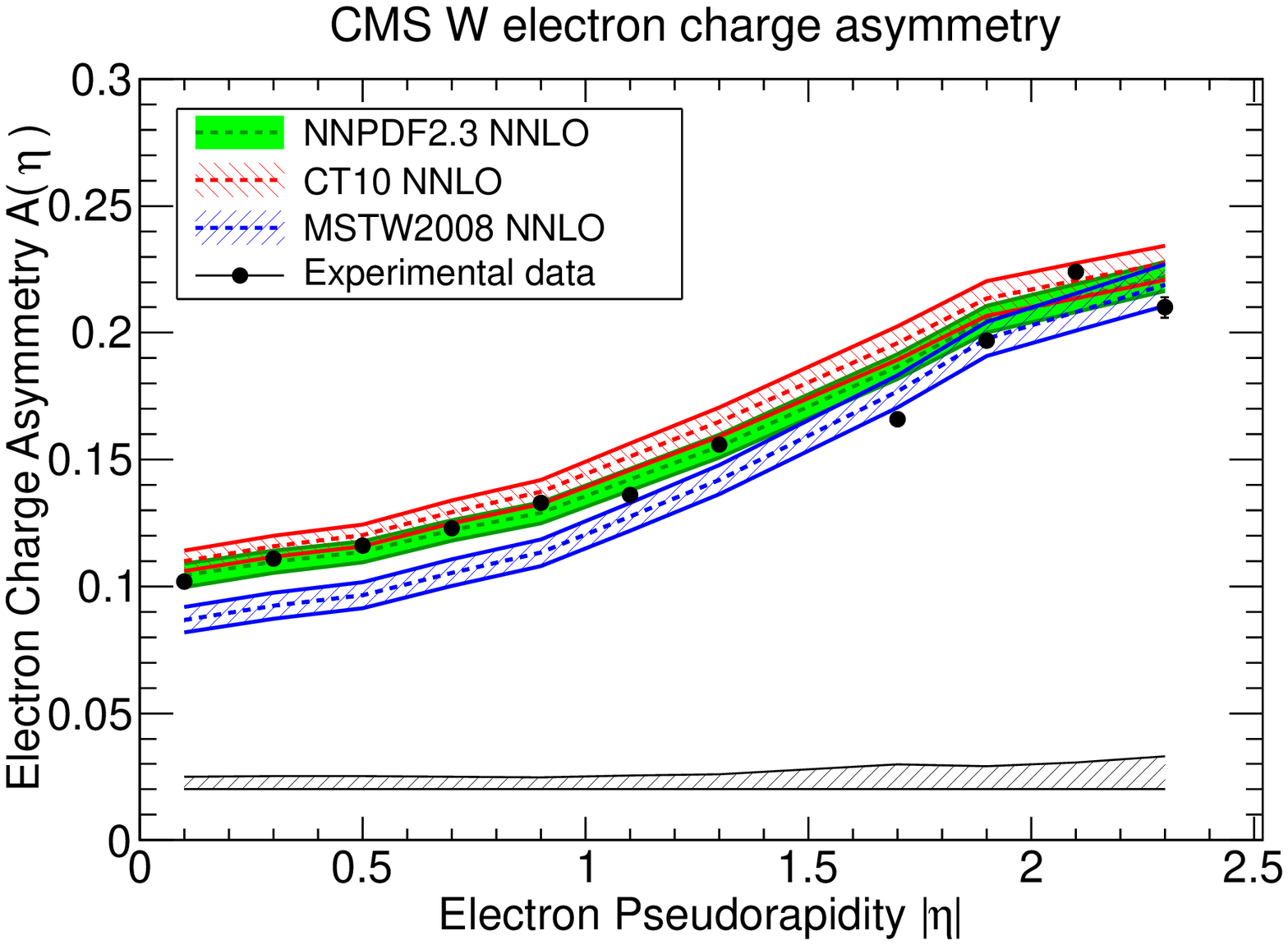}\quad
  \includegraphics[width=0.48\textwidth]{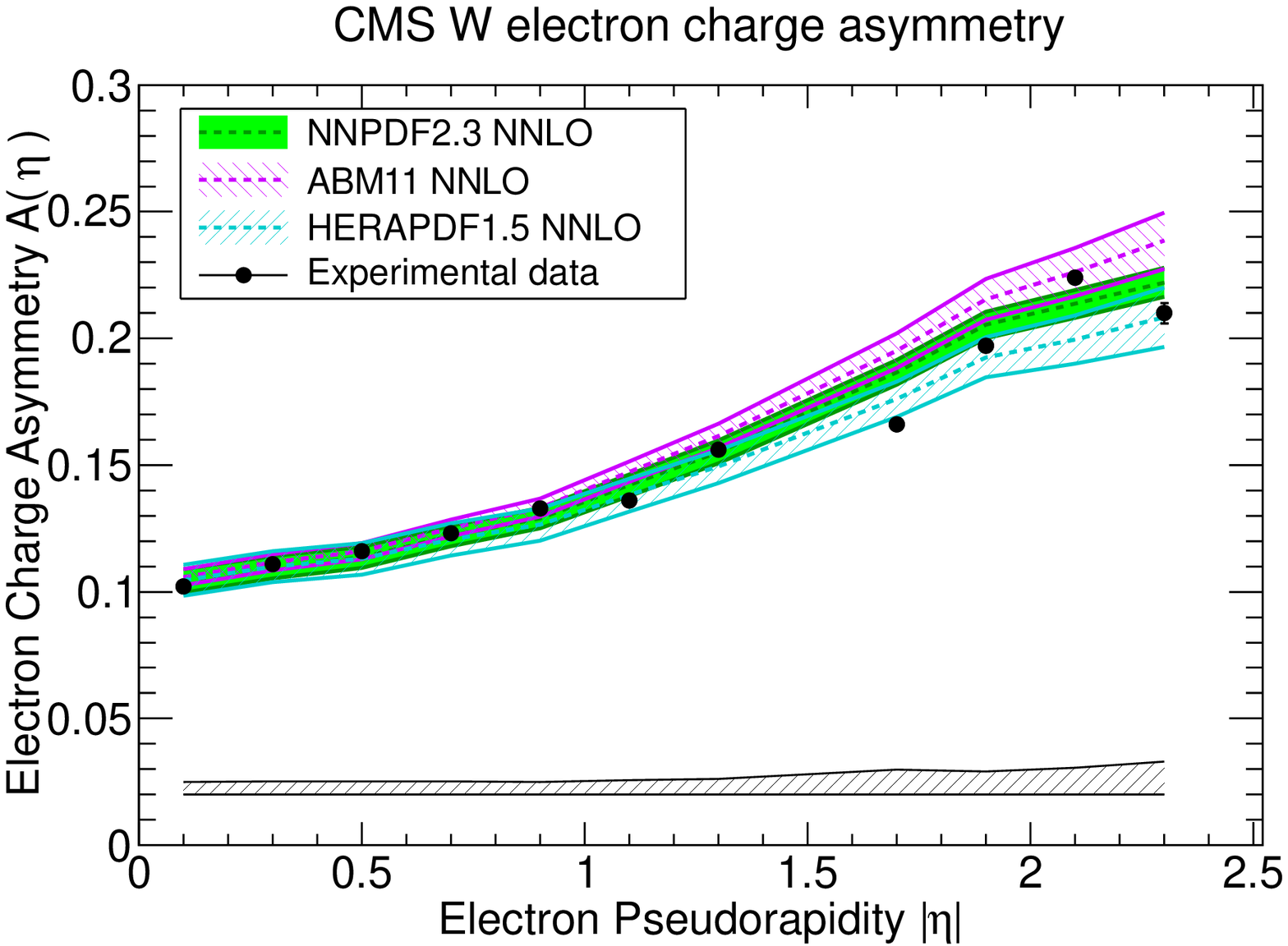}\\
      \includegraphics[width=0.48\textwidth]{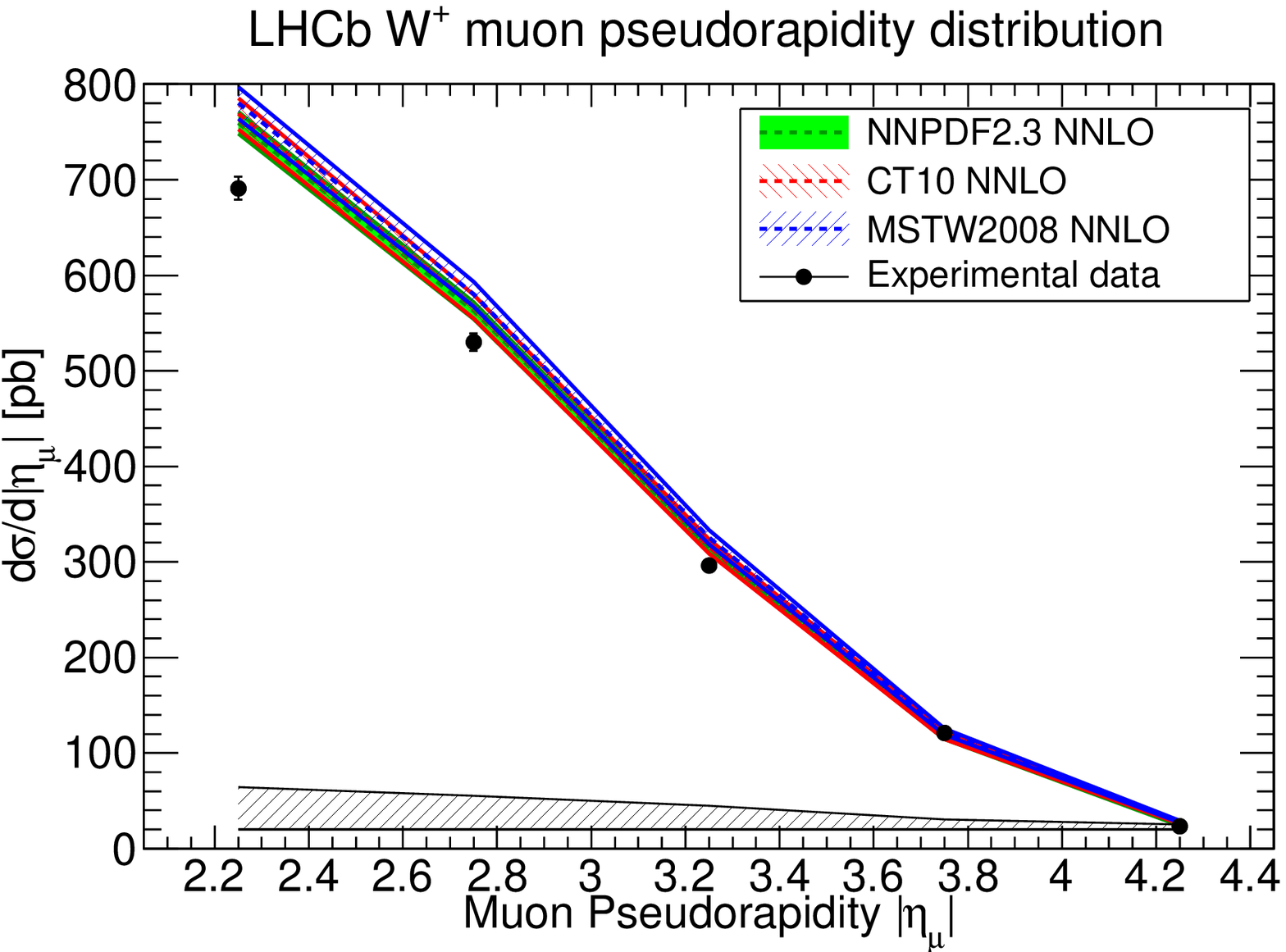}\quad
   \includegraphics[width=0.48\textwidth]{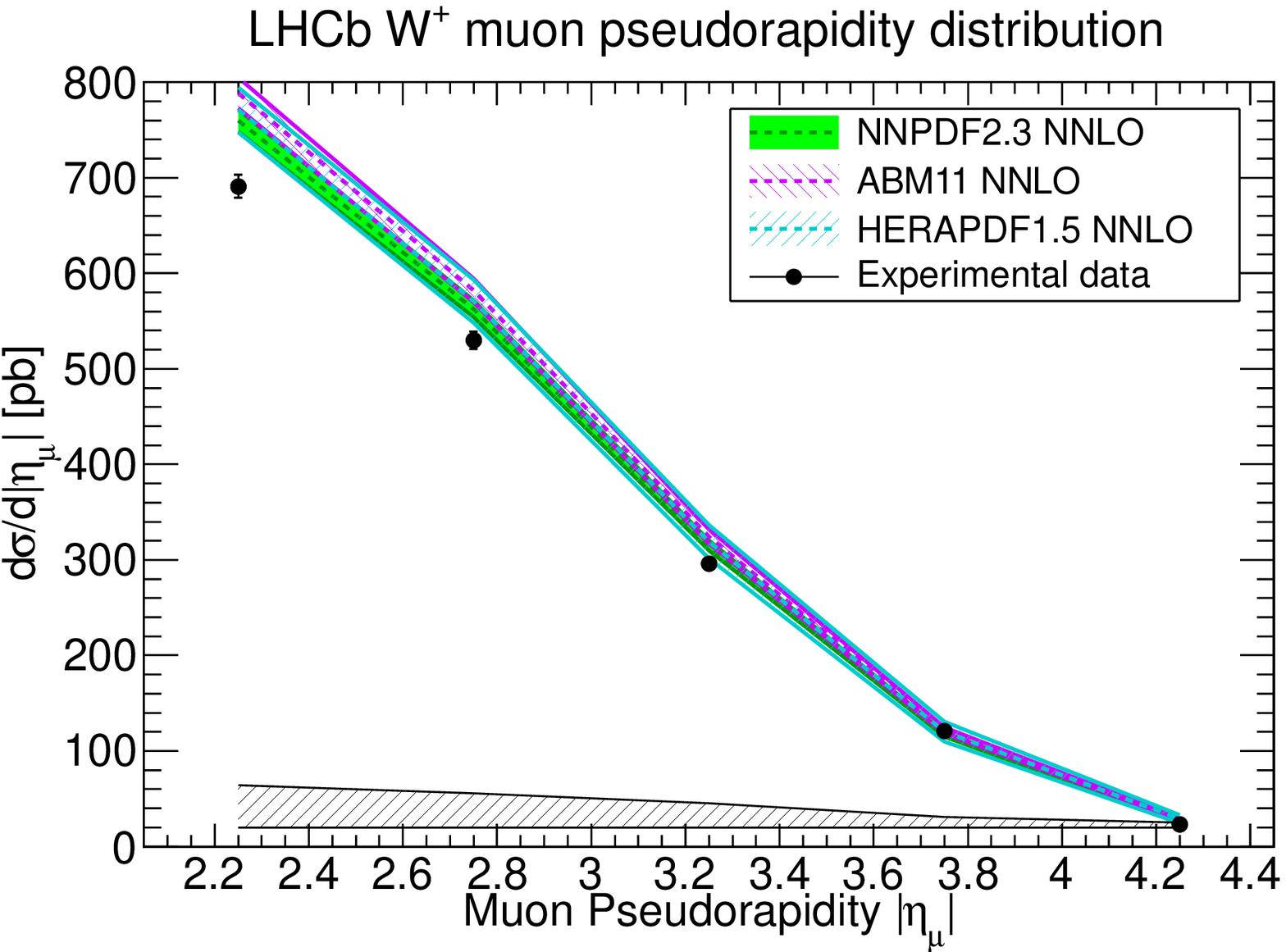}
    \end{center}
    \vskip-0.5cm
    \caption{\small Same as Fig.~\ref{fig:LHCdataplots-ewk1}
for CMS and LHCb $W$ production.
}
    \label{fig:LHCdataplots-ewk2}
\end{figure}

\begin{figure}[ht]
    \begin{center}
      \includegraphics[width=0.48\textwidth]{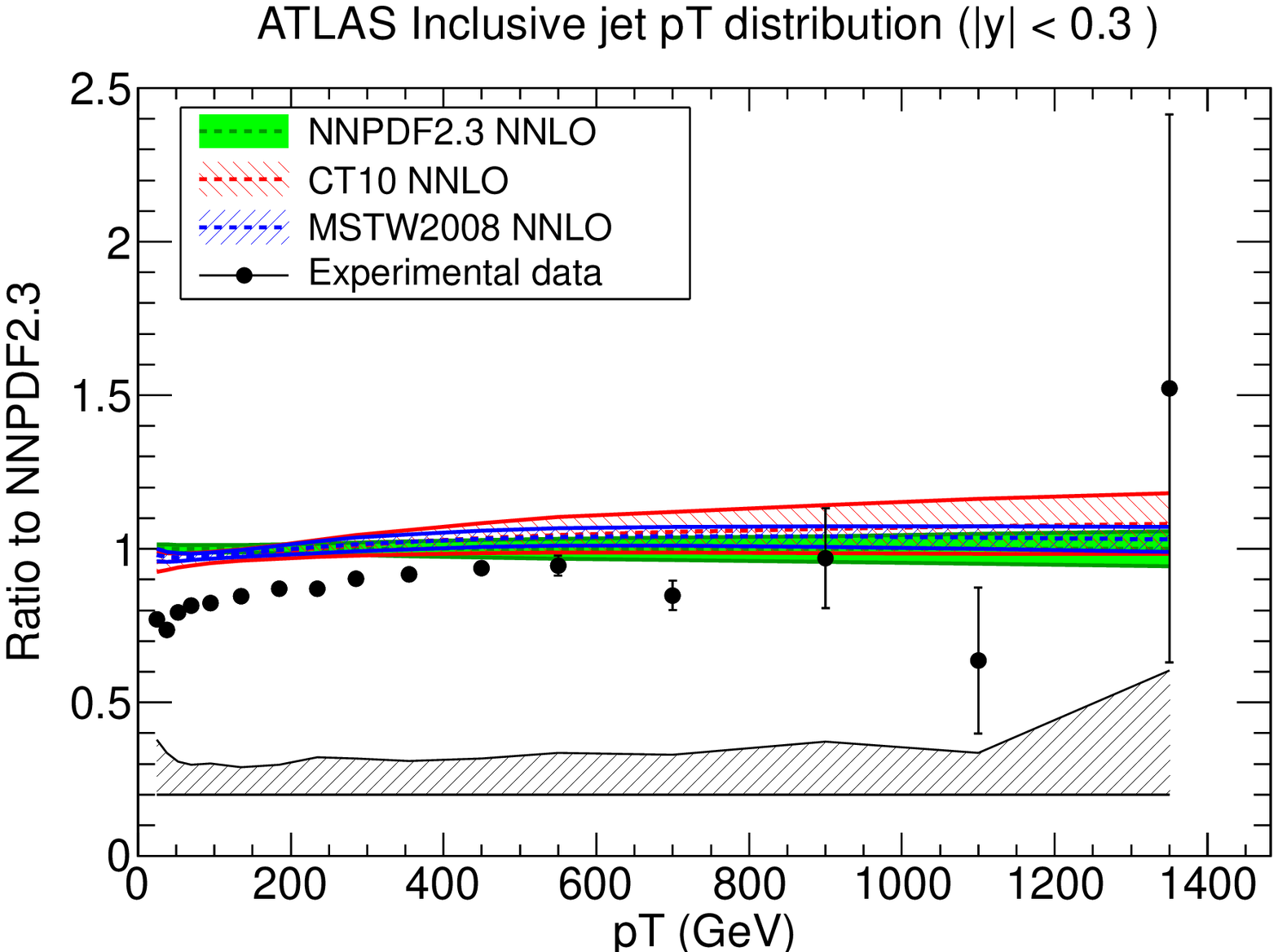}\quad
    \includegraphics[width=0.48\textwidth]{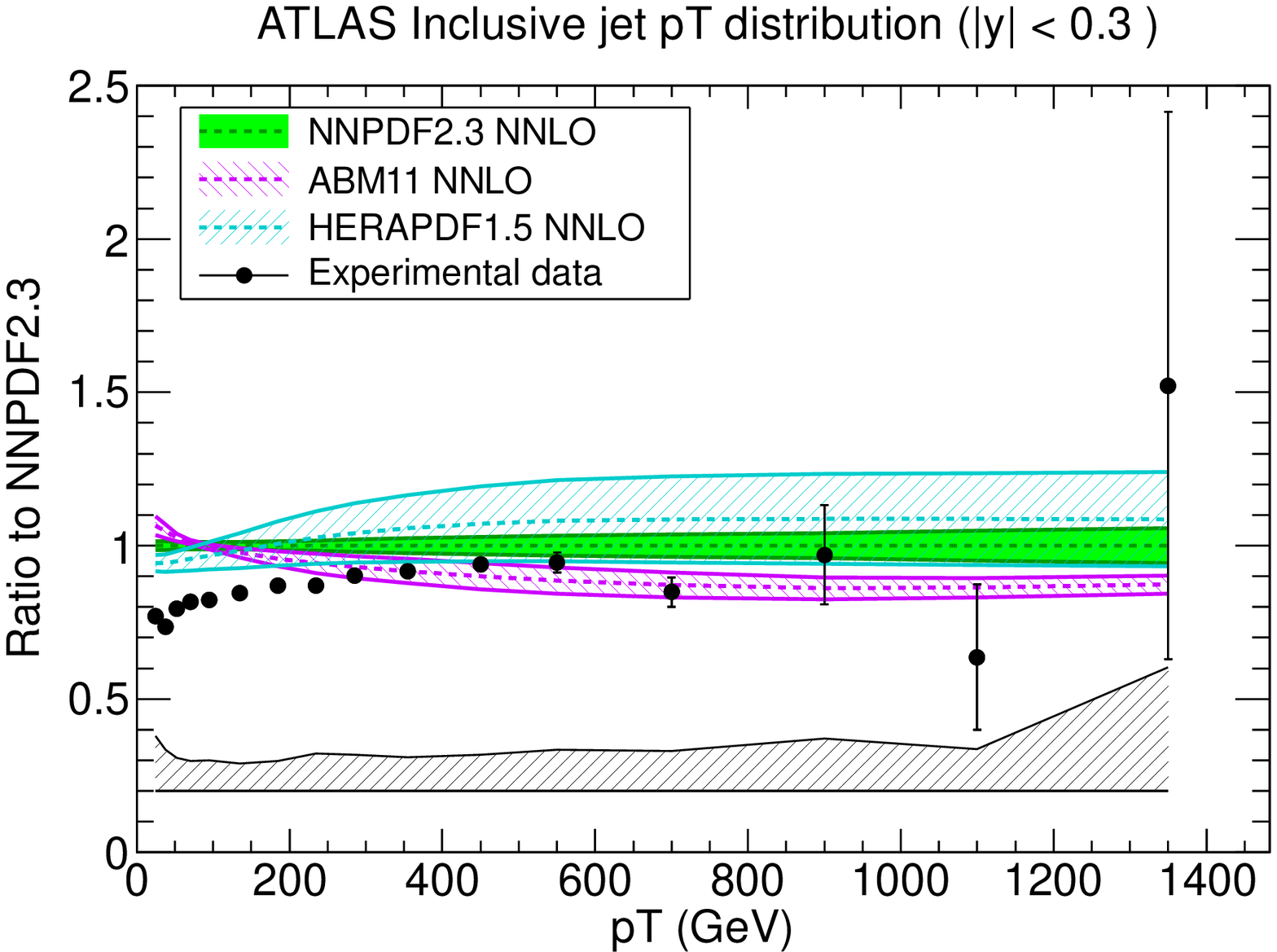}\\
      \includegraphics[width=0.48\textwidth]{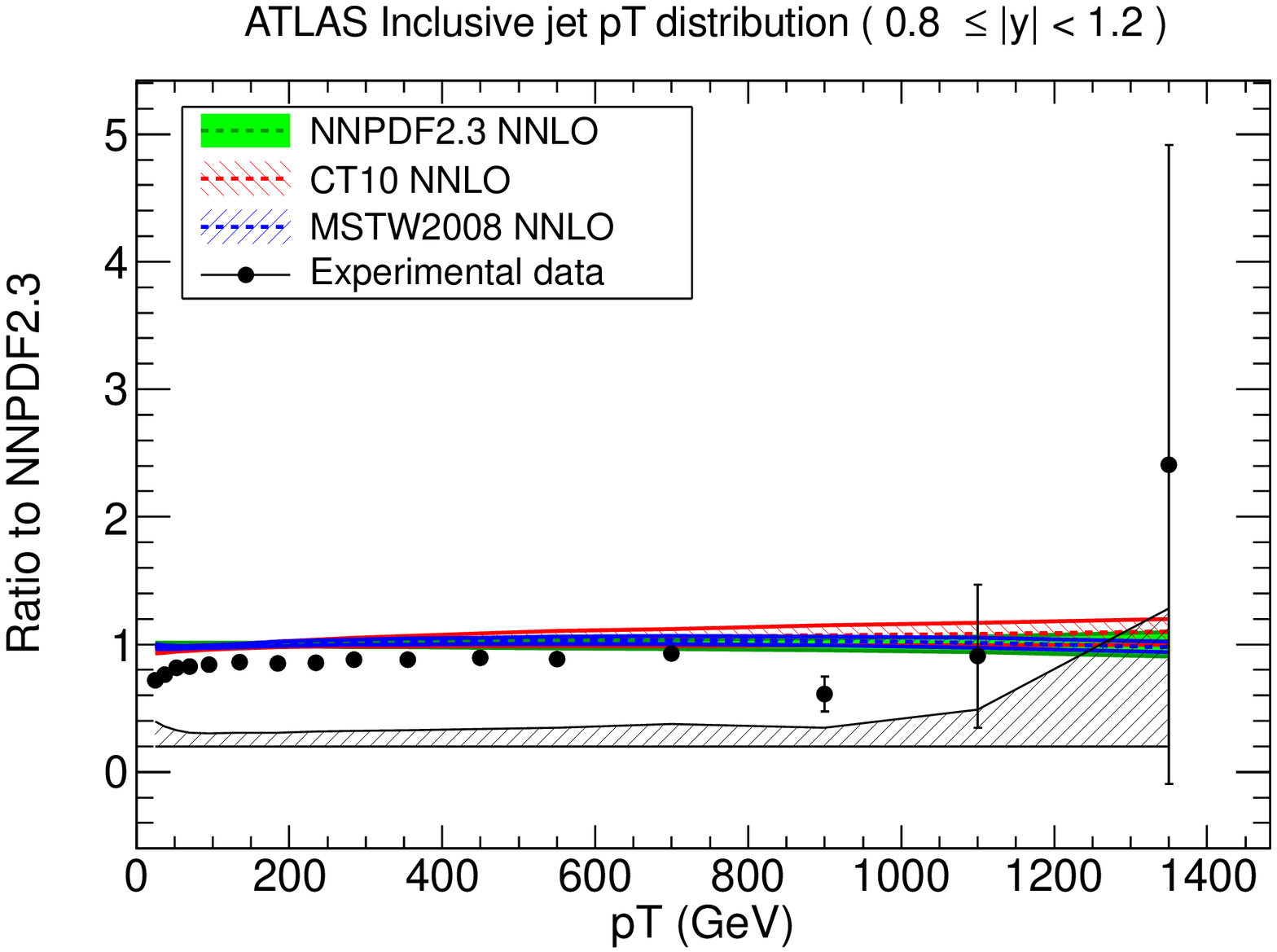}\quad
  \includegraphics[width=0.48\textwidth]{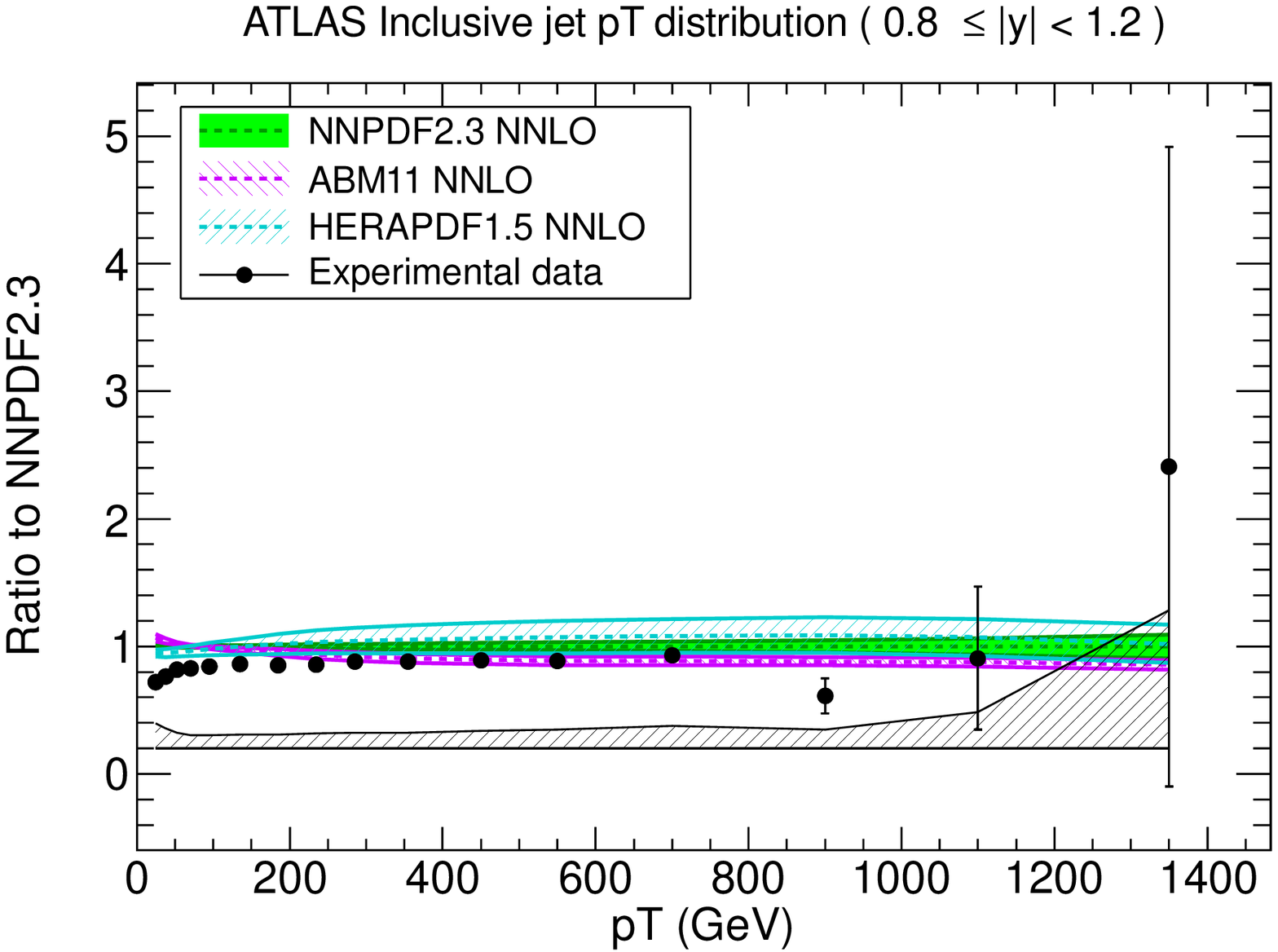}
    \end{center}
    \vskip-0.5cm
    \caption{\small Comparison of the ATLAS $R=0.4$ 
inclusive jet production data from the 2010 dataset
with the NNPDF2.3, CT10 and MSTW2008 NNLO PDF sets and 
$\alpha_{S} = 0.118$.
The error bars correspond to statistical
uncertainties, while the band in the bottom of the plot indicates the
correlated systematics (including normalization errors)
}
    \label{fig:LHCdataplots-jets}
\end{figure}

As before, we show  on the one hand a comparison of NNPDF2.3, CT10
and MSTW, and on the other of NNPDF2.3, ABM11 and
HERAPDF1.5. The comparison for the ATLAS  electroweak
boson production data is shown in  Fig.~\ref{fig:LHCdataplots-ewk1},
for CMS and LHCb $W$ production in Fig.~\ref{fig:LHCdataplots-ewk2},
and for ATLAS inclusive jet data 
in Fig.~\ref{fig:LHCdataplots-jets}. We show only a subset of all the possible 
comparisons, and only for $\alpha_s=0.118$; a fuller set of plots can
be found at the  {\tt HepForge} link mentioned previously.

%%%%%%%%%%%%%%%%%%%%%%%%%%%%%%%%%%%%%%%%%%%%%%%%%
%%%%%%%%%%%%%%%%%%%%%%%%%%%%%%%%%%%%%%%%%%%%%%%%%%

%%%%%%%%%%%%%%%%%%%%%%%%%%%%%%%%%%%%%%%%%%%%%%%%%
%%%%%%%%%%%%%%%%%%%%%%%%%%%%%%%%%%%%%%%%%%%%%%%%%%

%%%%%%%%%%%%%%%%%%%%%%%%%%%%%%%%%%%%%%%%%%%%%%%%%
%%%%%%%%%%%%%%%%%%%%%%%%%%%%%%%%%%%%%%%%%%%%%%%%%%

\clearpage
\section{ATLAS inclusive jet production at NLO}

\label{sec:atlasjets}

As outlined in the last section, jet production
is one of the cornerstone processes of the physics
program at the LHC. It has reached unprecedented statistical precision
and can serve both for detailed tests of perturbative QCD and searches
for hypothetical new interactions. Inclusive jet production measurements
impose direct constraints on the gluon PDF, and the LHC data can in principle
be sensitive to the gluon PDF in a very wide range of momentum fractions
$x$~\cite{GaoNadolskyCEMA}.
Inclusive jet production at the Tevatron and LHC can be used to
reduce the gluon uncertainty, and thus improve the predictions
for important processes like Higgs production in gluon fusion. 
The last section gave a brief outline 
of the current comparison of the QCD predictions with various PDF sets to the 
current ATLAS data, but here we point out some more detailed features of
the analysis, which will become more important as the precision of the data 
collected improves. 
 
There exist two independent computer programs for computing
single-inclusive jet and dijet production at NLO at the parton level,
{\tt EKS}~\cite{Ellis:1992en} and {\tt NLOjet++}~\cite{Nagy:2001fj,Nagy:2003tz}.
The {\tt EKS} code was written in the early 1990's and was used to tabulate
point-by-point NLO/LO K factors for jet production in previous
CTEQ global fits. As the precision of the jet data increased,
it became necessary to develop a new version of {\tt EKS} with enhanced
numerical stability and percent-level accuracy. It also became clear
that the PDFs that are constrained by the jet cross sections may depend
on the theoretical assumptions made in the computation of
NLO theoretical cross sections. To address this issue, a deeply revised
version of the {\tt EKS} code, designated as {\tt MEKS}~\cite{Gao:2012he}, was
recently released and compared against the other independent code,
{\tt NLOjet++}~\cite{Nagy:2001fj,Nagy:2003tz}. This study documented specific
settings in the two codes that bring them into agreement to within
1-2\% at both the Tevatron and LHC.

The {\tt MEKS} and {\tt NLOjet++} calculations are relatively slow and require
significant CPU time to reach acceptable accuracy, so that their direct
use in the PDF fits is impractical. Instead, the global PDF analyses
reproduce the NLO cross sections by fast numerical approximations.
Besides the interpolation of the tabulated NLO/LO K factors that was
utilized until recently by CTEQ, a more flexible approach is provided
by the programs {\tt FastNLO}~\cite{Kluge:2006xs,ftnlo:2010xy,Wobisch:2011ij} 
and {\tt APPLgrid}~\cite{Carli:2010rw}. They quickly and accurately
interpolate the tables of NLO jet cross sections initially computed in 
{\tt NLOjet++}.  The  threshold corrections
to inclusive jet production of ${\cal O}(\alpha_{s}^{2})$~\cite{Kidonakis:2000gi} are also available
as an estimate of the unknown NNLO terms.%
\footnote{Threshold corrections are not included in this study.%
} 

Besides  fixed-order QCD calculations, NLO event generators 
such as POWHEG~\cite{Alioli:2010xa} and SHERPA~\cite{Hoeche:2012fm}
combine the NLO hard cross section for inclusive jet production with
leading-log showering evaluated by HERWIG or PYTHIA. POWHEG predictions
for ATLAS jet production are different from the fixed-order 
predictions~\cite{Aad:2011fc} and also show quite a strong dependence on the parton 
showering, even at the highest $p_T$, while the SHERPA
results are in general closer to NLO.  The reasons
 of the differences between SHERPA and POWHEG
are still not well understood, and until this is settled by the
Monte Carlo authors, it is more reliable to stick to fixed order NLO
QCD calculations.
Thus only fixed-order calculations will 
be considered in the rest of this section. Electroweak
corrections to dijet production have also been studied
in Refs.~\cite{Dittmaier:2012kx,Moretti:2005ut}.

In their most recent PDF sets, 
{\tt FastNLO} is used by the CTEQ and MSTW groups,
while {\tt APPLgrid} is used by NNPDF.\footnote{
Since NNPDF2.0~\cite{Ball:2010de}, the Tevatron jet data is included
in NNPDF with the {\tt FastNLO} package. In the recent NNPDF2.3, the
 {\tt FastNLO} grid tables are accessed using the
 {\tt APPLgrid} wrapper. } Predictions from either program depend
significantly on the choices for the QCD renormalization and factorization
scales ($\mu_{R}$ and $\mu_{F}$), recombination scheme, and realization
of the jet algorithm~\cite{Gao:2012he}. In the case of inclusive
jet production, the default hard scale specifying the $\mu_{F}$ and
$\mu_{R}$ values in each event can be taken to be equal to ``$p_{T}$
of each individual jet'' ({\tt FastNLO} version 2), ``$p_{T}$ of the
hardest jet\textquotedblright{}, ``$p_{T}$ of the hardest jet in
each rapidity bin\textquotedblright{} ({\tt APPLgrid}), ``average $p_{T}$
in each $p_{T}$ bin ({\tt FastNLO} version 1)\textquotedblright{}. Differences
between these choices are relevant in modern comparisons, 
as will be shown below. Similar ambiguities are present
in computations for dijet production. We will explicitly distinguish
between these various scale prescriptions to avoid a common inaccuracy of
referring to all of them as ``the scales that are equal to jet $p_{T}$''.

\subsection{Comparison of computer programs and scale dependence}

\begin{figure}[h]
\begin{centering}
\includegraphics[width=\textwidth]{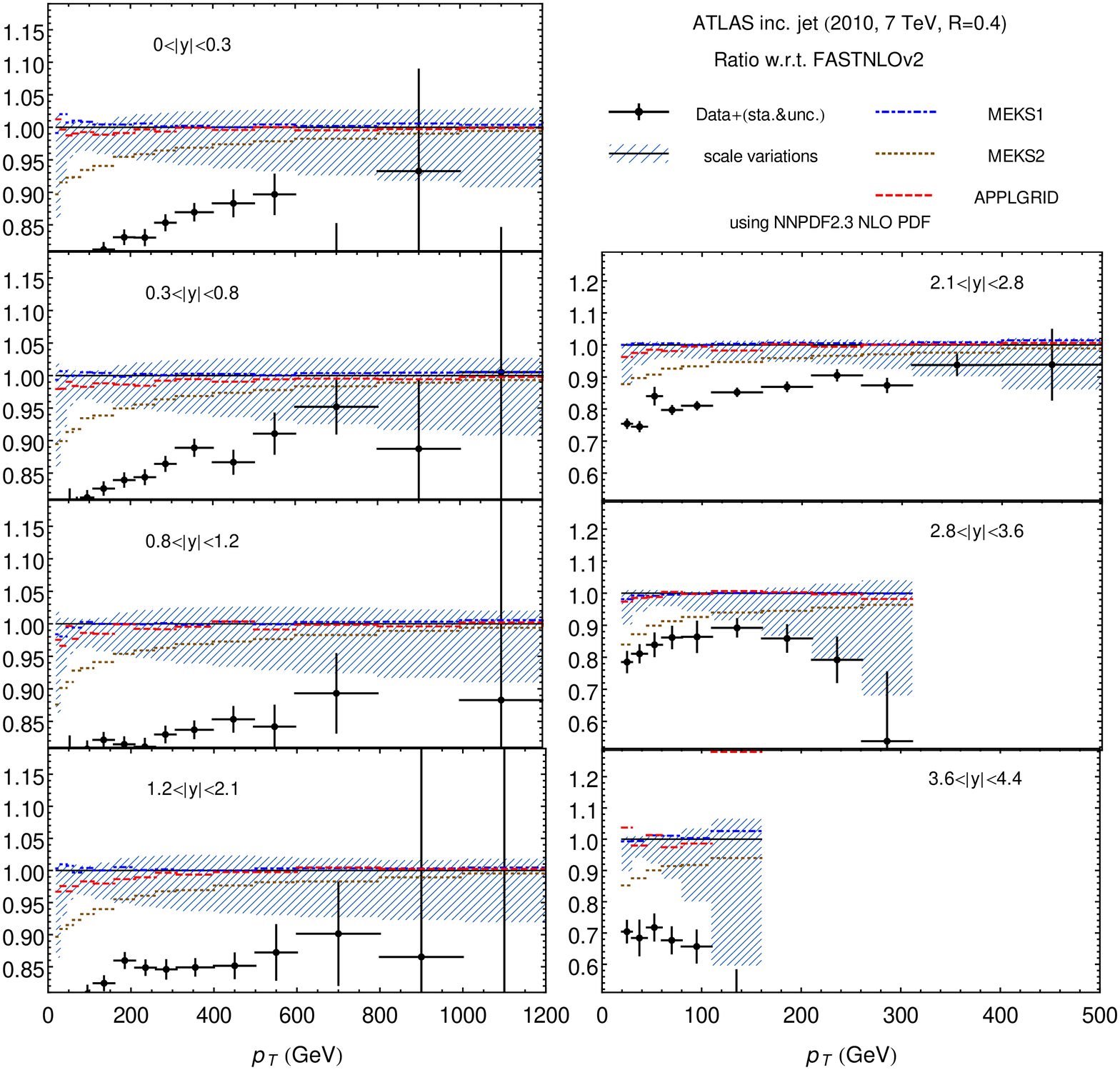} 
\par\end{centering}

\vspace{-3ex}
 \caption{\label{jetbench1} Comparison of NLO theoretical predictions obtained
with various numerical programs for the 2010 ATLAS measurement of
single-inclusive jet production \cite{Aad:2011fc}. NNPDF2.3 NLO PDFs and
$\alpha_{s}(M_{Z})=0.119$ are used with all programs.}
\end{figure}

In this section, we compare predictions
of {\tt APPLgrid}, {\tt FastNLO}, and {\tt MEKS} for inclusive jet production 
in ATLAS
at 7 TeV \cite{Aad:2011fc}. In Fig.~\ref{jetbench1}, the 2010 ATLAS
data set (with the jet cone size $R=0.4$) is compared to NLO predictions
from {\tt APPLgrid}, {\tt FastNLO} (version 2), and {\tt MEKS}, using the NNPDF2.3 NLO
PDF set. The cross sections are plotted
vs. jet transverse momentum, $p_{T}$, in seven bins of the magnitude
$|y|$ of jet rapidity. The error bars show the experimental data
with the statistical and uncorrelated systematic errors added in quadrature:
no correlated systematic shifts are included. In each $p_{T}$
bin, all cross sections are normalized to the corresponding prediction
from {\tt FastNLO}. The cross sections of the central {\tt FastNLO} prediction
are computed using the $p_{T}$ of each individual jet as the renormalization
and factorization scale, $\mu_{R}=\mu_{F}=p_{T}^{\rm ind}$. Hatched
bands represent a scale variation of the {\tt FastNLO} predictions, obtained
by varying $\mu_{R}$ and $\mu_{F}$ separately in the intervals $p_{T}^{\rm ind}/2\leq\mu_{R,F}\leq2p_{T}^{\rm ind}$.
Three colored lines correspond to two predictions from {\tt MEKS} and 
a prediction from {\tt APPLgrid}. 

The {\tt MEKS} cross sections are obtained with the scales equal to the
individual jet $p_{T}^{\rm ind}$ (same as in {\tt FastNLO} and denoted by
{\tt MEKS1}) or the hardest jet $p_{T}^{hard}$ in each event ({\tt MEKS2}). In
the {\tt MEKS1} convention, if the transverse momenta $\{p_{T}\}=\{p_{T}^{(1)},p_{T}^{(2)},p_{T}^{(3)}\}$
of the jets in a three-jet event are ordered as 
$p_{T}^{(1)} > p_{T}^{(2)} > p_{T}^{(3)}$,
the event contributes cross section weights $w(\{p_{T}\},\mu=p_{T}^{(1)}),$
$w(\{p_{T}\},\mu=p_{T}^{(2)}),$ and $w(\{p_{T}\},\mu=p_{T}^{(3)})$
into the $p_{T}$ bins around $p_{T}^{(1)},$ $p_{T}^{(2)}$, and
$p_{T}^{(3)}$, respectively. In the {\tt MEKS2} convention, the event contributes
the same cross section weight $w(\{p_{T}\},\mu=p_{T}^{(1)})$ into
all three bins.

The scale choice in {\tt APPLgrid} sets $\mu_{R}$ and $\mu_{F}$ equal
to the $p_{T}$ of the hardest jet in each rapidity bin. It coincides
with the {\tt MEKS1} convention if all $p_{T}$ values fall into different
rapidity bins, but will select the larger of the two $p_{T}$ values as
the scale if two jets are in the same rapidity bin. 

In Fig.~\ref{jetbench1}, we can see that, at the largest $p_{T}$
values, all four predictions agree to within 1\%. {\tt FastNLO} and {\tt MEKS1}
agree to about 1\% even at low $p_{T}$, apart from minor fluctuations
caused by Monte-Carlo integration errors. Their agreement is not surprising,
since {\tt FastNLO} and {\tt MEKS1} follow the same scale choice.

At low $p_{T}$, the {\tt APPLgrid} event rate 
shows a systematic deficit of up to 4\% compared
to {\tt FastNLO}, while the {\tt MEKS2} rate is even smaller in this region.
This is the consequence of using the QCD scale that is equal or close
to the hardest jet $p_{T}$, which suppresses the cross section
compared to other scale choices. The {\tt MEKS2} curve lies, for the most
part, within the scale uncertainty band of the {\tt FastNLO} prediction,
with the exception of the $p_{T}<200$ GeV region. We conclude 
that the most up-to-date versions of the parton-level NLO programs show a
very good agreement for the same scale choice. However, the scale
dependence of the NLO cross section is an important systematic uncertainty,
its magnitude is of the same order as the experimental correlated 
systematic errors. In Fig.~\ref{jetbench1}, which shows the
experimental data without the correlated systematic errors, 
the difference between the
theoretical predictions $T_{k}$ and the unshifted central data values $D_{k}$
provides a crude estimate of the size
of the correlated systematic error. As seen in the last section, the 
quality of the fit is very good, so the data and theory predictions can be 
brought into line using shifts of data corresponding to the size of the 
correlated errors, or less. In fact, it can be checked 
from the results in~\cite{Aad:2011fc} that 
this is a reasonable approximation, especially at 
the highest $p_T$ values and the highest rapidity bins, where the 
systematic uncertainty is larger than the difference between
$T_{k}$ and $D_{k}$ in Fig.~\ref{jetbench1}. 
The scale uncertainty, defined as above,
varies from about 15\% of $T_{k}-D_{k}$ in the bins with
the small rapidity to 40\% at the largest $\left|y\right|$. Hence, 
the contribution of the scale uncertainty is significant compared to
the experimental systematic uncertainty, and 
reduces the sensitivity of LHC inclusive (di)jet production to different
PDF models, particularly at the highest rapidities.\footnote{Theoretical uncertainties should be treated in a completely
different footing of experimental uncertainties, and in particular,
should not be included in the $\chi^2$ definition. The issue of the
proper
inclusion of theory uncertainties in PDF analysis is beyond
the scope of this paper, and thus we do not explore it further.}

\begin{figure}[h]
\begin{centering}
\includegraphics[width=\textwidth]{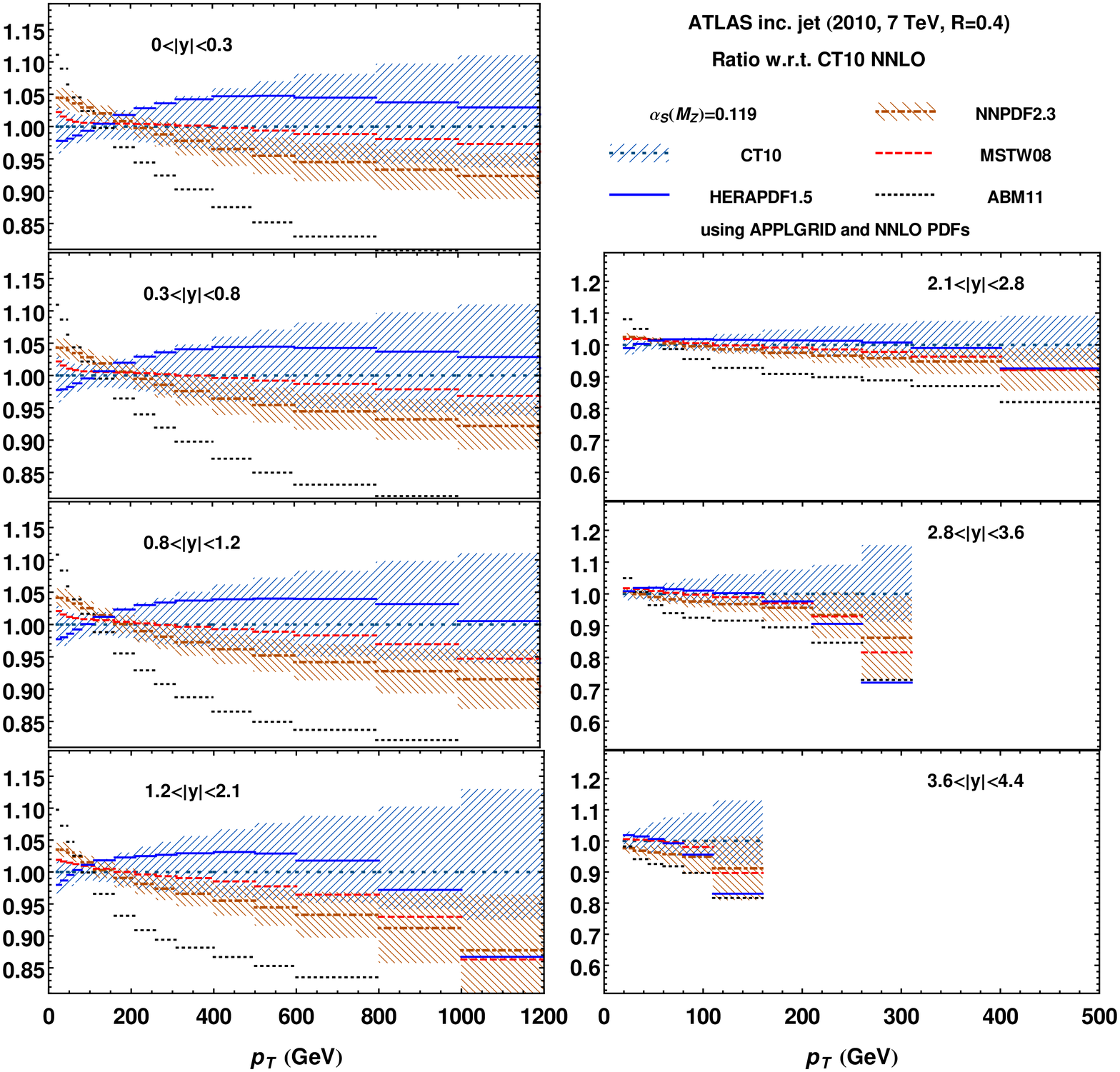} 
\par\end{centering}

\vspace{-3ex}
 \caption{\label{jetbench2} Comparison of NLO theoretical predictions obtained
with various NNLO PDF sets for the 2010 ATLAS measurement of single-inclusive
jet production~\cite{Aad:2011fc}. {\tt APPLgrid} and $\alpha_{s}(M_{Z})=0.119$
are used with all PDF sets.}
\end{figure}

\subsection{PDF dependence}

As already seen in the previous section, 
all available PDF sets can fit well the current ATLAS jet data, 
which therefore does not provide much discrimination. 
However, there are still interesting features to pick out which will 
become more important for future data.  
Fig.~\ref{jetbench2} compares the corresponding NLO predictions 
made using {\tt APPLgrid}
and various NNLO PDFs: ABM11, CT10, HERA1.5, MSTW08, and NNPDF2.3.
We take $\alpha_{s}(M_{Z})=0.119$ both in the hard cross sections
and PDFs for all PDF sets. All the predictions are normalized to the
central prediction based on the CT10 NNLO PDF set (with $\alpha_{s}=0.119$).
For the NNPDF2.3 and CT10 sets, we show the 68\% C.L. PDF uncertainties
by the hatched bands. The CT10 central predictions
are larger than NNPDF2.3 or MSTW2008, mainly due
to the harder gluon distribution in the CT10 set. In general, predictions
from different PDFs agree with each other within the range of PDF uncertainties,
apart from ABM11, particularly at low rapidities. 
It is also instructive to compare the scale uncertainties shown
in Fig.~\ref{jetbench1} with the PDF uncertainties shown in Fig.~\ref{jetbench2}.
In the low $p_{T}$ region, i.e. less than $p_T\sim 200$~GeV, 
the scale uncertainty of NLO predictions
is comparable to, or even larger than, the PDF uncertainties from
CT10. This is another indication that the scale uncertainty presents
a limiting factor in the discrimination between the PDF sets, especially for
PDFs which are already well-constrained, in this case by HERA data.

\begin{table}[h!]
\begin{centering}
\begin{tabular}{c|cccc}
\hline 
PDF set & $\chi^{2}/N_{pt}$  & $\chi_{D}^{2}$  & $\chi_{\lambda}^{2}$  & $\lambda_{0,\rm lum}$ \tabularnewline
\hline 
\hline 
ABM11   & 0.81  & 44.4  & 28.5  & -1.12 \tabularnewline
\hline 
CT10  & 0.81  & 47.4  & 25.5  & -1.76 \tabularnewline
\hline 
CT10 NLO  & 0.94  & 54.0  & 30.6  & -1.18 \tabularnewline
\hline 
HERA1.5   & 0.85  & 50.7  & 25.8  & -2.36 \tabularnewline
\hline 
MSTW08   & 0.79  & 45.7  & 25.1  & -2.00 \tabularnewline
\hline 
NNPDF2.3  & 0.79  & 42.4  & 29.1  & -1.88 \tabularnewline
\hline 
\end{tabular}
\par\end{centering}

\caption{\label{jetchi2} $\chi^{2}/N_{\rm pt}$ values for the 2010 ATLAS single-inclusive
jet data ($R=0.4$) computed according to Eq.~(\ref{Chi2sys}) using
{\tt FastNLO} (version 2) and various NNLO PDF sets and the CT10 NLO
set. The 
$\chi_{D}^{2}$ and $\chi_{\lambda}^{2}$ contributions to $\chi^{2}$
from the data residuals and penalties for systematic shifts 
defined in Eqs.~(\ref{Chi2D}) and (\ref{Chi2lambda}) are shown. The last
column contains the best-fit luminosity parameter shift $\lambda_{0,\rm lum}$ for
each PDF set. We have used 
 $\alpha_s=0.119$ for all sets. }
\end{table}

\subsection{Systematic shifts in a fit to the ATLAS jet data}

When the NLO theoretical predictions are compared to the ATLAS inclusive
jet data without including the systematic errors, as in Fig.~\ref{jetbench1},
one generally finds a very poor agreement for any PDF set. In this
case, the $\chi^{2}$ value can reach several thousand units for a
total of $N_{\rm pt}=90$ data points. The agreement is improved dramatically
after the correlated systematic errors are considered. This can be 
done, {\it e.g.}, 
by including a term with a correlation matrix $\beta_{k\alpha}$
into the log-likelihood function $\chi^{2}$  \cite{Pumplin:2002vw}, 
as described in the appendix.
We will use the definition of $\chi^{2}$ provided by Eq.~(\ref{Chi2sys}),
which introduces a normally distributed nuisance parameter $\lambda_{\alpha}$
(with the central value of zero and standard deviation of one) to characterize
each of $N_{\lambda}$ correlated errors. 

The ATLAS measurement provides
$88$ sources of correlated systematic errors, including the luminosity
error and the uncertainty in the nonperturbative correction. Each
of these errors can cause variations (shifts) of the experimental
points from their central values. In addition, each
data point is affected by an uncorrelated systematic error, which
is significant compared to the statistical error. 
When both uncorrelated and correlated systematic uncertainties
are included into $\chi^{2}$, the resulting $\chi^{2}/N_{\rm pt}$ values
are less than $1$ for all considered NNLO PDFs, 
as shown in Table~\ref{jetchi2}.
In this comparison, $\chi^{2}$ is computed according to the procedure
summarized in Sec.~\ref{app:shifts} and numerically equivalent 
to Eq.~(\ref{eq:covmat}). None of the PDF sets is preferred by these
$\chi^{2}$ values.
As one can see the $\chi^2$ values are extremely similar 
to those in the previous section, 
Tables~\ref{tab:chi2-nnlo-as0117} and \ref{tab:chi2-nnlo-as0119}, 
even though they are computed with a different code ({\tt FastNLO}).

\begin{figure}[h]
\begin{centering}
\includegraphics[width=\textwidth]{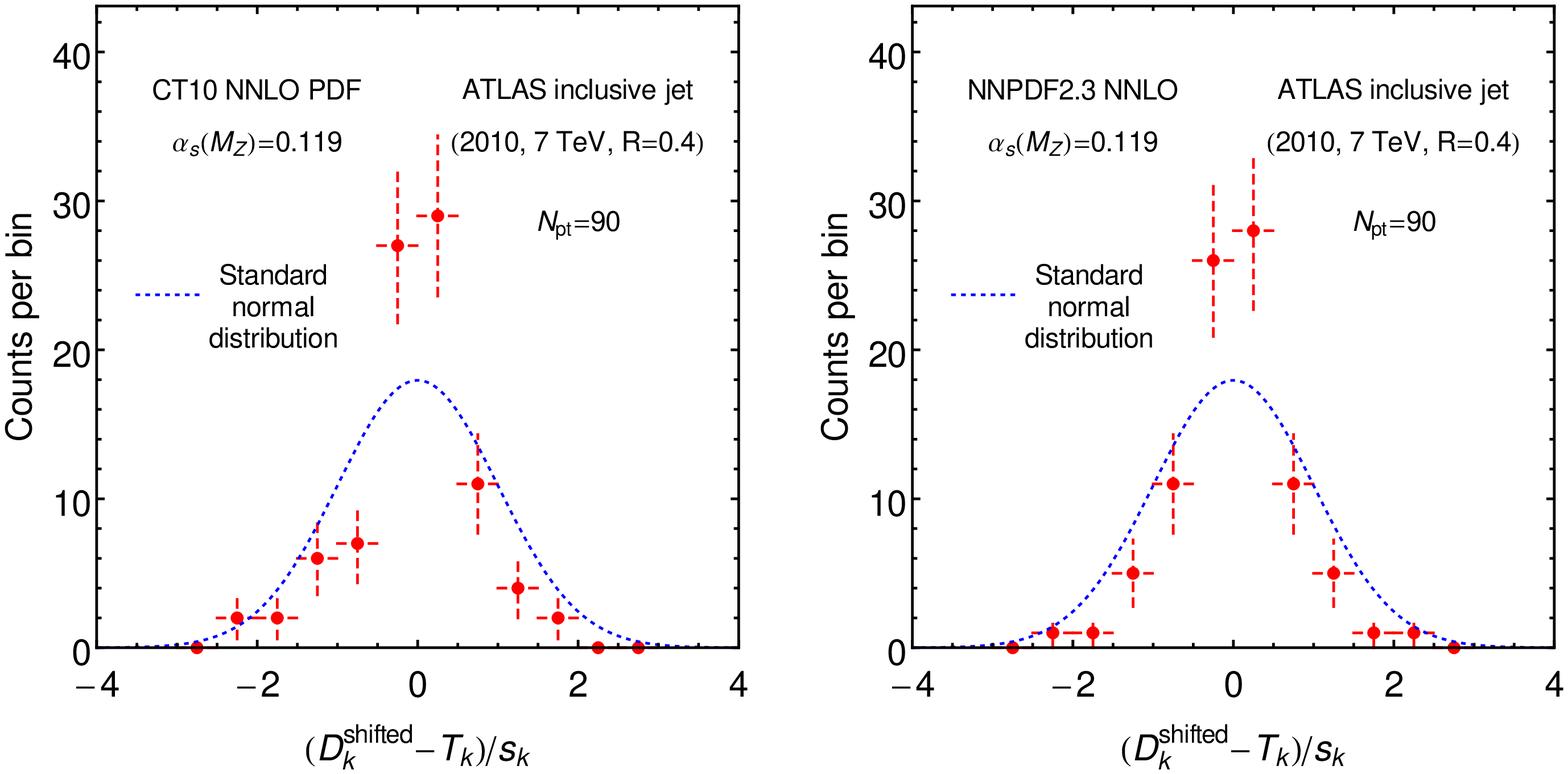} 
\par\end{centering}
\caption{\label{jetres} Distribution of residuals for the fit of 2010 ATLAS
single-inclusive jet data ($R=0.4$). Left (right) plot corresponds
to using NLO theoretical predictions from {\tt FastNLO} v.2 with CT10 (NNPDF2.3)
NNLO PDFs and $\alpha_{s}(M_{Z})=0.119$.}
\end{figure}

\begin{figure}[h]
\begin{centering}
\includegraphics[width=\textwidth]{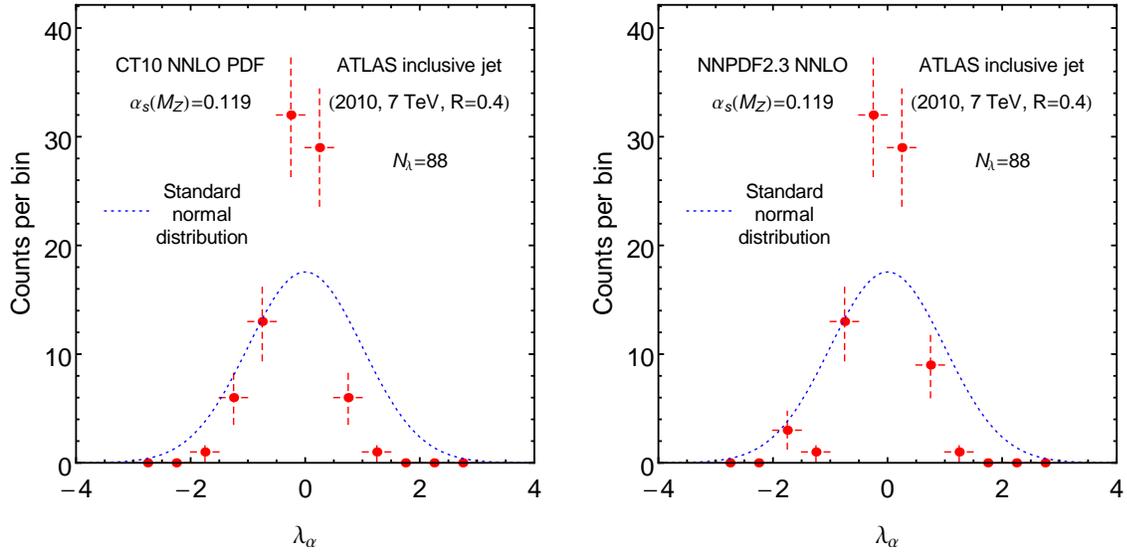} 
\par\end{centering}
\caption{\label{jetlam} Distribution of best-fit nuisance parameters $\lambda_{\alpha}$
for the fit to the 2010 ATLAS single-inclusive jet data ($R=0.4$).
Left (right) plot corresponds to using NLO theoretical predictions
from {\tt FastNLO} v.2 with CT10 (NNPDF2.3) NNLO PDFs and $\alpha_{s}(M_{Z})=0.119$.}
\end{figure}

For each set of theoretical predictions $\{T_{k}\}$, we can also
determine the value $\lambda_{0\alpha}$ of each nuisance parameter
that gives the best description of the data. It is found according
to Eq.~(\ref{lambda0}) once the $\{T_{k}\}$ values are known. In
Eq.~(\ref{Chi2sys}) for the total $\chi^{2}$, we can identify two
parts: $\chi_{D}^{2}$ containing contributions from the data residuals
$d_{k}=(D_{k}^{\rm shifted}-T_{k})/s_{k}$, where $D_{k}^{\rm shifted}=D_{k}-\sum_{\alpha}\beta_{k\alpha}\lambda_{0\alpha}$;
and $\chi_{\lambda}^{2},$ which is a quadrature sum $\sum_{\alpha}\lambda_{\alpha}^{2}$
of the shifted nuisance parameters. We list $\chi_{D}^{2}$ and $\chi_{\lambda}^{2}$
separately in Table~\ref{jetchi2} and include histograms of the
data residuals $d_{k}$ and best-fit parameters $\lambda_{0\alpha}$
in Figs.~\ref{jetres} and~\ref{jetlam}. In the histograms (which
are shown here for CT10 NNLO and NNPDF2.3 NNLO PDFs, but are also
representative of the histograms for the other NNLO PDF sets), the
observed $d_{k}$ and $\lambda_{0\alpha}$ distributions are narrower
than the standard normal distributions shown by the dotted curves.
In other words, the fit to the 2010 ATLAS data is too good and can't
distinguish between the PDF sets. Most  of 88 best-fit parameters
$\lambda_{\alpha0}$ are close to zero, i.e., they don't contribute
much to the improvement of $\chi^{2}$. None of the best-fit parameters
included in Fig.~\ref{jetlam} has changed by more than 2.5 standard
deviations. 

At the Tevatron, some PDF sets required a shift in the data downwards
due to the  luminosity uncertainty  by as much as 3-4 standard deviations in order
to agree with the single-inclusive jet production data, cf. the appendix
in Ref.~\cite{Thorne:2011kq}. In that paper, it was argued that such
shifts are not strictly allowed. The luminosity is common to the 
data on the $Z$ and $W$ total cross sections and the $Z$ rapidity 
distribution, which are rather constraining, 
and for which the PDF predictions are consistent 
with the nominal luminosity, or even a shift in the data
upwards due to the luminosity uncertainty.
 It should be a mandatory test of PDFs that they fit the Tevatron
and LHC 
jet and vector boson production data simultaneously, while the luminosity
uncertainty is treated as completely correlated between the two types of 
measurement coming from the same experiment and the same data taking 
period. This has not been checked for all PDF sets and 
could help explain how some inconsistencies may arise.  
Note that Fig.~17 in Ref.~\cite{Thorne:2011kq} is of the same form as
Fig.~\ref{jetlam}, but for Tevatron jet data. For the Tevatron inclusive 
jet data the distribution of the $\lambda_{0\alpha}$ is as expected, or even 
wider for poorly fitting PDFs, in contrast to those for ATLAS data.

The last column of Table~\ref{jetchi2} lists the best-fit values
of the luminosity shift parameter in the ATLAS measurement, computed
with the {\tt FastNLO} code. Only one PDF set (HERA1.5 NNLO) requires a
$2.4\sigma$ shift in the ATLAS luminosity. 
However, none of the PDF sets requires a luminosity shift by more
than $3\sigma,$ suggesting that they are all compatible with the
2010 ATLAS jet data. This is despite the wide variety of predictions 
exhibited in Fig.~\ref{jetbench2}. Clearly the improvement 
of the correlated systematic errors will be a priority for future data, 
since at present the shifts in data 
can accommodate quite dramatic differences in predictions
without a large penalty in $\chi^2$.

\clearpage

\section{Combined uncertainties in
Higgs production}
\label{sec:higgs}

In this section we discuss in somewhat greater detail 
PDF and $\alpha_s$ uncertainties for Higgs production
via gluon fusion
at the LHC, and specifically how PDF updates affect results obtained
using the PDF4LHC 
recommendation~\cite{Botje:2011sn} for the determination of
 PDF+$\alpha_s$ uncertainties. At NLO this prescription 
entails finding the envelope of CT, MSTW and NNPDF  
PDF+$\alpha_s$ uncertainty bands, each obtained with a different
choice for the central value of $\alpha_s$. The outer bands of the envelope
are taken as the upper and lower limits of uncertainty, and the
midpoint value as 
the best prediction. When the prescription was published,  of 
the three PDF sets included in the prescription, only MSTW
was available at NNLO. The NNLO prescription recommended  taking 
the MSTW08 prediction as the central value, while rescaling the MSTW08
uncertainty by a factor determined comparing at NLO the MSTW08
uncertainty to the envelope uncertainty. 

The  NNLO cross section for Higgs production at LHC (8 TeV)
is currently quoted by the Higgs Cross Section Working Group
(HXSWG)\footnote{https://twiki.cern.ch/twiki/bin/view/LHCPhysics/CERNYellowReportPageAt8TeV} as
\be
\sigma_H^{\rm NNLO} = 19.52 \pm 1.41~{\rm pb}, \qquad (\pm7.2\%~"{\rm PDF}+\alpha_s").
\ee
The HXSWG cross section numbers have
been  
computed with the current (2010) PDF4LHC prescription, $m_H=125$ GeV, 
and de Florian-Grazzini
code~\cite{deFlorian:2012yg}, which incorporates soft-gluon
effects up to next-to-next-to-leading logarithmic accuracy on top of the
exact NNLO calculation. Since in this work we use fixed order NNLO
calculations as implemented in {\tt iHixs}, the central values that
we will quote cannot be compared directly to the HXSWG numbers.
However, this should have a minimal effect on the percentage 
PDF+$\alpha_s$ uncertainty.

We can thus investigate how the combined  PDF+$\alpha_s$ uncertainties would
change if computed using an envelope prescription based on 
the most updated NNLO PDFs from the three global sets: 
NNPDF2.3, MSTW08 and CT10. Instead of the exact
implementation of the PDF4LHC envelope, see {\it e.g.}, 
Refs.~\cite{Demartin:2010er,Watt:2011kp}, for simplicity we 
use the following definition: 
we compute the combined PDF+$\alpha_s$ uncertainties for
the three PDF sets for $\alpha_s=0.117$ and $\alpha_s=0.119$ and
let the maximum and minimum values of the cross section in this range
define the envelope. Combined  PDF and $\alpha_s$ uncertainties 
are obtained  adding the two uncertainties  in quadrature.
The uncertainty on $\alpha_s$ is taken to be $\delta {\alpha_s}=0.0012$
at the 68\% confidence level. 
The central
value is taken as the midpoint of the envelope defined in this way.

This differs from the 2010 PDF4LHC prescription because in the latter the
prediction from each of the three sets is obtained using a different
value of $\alpha_s$ ($\alpha_s=0.118$ for CTEQ, $\alpha_s=0.119$ for
NNPDF and $\alpha_s=0.120$ for MSTW), and also because $\alpha_s$ and PDF
uncertainties are added in quadrature instead of being determined
exactly in the Hessian or Monte Carlo method (though
in the Hessian  method the two procedures are equivalent~\cite{cteqas}).
The change in $\alpha_s$ range moves the central value a little,
however, because the width of the $\alpha_s$ range is unchanged
the uncertainty is not affected significantly.
Adding the PDF and 
$\alpha_s$ uncertainties in 
quadrature reduces somewhat the MSTW08 uncertainty. 
Note also that the addition in quadrature was a
simplification in the original PDF4LHC prescription.
We used it  because we think
it is more suitable for benchmarking (which is the goal of this paper)
while  asymmetric $\alpha_s$ uncertainties
 may be more accurate and thus better for
phenomenology.

As in Sect.~\ref{sec:LHCincl}, the cross sections are
computed at NNLO with the {\tt iHixs} 
code~\cite{Anastasiou:2011pi}.
 The central scale has been taken to be
$Q=m_H$, which is the same choice used for the default
predictions for Higgs production adopted by
the Higgs Cross Section Working Group~\cite{Dittmaier:2011ti}.

We begin by computing the envelope defined as above at
NLO with the same NLO PDF sets of 2010 
PDF4LHC prescription: CTEQ6.6, MSTW08, and NNPDF2.0. 
The corresponding results for $\alpha_s=0.117$ and $0.119$ are summarized 
in   
Table~\ref{tab:higgstab}. The envelope is
\be
\sigma^{\rm NLO}_H = 13.98 \pm 0.85~{\rm pb}, 
\qquad (\pm6.1\%~"{\rm PDF}+\alpha_s"),
\ee
so the uncertainty is a bit smaller than the current HXSWG result.

Next, we repeat the computation
of the NLO envelope, but now with the most up-to-date PDF sets:
CT10, MSTW08, and NNPDF2.3. Results are also summarized in
Table~\ref{tab:higgstab}, and lead to the 
envelope:
\be
\sigma^{\rm NLO}_H = 14.05 \pm 0.86~{\rm pb}, 
\qquad (\pm6.1\%~"{\rm PDF}+\alpha_s").
\ee
so neither the central value nor the uncertainty change significantly.
Note that the 
increase in the Higgs cross section using NNPDF2.3, as compared to
NNPDF2.0, does not lead to an increase
of the combined PDF+$\alpha_s$ error since the CT10 prediction
also increases by a similar amount. 

Finally, using the NNLO
cross sections from the most updated NNLO PDF sets, but otherwise 
using the same prescription as at NLO, we obtain
\be
\sigma_H^{NNLO} = 18.75 \pm 1.24~~{\rm pb}, \qquad (6.6\%~"{\rm PDF}+\alpha_s").
\ee
The combined PDF+$\alpha_s$ error is thus essentially unchanged when
going from NLO to NNLO, while the central value is within 
2\% from the MSTW2008 NNLO value of $18.45$~pb, which in the 2010
PDF4LHC prescription was taken 
as the central value.

%%%%%%%%%%%%%%%%%%%%%%%%%%%%%%%%%%%%%%%%%%%%%%%%%%%%%%%%%%%%%%%%
\begin{table}[t]
  \centering

 \begin{tabular}{c||c|c|c}
\hline
\multicolumn{4}{c}{{\bf 2010 NLO PDFs}}\\
 \hline
 $\alpha_S(M_Z)$ & NNPDF2.0 & MSTW08& CTEQ6.6  \\
 \hline
 \hline
0.117         &  14.04 $\pm$ 0.20 $\pm$ 0.27   & 13.94   $\pm$ 0.22 $\pm$ 0.27    &  13.49 $\pm$ 0.27 $\pm$ 0.24   \\ 
0.119         &  14.49  $\pm$ 0.21 $\pm$ 0.27   &   14.38 $\pm$ 0.23 $\pm$ 0.27   &  13.88 $\pm$ 0.28 $\pm$ 0.24  \\ 
 \hline
 \end{tabular}  

\vspace{0.4cm}

\begin{tabular}{c||c|c|c}
 \hline
\multicolumn{4}{c}{{\bf 2012 NLO PDFs}}\\
 \hline
 $\alpha_S(M_Z)$ & NNPDF2.3 & MSTW08& CT10  \\
 \hline
 \hline
0.117         &  14.21 $\pm$ 0.20 $\pm$ 0.25   & 13.94   $\pm$ 0.22 $\pm$ 0.27    &  13.57 $\pm$ 0.28  $\pm$ 0.26   \\ 
0.119         &  14.61  $\pm$ 0.17 $\pm$ 0.25   &   14.38 $\pm$ 0.23 $\pm$ 0.27   &  14.00 $\pm$ 0.29  $\pm$ 0.26  \\ 
 \hline
 \end{tabular} 

\vspace{0.4cm}

\begin{tabular}{c||c|c|c}
 \hline
\multicolumn{4}{c}{{\bf 2012 NNLO PDFs}}\\
 \hline
 $\alpha_S(M_Z)$ & NNPDF2.3 & MSTW08& CT10  \\
 \hline
 \hline
0.117         &  18.90 $\pm$ 0.20 $\pm$ 0.38  &  18.45  $\pm$ 0.24 $\pm$ 0.40   &  18.05 $\pm$ 0.36 $\pm$ 0.41  \\ 
0.119         &  19.54 $\pm$ 0.25 $\pm$ 0.38  &  19.12 $\pm$ 0.25 $\pm$ 0.40  &   18.73 $\pm$ 0.37 $\pm$ 0.41    \\ 
 \hline
 \end{tabular}  
  \caption{\label{tab:higgstab}  
\small The Higgs boson production 
cross section (in pb) in the gluon fusion channel, for $m_H=125$~GeV
at LHC 8 TeV. The two uncertainties shown in
each case are the PDF and $\alpha_s$ uncertainty.
}
\end{table}
%%%%%%%%%%%%%%%%%%%%%%%%%%%%%%%%%%%%%%%%%%%%%%%%%%%%%%%%%%%

%%%%%%%%%%%%%%%%%%%%%%%%%%%%%%%%%%%%%%%%%%%%%%%%%
\begin{figure}[t]
    \begin{center}
\includegraphics[width=0.48\textwidth]{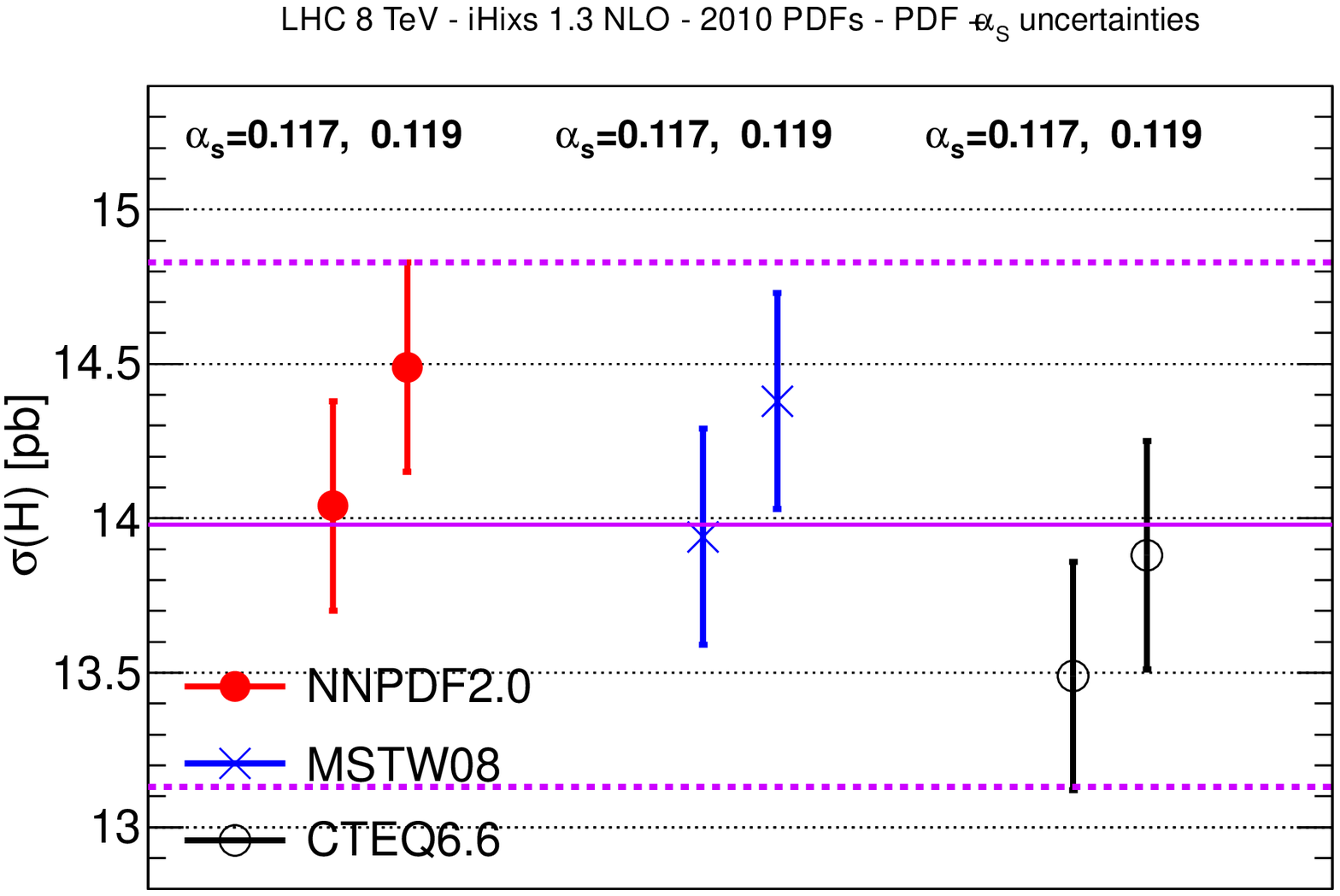}\quad
\includegraphics[width=0.48\textwidth]{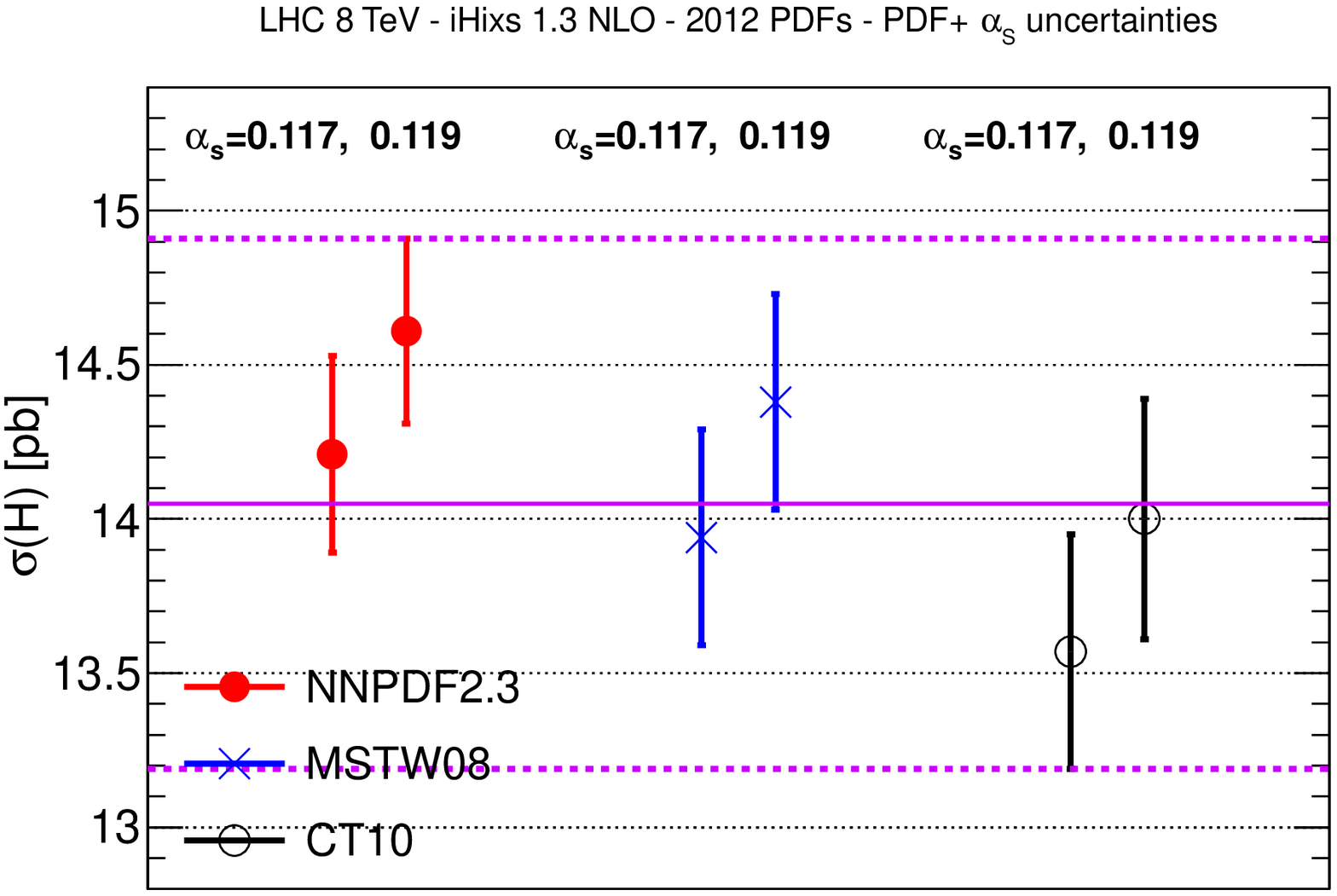}
      \end{center}
     \caption{\small The Higgs boson production
cross section in the gluon fusion channel using the 
NLO PDF sets included in the PDF4LHC prescription for $\alpha_s=0.117$ and
0.119. The left plot has been computed with 2010 PDFs and
the right plot with 2012 PDF sets. The envelope (dashed violet
horizontal lines) is defined by the upper and lower values
of the predictions from all the three PDF sets and the two values
of $\alpha_s$. The solid violet horizontal line is the midpoint
of the envelope.
    \label{fig:h8nlo} }
\end{figure}
%%%%%%%%%%%%%%%%%%%%%%%%%%%%%%%%%%%%%%%%%%%%%%%%%%

\begin{figure}[h]
    \begin{center}
\includegraphics[width=0.49\textwidth]{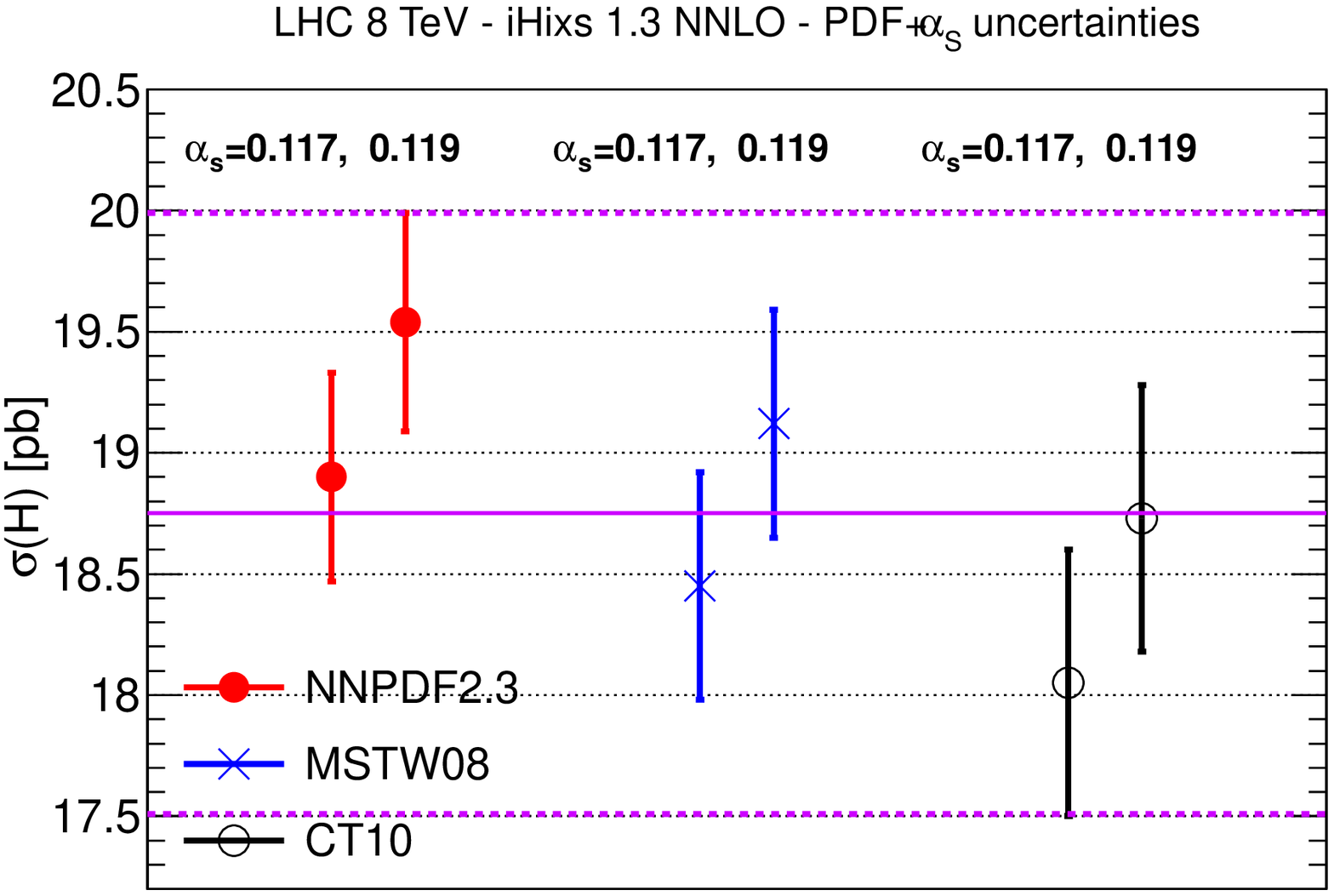}
      \end{center}
     \caption{\small Same as Fig.~\ref{fig:h8nlo}, but using 2012 NNLO PDFs.
    \label{fig:h8nnlo} }
\end{figure}

These cross sections are plotted  in Fig.~\ref{fig:h8nlo} and 
Fig.~\ref{fig:h8nnlo}, showing both the cross sections 
from each individual PDF set and the  envelope.

In summary, neither the central value nor the uncertainty on the NLO
prediction are significantly affected when replacing 2010 PDF with
2012 PDFs, and if the NLO PDF4LHC prescription is also used at NNLO, 
the combined PDF+$\alpha_s$ uncertainty for the 
Higgs cross section 
moderately rises from 6.1\% to 6.6\% when going from NLO to NNLO.

In this respect, the gluon fusion channel with $m_H=125$ GeV is an
unusually unlucky case: for most standard candle processes, as well
as for other Higgs production modes, and even for gluon fusion, but
with other values of the Higgs mass, the uncertainties decrease when going
from 2010 NLO PDFs to 2012 NNLO PDFs, as it is clear from comparing the
luminosity plots of Section~\ref{sec:pdfs} with analogous plots from previous
benchmarks~\cite{Watt:2011kp,Watt:2012np}.

%%%%%%%%%%%%%%%%%%%%%%%%%%%%%%%%%%%%%%%%%%%%%%%%%
%%%%%%%%%%%%%%%%%%%%%%%%%%%%%%%%%%%%%%%%%%%%%%%%%%

\begin{figure}[h]
    \begin{center}
\includegraphics[width=0.48\textwidth]{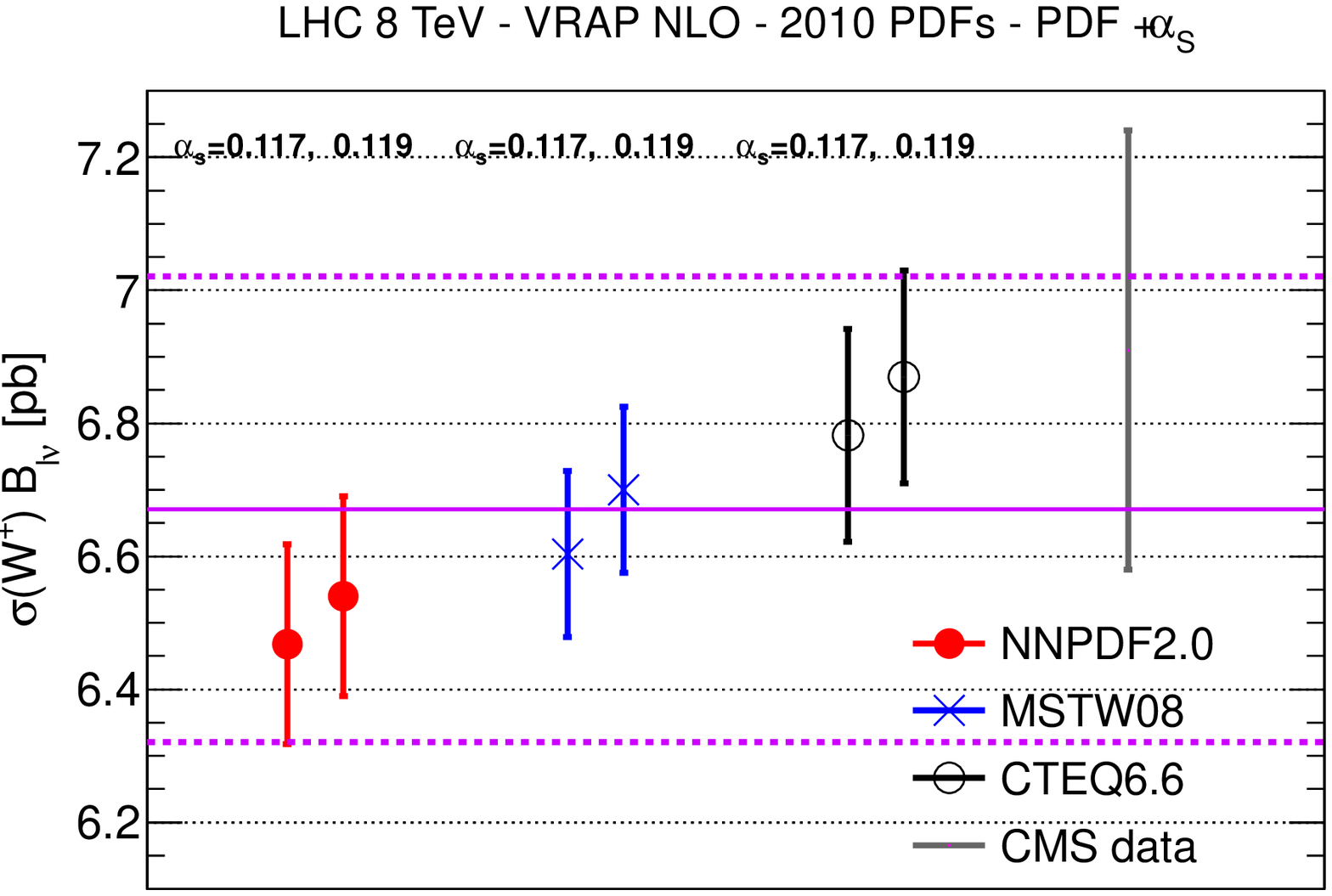}
\includegraphics[width=0.48\textwidth]{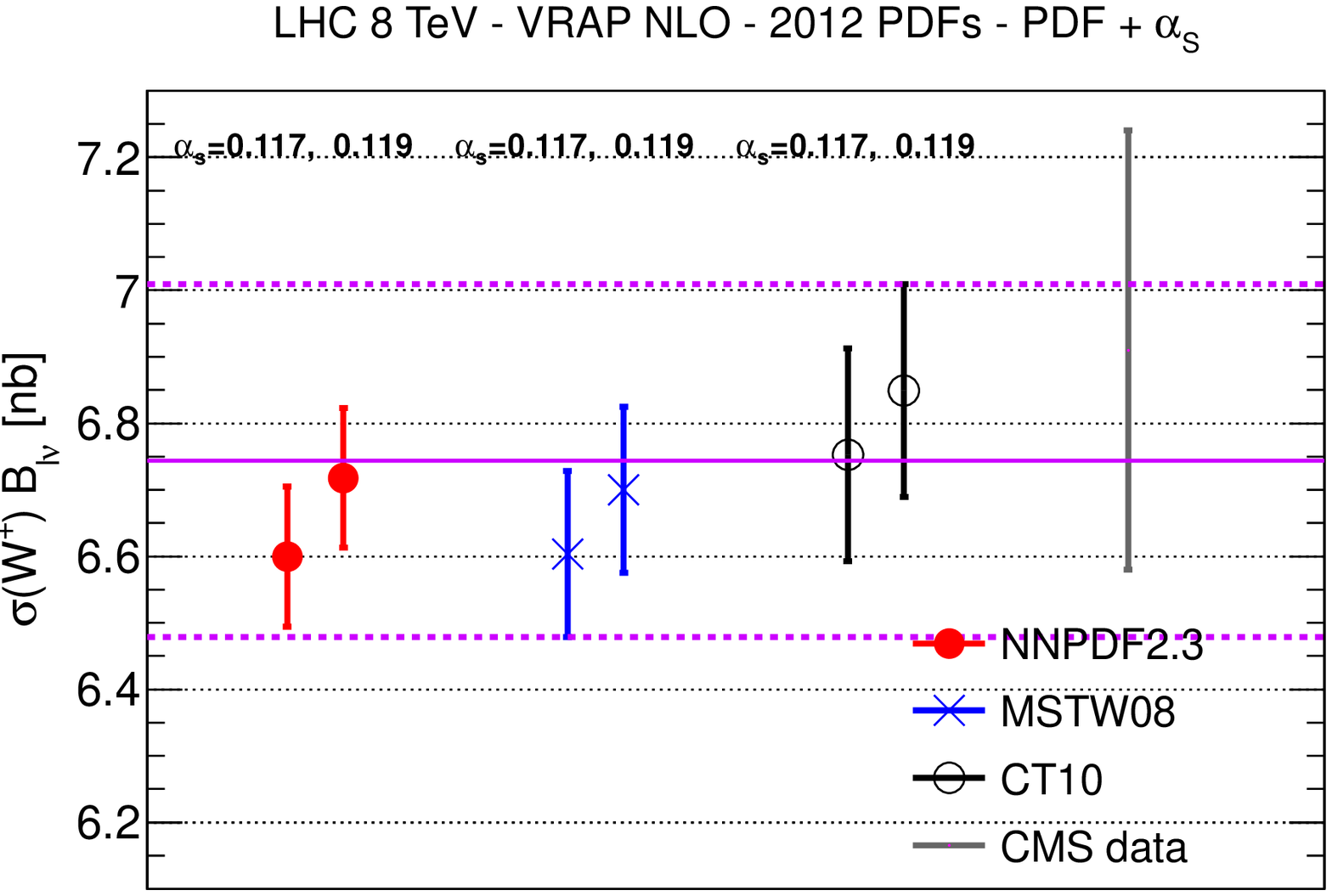}\\
\includegraphics[width=0.48\textwidth]{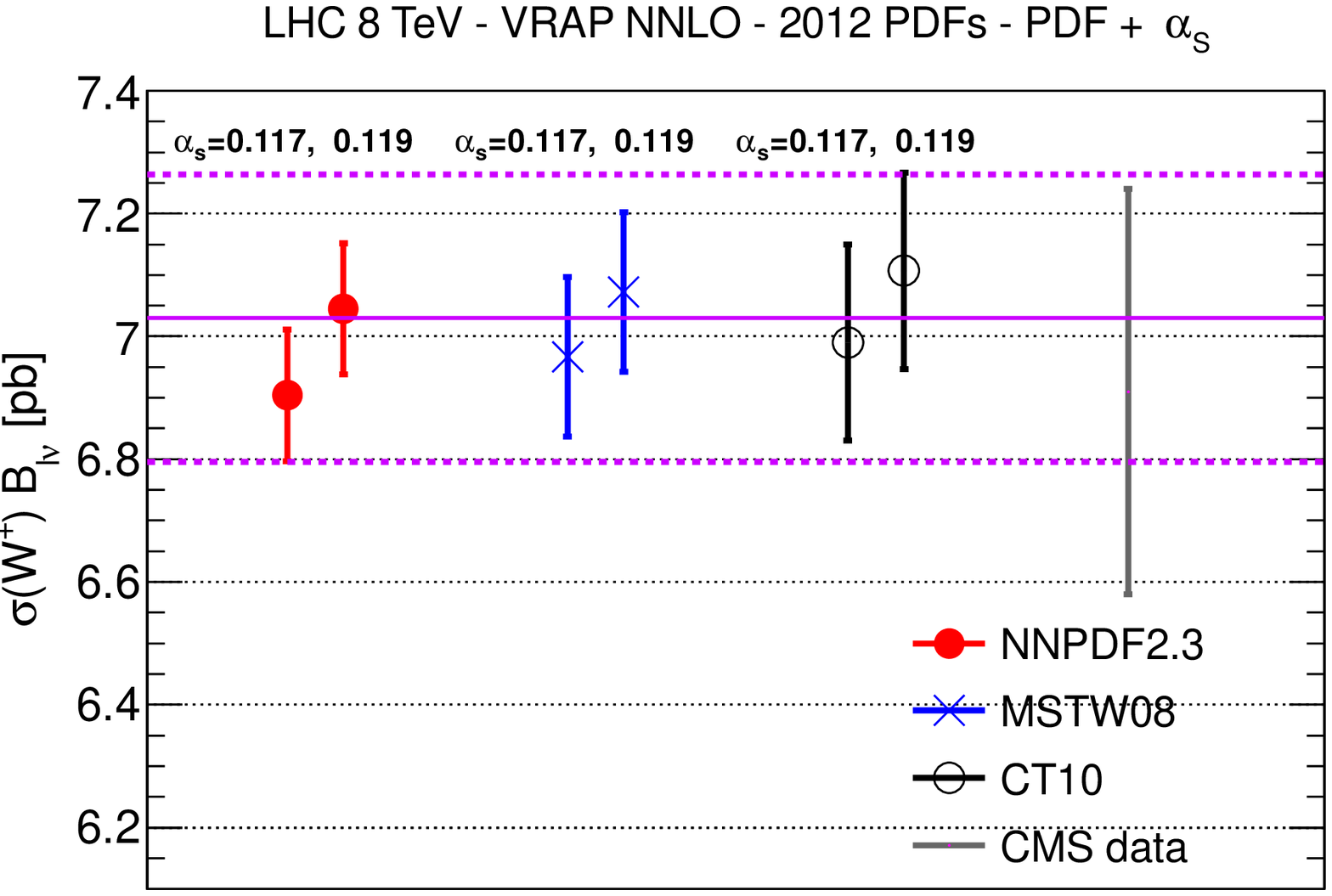}
      \end{center}
     \caption{\small The
$W^+$ production
cross sections determined using the same PDFs and envelope as in
Figs.~\ref{fig:h8nlo}-\ref{fig:h8nnlo}. 
The upper plots
show the NLO comparison based on 2010 PDFs (left plot) and
on 2012 PDFs (right plot). 
The lower plot show the
comparison with the  2012 NNLO PDFs. 
The recent 8 TeV CMS measurement
is also shown.
    \label{fig:wpbench} }
\end{figure}

To illustrate this explicitly, we compare
in Fig.~\ref{fig:wpbench}
 predictions  for
$W^+$ boson production 
based on NLO PDFs, both from 2010
and from 2012, and 2012 NNLO PDFs from CT, MSTW and NNPDF. 
The improved agreement of the PDF sets when going
from 2010 to 2012 PDFs
is clear: the relative PDF+$\alpha_s$ uncertainty, defined with the
same prescription as for the Higgs cross section, goes
down from  $\Delta_{\rm PDF+\alpha_s}=5.3\%$ to  $\Delta_{\rm PDF+\alpha_s}
=3.3\%$, {\it i.e.} from more than twice the MSTW2008 uncertainty 
(sometimes used as a simple approximation to the full envelope) to
about 1.5 times the MSTW2008 uncertainty. 
Several factors contribute
to this improvement, which include for instance  the adoption of 
a GM-VFN scheme in NNPDF2.1 and a more similar choice of data sets
in the different fits.
Similar improvements are expected in all quark-initiated cross sections.

%%%%%%%%%%%%%%%%%%%%%%%%%%%%%%%%%%%%%%%%%%%%%%%%%
%%%%%%%%%%%%%%%%%%%%%%%%%%%%%%%%%%%%%%%%%%%%%%%%%%

\clearpage
\section{Conclusions and outlook}
\label{sec:conclusions}

In this paper we have presented an updated benchmark comparison
of the most recent NNLO PDF sets from the ABM, CT,
HERAPDF, MSTW and NNPDF collaborations. We have compared
PDFs, parton luminosities, LHC inclusive cross sections
and differential distributions, always consistently for
a common value of $\alpha_s$.

Our main result is that the agreement between the most recent 
CT, MSTW and NNPDF NNLO parton distributions is  at least as good
as it was at NLO, and in many  cases there is a clear
improvement, in that the spread of predictions from different groups
is reduced significantly. 
The HERAPDF1.5 NNLO central values are generally
in good agreement with those of CT, MSTW and NNPDF, but
with rather larger uncertainties due to the smaller dataset that HERAPDF uses. 
We find no evidence for tension between the
HERA-only PDF sets and the PDF sets based on global data sets.
It is interesting to observe that at NLO the HERAPDF1.5 set
has   smaller uncertainty and a 
more significant  disagreement with other sets. 
The improvement in methodology in 
the HERAPDF1.5 NNLO analysis seems to not only to enlarge the uncertainty, but 
also to bring the central values more in line with the other sets. 

We find that in several cases ABM11 disagrees with CT, MSTW and NNPDF both
for  PDFs and LHC cross sections,
even when a common value of $\alpha_s$ is used. For the ABM11 default 
$\alpha_s(M_Z)=0.1134$ value, many of these differences with other sets
would further increase (though the vector boson production predictions 
would become more similar). 
We have discussed some of the possible explanations of these differences. 
A plausible
 explanation seems to  be the use of the FFN scheme instead of the 
GM-VFN scheme used by the other groups, together with the absence of 
collider data in the ABM11 fit~\cite{Thorne:2011kq}.
Other, perhaps less likely explanations, include the presence of higher
twist contributions in the ABM PDF determination.
We have also shown (cf. the end of Sect.~3) that the 
8 TeV LHC data on total inclusive cross sections 
tend to disfavor ABM11, especially in
top quark pair production for the default ABM11 $\alpha_s$ value, 
though experimental uncertainties are not
yet precise enough to allow for a decisive discrimination.

For Higgs production via gluon-gluon fusion, 
we have shown  that the combined PDF+$\alpha_s$ uncertainties 
obtained from the envelope of CT, MSTW and NNPDF sets at NNLO are 
very similar to those obtained at NLO, which in turn are unchanged if
2012
instead of 2010 PDFs are used.
For several other LHC processes (in particular quark-initiated
processes) the NNLO
combined PDF+$\alpha_s$ uncertainty is smaller than  
the 2010 NLO result.

We would like to emphasize that we are not advocating here any
new prescription to combine PDF sets, but only exploring 
the robustness of the original (2010) 
recommendation with respect to the update of its PDF sets.
It is the task of the PDF4LHC Steering Committee to provide
official
updated recommendations for the use of PDFs in the comparisons
with LHC data.

Available LHC data is already providing important information
on PDFs, and future LHC data will provide even more stringent
constraints. Such constraints will come from more precise
measurements of already available processes (such as vector
boson production and jet production), measurements
of new PDF sensitive differential distributions
(such as low-mass Drell-Yan pair, $W+$charm, $t\bar t$, or
single-top production), as well as new ways 
of combining the existing data 
(such as ratios of LHC cross sections at different 
center-of-mass energies~\cite{Mangano:2012mh}).

Here we have presented only a small subset of all
the available plots. A complete repository of all available plots
is
\begin{center}
{\bf \url{http://nnpdf.hepforge.org/html/pdfbench/catalog}}\, ,
\end{center}
where in particular we provide
\begin{itemize}
\item Comparisons of PDFs and parton luminosities at NLO and NNLO,
for $\alpha_s(M_Z)=0.117$ and 0.119.
\item Comparisons of PDFs at a low scale of 2 GeV$^2$, and 
as ratios with respect to a reference set for an LHC scale
of $10^4$ GeV$^2$.
\item Comparison of PDFs to all the relevant LHC data 
from ATLAS, CMS and LHCb at NNLO, for $\alpha_s(M_Z)=0.117$ and 0.119.
\item PDF dependence of benchmark cross sections.
\end{itemize}

\bigskip
\bigskip
\begin{center}
\rule{5cm}{.1pt}
\end{center}
\bigskip
\bigskip

{\bf\noindent  Acknowledgments \\}
We are grateful to Marco Zaro for
providing the VBF results with the {\tt VBF@NNLO} program. 
We would like to thank Alan Martin, Jon Pumplin, 
James Stirling, and Graeme Watt for discussions. 
This work is supported 
by the U.S. National Science Foundation under Grant No. PHY-0855561,
the U.S. DOE Early Career Research Award DE-SC0003870, 
and by Lightner-Sams Foundation. 
J.~R. is supported by a Marie Curie 
Intra--European Fellowship of the European Community's 7th Framework 
Programme under contract number PIEF-GA-2010-272515.
 The work of R.~S.~T. is 
supported partly by the London Centre for Terauniverse
Studies (LCTS), using funding from the European Research Council 
via the Advanced Investigator Grant 267352.

\clearpage

\appendix
\section{Definitions of $\chi^2$}

\label{sec:chi2def}
\def\half{{\textstyle{1\over2}}}

The value of the $\chi^2$ estimator depends 
on the assumed functional form for $\chi^2$ 
in the presence of experimental correlated systematic
uncertainties. In this appendix, 
we document the various definitions of the $\chi^{2}$ 
function adopted in this paper and the numerical inputs that
were used to obtain our results.

Statistical experimental errors 
are usually reported in the form of a list containing 
their absolute values, 
while for systematic errors the list gives relative values
expressed as percentages of the central value. Often 
the systematic errors are asymmetric, {\it i.e.} they have different 
positive and negative deviations. The covariance matrix $({\rm cov})_{ij}$ 
is calculated from this published information by
following one of the methods described below. 
Needless to say it is important, when benchmarking the 
various PDF predictions, to state precisely how the covariance matrix
was computed. On the other hand some experiments directly provide
the covariance matrix rather than the list of systematic
errors, and in this case no ambiguity is possible.

\subsection{Definitions of $\chi^2$ 
with the covariance matrix \label{app:covmat}}

We can define the $\chi^2$ for a specific experiment with 
$N_{\rm pt}$ data points by
\begin{equation}
\chi^2 = \sum_{i,j}^{N_{\rm pt}} (T_i - D_i) ({\rm cov^{-1}})_{ij} (T_j - D_j),
\label{eq:chi2}
\end{equation}
and use it as a figure of merit to judge the agreement between theory
and data. The covariance matrix $({\rm cov})_{ij}$ used in this definition 
may be written as
\be
\label{eq:covmat}
({\rm cov})_{ij}=
\delta_{ij} s_{i}^2 + 
\lp\sum_{\alpha=1}^{N_c}\sigma^{(c)}_{i,\alpha}\sigma^{(c)}_{j,\alpha}
+ \sum_{\alpha=1}^{N_{\cal L}} \sigma_{i,\alpha}^{({\cal L})}\sigma_{j,\alpha}^{({\cal L})}
\rp D_{i} D_{j},
\ee
where $i$ and $j$ run over the experimental points ($i,j =1,...,N_{\rm pt}$),
$D_{i}$ are the measured central values, and $T_{i}$ the corresponding 
theoretical predictions computed with a given set of PDFs. 
This covariance matrix depends on uncorrelated uncertainties $s_{i}$, 
constructed by adding the statistical and uncorrelated systematic 
uncertainties in quadrature; $N_{\cal L}$ multiplicative normalization
uncertainties, $\sigma_{i,\alpha}^{({\cal L})}$; 
and  $N_c$ other correlated systematic uncertainties, 
expressed for convenience in the above equation in terms of 
their relative values $\sigma^{(c)}_{i,\alpha}$.
The total number of correlated uncertainties is thus $N_\lambda=N_{\cal L}+N_c$. 
Asymmetric systematic uncertainties provided by the experiments must be 
symmetrized to use this expression. We symmetrize them by averaging, 
$\sigma^{(c)}_{i,\alpha}=\half(\sigma^{(c),+}_{i,\alpha}+\sigma^{(c),-}_{i,\alpha})$.

Note that it is important when fitting to distinguish between additive 
uncertainties (where the experimentalists have determined a absolute 
shift in the observable due to a systematic uncertainty) 
and multiplicative uncertainties (where the experimentalists have determined 
a relative shift, as a fraction of the measured observable). In 
particular it is important not to mistake an additive 
uncertainty for a multiplicative one just because it is 
presented multiplicatively (as are the correlated systematics 
in Eq.~(\ref{eq:covmat}), where the absolute shift in data point $i$ from 
systematic uncertainty $\alpha$ is written as 
$\sigma^{(c)}_{i,\alpha}D_i$). Correlated systematics which are truly 
multiplicative should of course be treated in the same way as the normalization 
uncertainty.

This distinction is important because if Eq.~(\ref{eq:covmat}) were used 
as a figure of merit in an actual PDF fit, it would result in a D'Agostini 
bias of the multiplicative uncertainties~\cite{Ball:2009qv}. However 
it is a suitable objective criteria for comparing {\it a posteriori} 
the various predictions from the different 
PDF sets that are discussed here, and we have used it as such throughout 
the body of this paper.

An alternative definition of the covariance matrix is the 
 $t_{0}$-prescription~\cite{Ball:2009qv}, 
where a fixed theory prediction $T_{i}^{(0)}$
({\it e.g.}, the final theory prediction from a previous fit) 
is used to define the normalization contribution to the $\chi^2$.
In the $t_{0}$-prescription the covariance matrix is thus
\be
\label{eq:covmat_t00}
({\rm cov})_{ij}=
\delta_{ij} s_{i}^2 + 
\sum_{\alpha=1}^{N_c}\sigma^{(c)}_{i,\alpha}\sigma^{(c)}_{j,\alpha}D_{i} D_{j}
+ \sum_{\alpha=1}^{N_{\cal L}} \sigma_{i,\alpha}^{({\cal L})}\sigma_{j,\alpha}^{({\cal L})}
T^{(0)}_{i} T^{(0)}_{j}.
\ee
This definition has the advantage of avoiding the D'Agostini bias from 
multiplicative normalization uncertainties when performing
a PDF fit.

When the breakdown into additive and multiplicative uncertainties is
not provided by the experiment, one  may use $T_{i}^{(0)}$ to
compute all systematic uncertainties,
to give an `extended-$t_0$' prescription:
\be
\label{eq:covmat_et00}
({\rm cov})_{ij}=
\delta_{ij} s_{i}^2 + 
\lp\sum_{\alpha=1}^{N_c}\sigma^{(c)}_{i,\alpha}\sigma^{(c)}_{j,\alpha}
+\sum_{\alpha=1}^{N_{\cal L}} \sigma_{i,\alpha}^{({\cal L})}\sigma_{j,\alpha}^{({\cal L})}
\rp T^{(0)}_{i} T^{(0)}_{j}.
\ee
This prescription rescales by $T_{i}^{(0)}$ all multiplicative uncertainties
(associated with the normalization or not), but also modifies the additive
uncertainties given by the experiment in a 
mild way consistent with their overall uncertainty. 
We will see below that the $t_0$ covariance matrix Eq.~(\ref{eq:covmat_t00}) 
and  the extended-$t_0$ covariance matrix Eq.~(\ref{eq:covmat_et00}) 
generally produce lower $\chi^2$ values than the experimental definition 
in Eqs.~(\ref{eq:covmat}) for datasets with substantial systematic 
uncertainties. 

In summary, we consider in this appendix three possible definitions
of the covariance matrix:

\vspace{.2cm}

\fbox{%
\vspace{1cm}

\begin{minipage}{5.5 in}
\begin{center}
\bea
({\rm cov})_{ij}&=&
\delta_{ij} s_{i}^2 + 
\lp\sum_{\alpha=1}^{N_c}\sigma^{(c)}_{i,\alpha}\sigma^{(c)}_{j,\alpha}
+ \sum_{\alpha=1}^{N_{\cal L}} \sigma_{i,\alpha}^{({\cal L})}\sigma_{j,\alpha}^{({\cal L})}
\rp D_{i} D_{j}\, , \quad "{\rm Exp}"\nonumber \\
({\rm cov})_{ij}&=&
\delta_{ij} s_{i}^2 + 
\sum_{\alpha=1}^{N_c}\sigma^{(c)}_{i,\alpha}\sigma^{(c)}_{j,\alpha}D_{i} D_{j}
+ \sum_{\alpha=1}^{N_{\cal L}} \sigma_{i,\alpha}^{({\cal L})}\sigma_{j,\alpha}^{({\cal L})}
T^{(0)}_{i} T^{(0)}_{j}\, , \quad "t_0" \nonumber \\
({\rm cov})_{ij}&=&
\delta_{ij} s_{i}^2 + 
\lp\sum_{\alpha=1}^{N_c}\sigma^{(c)}_{i,\alpha}\sigma^{(c)}_{j,\alpha}
+\sum_{\alpha=1}^{N_{\cal L}} \sigma_{i,\alpha}^{({\cal L})}\sigma_{j,\alpha}^{({\cal L})}
\rp T^{(0)}_{i} T^{(0)}_{j}\ , \quad "{\rm Extended}-t_0"  \nonumber
\eea
\end{center}
\vspace{0.cm}
\end{minipage}

}
\vspace{.2cm}

\subsection{Definitions of  $\chi^{2}$ with 
shift parameters\label{app:shifts}}

An alternative,  yet numerically equivalent, 
representation for the $\chi^2$ function has been used
in the jet benchmarking exercise of Sec.~\ref{sec:atlasjets}, following
the method traditionally adopted in the CTEQ and MSTW PDF fits 
for jet and some other data sets. 
In this representation, the $\chi^2$ figure of merit for goodness-of-fit 
to an experiment with correlated systematic uncertainties 
is expressed as \cite{Pumplin:2002vw}
\begin{equation}
\chi^{2}(\{a\},\{\lambda\})=\chi^{2}_D+\chi^{2}_\lambda\label{Chi2sys},
\end{equation}
where
\begin{equation}
\chi^{2}_D\equiv\sum_{k=1}^{N_{\rm pt}}\frac{1}{s_{k}^{2}}\left(D_{k}-T_{k}-
\sum_{\alpha=1}^{N_{\lambda}}\beta_{k,\alpha}\lambda_{\alpha}\right)^{2}\label{Chi2D},
\end{equation}
and
\begin{equation}
\chi^{2}_\lambda\equiv \sum_{\alpha=1}^{N_{\lambda}}\lambda_{\alpha}^{2},
\label{Chi2lambda}
\end{equation}
using the same notation as in the previous section, where the
$\beta_{k,\alpha}$ are the absolute correlated uncertainties.
Systematic uncertainties associated with $N_{\lambda}$ sources 
may now induce correlated variations (shifts) in the experimental data points. 
Their effect 
is approximated by including a sum $\sum_{\alpha}\beta_{k,\alpha}\lambda_{\alpha}$
dependent on the correlation matrix $\beta_{k,\alpha}$ ($k=1,...,N_{\rm pt}$;
$\alpha=1,...,N_{\lambda}$) and stochastic nuisance parameters $\lambda_{\alpha},$
with one nuisance parameter assigned to every source of the systematic
uncertainty. By a common assumption, each $\lambda_{\alpha}$ follows the
standard normal distribution. Its deviation from $\lambda_{\alpha}=0$ 
incurs a penalty contribution $\lambda_{\alpha}^{2}$ to $\chi^{2}$. 
Under this assumption the minimum of $\chi^{2}$ 
with respect to $\lambda_{\alpha}$
can be found algebraically, since the dependence on $\lambda_{\alpha}$ is 
quadratic \cite{Pumplin:2002vw}. 

We can solve for the best-fit values $\lambda_{0\alpha}$ of the nuisance
parameters to find
\begin{equation}
\lambda_{0\alpha}=\sum_{i=1}^{N_{\rm pt}}\frac{D_{i}-T_{i}}{s_{i}}
\sum_{\delta=1}^{N_{\lambda}}
\mathcal{A}_{\alpha\delta}^{-1}\frac{\beta_{i,\delta}}{s_{i}},
\label{lambda0}
\end{equation}
with 
\begin{equation}
\mathcal{A_{\alpha\beta}}=
\delta_{\alpha\beta}+\sum_{k=1}^{N_{\rm pt}}
\frac{\mathbb{\beta}_{k,\alpha}\beta_{k,\beta}}{s_{k}^{2}}.\label{A}
\end{equation}
When these $\lambda_{0\alpha}$ values are substituted into 
Eq.~(\ref{Chi2lambda}), one obtains the usual expression Eq.~(\ref{eq:chi2}) 
for the $\chi^{2}$, with
\begin{equation}
({\rm cov})_{ij}^{-1}=\left[\frac{\delta_{ij}}
{s_{i}^{2}}-\sum_{\alpha,\beta=1}^{N_{\lambda}}\frac{\beta_{i,\alpha}}
{s_{i}^{2}}\mathcal{A}_{\alpha\beta}^{-1}\frac{\beta_{j,\beta}}{s_{j}^{2}}\right],
\end{equation}
the inverse of 
\begin{equation}
({\rm cov})_{ij}\equiv s_{i}^{2}\delta_{ij}+\sum_{\alpha=1}^{N_{\lambda}}
\beta_{i,\alpha}\beta_{j,\alpha}.
\label{eq:covmat_cteq} 
\end{equation}

If the absolute correlation $\beta_{i,\alpha}$ is related to 
the relative correlation $\sigma_{i,\alpha}$ by multiplying by 
the experimental central values for {\it both}
$\sigma_{i\alpha}^{(c)}$ and $\sigma_{i\alpha}^{(\cal L)}$, 
\begin{equation}
 \beta_{i,\alpha} = \sigma_{i,\alpha} D_i,
\label{eq:betaD}
\end{equation}
the expression in Eq.~(\ref{eq:covmat_cteq}) 
coincides with the covariance matrix introduced earlier 
in Eq.~(\ref{eq:covmat}).
It is equivalent to the 
usual definition Eq.~(\ref{eq:covmat}), but also contains 
explicit information about the values 
of the systematic parameters $\lambda_{0\alpha}$
at the best fit.

If instead of Eq.~(\ref{eq:betaD}) we set   
\begin{equation}
 \beta_{i,\alpha} = \sigma_{i,\alpha} T^{(0)}_i,
\label{eq:betaT0}
\end{equation}
we recover the extended-$t_0$ $\chi^2$ in Eq.~(\ref{eq:covmat_et00}). 
Finally, using Eq.~(\ref{eq:betaD}) to find $\sigma_{i\alpha}^{(c)}$ and 
Eq.~(\ref{eq:betaT0}) to find $\sigma_{i\alpha}^{(\cal L)}$,
we recover the $t_0$ definition in Eq.~(\ref{eq:covmat_t00}). 
Thus the $\chi^2$ values in the shift method as defined 
here are entirely equivalent 
to the methods based on direct inversion of the covariance matrix
in Sec.~\ref{app:covmat}.

\subsection{Impact on LHC cross sections}

Numerical comparisons of the different $\chi^2$ prescriptions will 
depend on the exact procedure used to determine $s_{i}$ and 
$\sigma_{i,\alpha}$. For example, 
in the comparisons to the ATLAS jet data 
in Sec.~\ref{sec:atlasjets}, we compute $\beta_{k,\alpha}$ using 
Eq.~(\ref{eq:betaD}) (equivalent to Eq.~(\ref{eq:covmat})), averaging 
any asymmetric errors. Given the 
large number of independent systematic parameters ($N_\lambda=88$), 
the asymmetry of some nuisance
parameters is not expected to significantly bias the resulting PDFs,
which has been confirmed by computing the $\chi^2$ tables 
using the same $\chi^2$ definition, but following
alternative error symmetrization procedures. In all cases examined,
the choice of the symmetrization procedure had a smaller effect 
on $\chi^2$ for the ATLAS jet data than the choice of 
the $\chi^2$ definition. 

We have also checked numerically that the covariance matrix definitions 
described in Sec.~A.1 and the corresponding shift definitions described in 
Sec.~A.2 give the same results when implemented numerically (as they should).
Thus for the remainder of this section we will focus on the difference 
between the three definitions of the covariance matrix described in Sec.~A.1.

\begin{table}[t]
\small
\begin{center}

\begin{tabular}{c||c|c|c|c|c}
 \hline
& \multicolumn{4}{c}{NLO PDFs, $\alpha_s=0.119$  } \\
\hline
 Dataset & NNPDF2.3 & MSTW08& CT10& ABM11 & HERAPDF1.5 \\
 \hline
 \hline
ATLAS $W,Z$  (Exp)         &     1.268   &  2.004   &  1.062   &  1.558 &    1.747 \\
ATLAS $W,Z$  ($t_0$)         &       1.292  &   2.024  &   1.026  &   1.487  &   1.676 \\
\hline
CMS $W$ el asy (Exp)       &     0.820   &  4.690   &  1.419   &  1.915   &  0.687 \\
CMS $W$ el asy  ($t_0$)      &      0.820  &   4.690  &   1.419  &   1.915  &   0.687 \\
\hline
LHCb $W$     (Exp)       &      0.670   &  0.907  &   1.064 &    2.328  &   4.125 \\
LHCb $W$     ($t_0$)       &       0.662    & 0.896 &    1.046 &    2.298 &    4.100 \\
\hline
ATLAS jets   (Exp)         &     0.999   &  0.974   &  1.350  &   1.342  &   1.106 \\
ATLAS jets    ($t_0$)        &      0.836  &   0.825 &    1.234 &    1.317 &    1.032 \\
 \hline
 \end{tabular}

\caption{\small The $\chi^2/N_{\rm pt}$ values for
the available LHC data with published correlated uncertainties,
computed using  the five PDF sets considered.
The experimental (``Exp'') definition 
of $({\rm cov})_{ij}$ in Eq.~(\ref{eq:covmat}) 
is compared to the $t_0$ definition in Eq.~(\ref{eq:covmat_t00}). 
Theoretical predictions have been computed
at NLO with {\tt APPLgrid} for a common value of the strong coupling
$\alpha_s\lp M_Z\rp=0.119$. The central PDF set of each
collaboration has been used to compute the $t_0$ matrix
for the corresponding set.
\label{tab:chi2-nlo-as0119-t0}}
\end{center}
\end{table}

In Table~\ref{tab:chi2-nlo-as0119-t0}, we compare the default 
'experimental' definition of the covariance matrix 
used in the paper (cf. Eq.~(\ref{eq:covmat})) 
and the $t_0$ definition of Eq.~(\ref{eq:covmat_t00}). 
In this case, recent LHC measurements for $W$, $Z$, and jet production 
are compared to NLO predictions with five PDF sets and $\alpha_s=0.119$.
Results at NNLO and for other values of the strong coupling
are qualitatively similar. 
One can see that the $t_0$ definition leads to smaller
numerical values of $\chi^2$ for all PDF sets considered, especially
in experiments with sizable normalization contributions,
though it is also clear that the qualitative comparison between
PDF sets in Sect.~\ref{sec:LHCdist} is not affected by this alternative
definition.

Similarly, the experimental definition is compared with 
the extended-$t_0$ definition in the case of ATLAS jet production with 
$R=0.4$ in Table~\ref{tab:ATL7jtR4}. The comparisons are made 
for the NLO PDF sets, $\alpha_s$ values, and computer
codes specified in the table. Three columns of $\chi^2/N_{\rm pt}$ are shown, 
corresponding to the 'experimental' definition 
realized according to Eqs.~(\ref{eq:covmat_cteq}) and~(\ref{eq:betaD}) 
in column 1; and the extended-$t_0$ 
definition based on  Eqs.~(\ref{eq:covmat_cteq}) and~(\ref{eq:betaT0}), 
with the reference
cross sections $T^{(0)}_i$ found using the central 
CT10 NLO in column 2 and NNPDF2.3 NLO PDFs in column 
3.\footnote{The ``exp'' NNPDF2.3 entry with {\tt APPLgrid} in 
this table is numerically equivalent
to the corresponding ``exp'' entry in the 
next-to-the last row of Table~\ref{tab:chi2-nlo-as0119-t0}.} 
In this case, the 
the $\chi^{2}/N_{\rm pt}$ values in columns 2 and 3 are noticeably lower 
than in column 1. They are not exactly the same in columns 2 and 3, 
indicating that $\chi^2$ also depends to some extent 
on the PDF that was used to compute $T^{(0)}$. However this 
difference is much smaller than the difference between results 
using different codes, or different scale choices.

\begin{table}[t]
\small
\begin{center}
\begin{tabular}{c|c|c|c|c|c}
\hline 
NLO PDF &  $\alpha_s$ & Code & \multicolumn{3}{c}{$({\rm cov})_{ij}$ definition}\tabularnewline
\hline 
  & &  & Exp & Ext. $t_0$ & 
Ext. $t_0$ \tabularnewline
 & &  &  & CT10 &NNPDF2.3 \tabularnewline
\hline
\hline 
CT10 & 0.118 & {\tt FastNLO} & 0.95 & 0.55 & 0.60\tabularnewline
\hline 
CT10 & 0.118 & {\tt MEKS1} & 1.00 & 0.57 & 0.61\tabularnewline
\hline 
CT10 & 0.118 & {\tt MEKS2} & 0.89 & 0.55 & 0.59\tabularnewline
\hline 
NNPDF2.3  & 0.119 & {\tt FastNLO} & 0.87 & 0.60  & 0.57\tabularnewline
\hline 
NNPDF2.3 & 0.119 & {\tt MEKS1} & 0.90 & 0.58 & 0.55\tabularnewline
\hline 
NNPDF2.3 & 0.119 & {\tt MEKS2} & 0.78 & 0.54 & 0.53\tabularnewline
\hline 
NNPDF2.3 & 0.119 & {\tt APPLgrid} & 1.00 & 0.64 & 0.62\tabularnewline
\hline
\end{tabular}
\caption{\small The $\chi^{2}/N_{\rm pt}$ values for 
the ATLAS inclusive jet production data obtained with the experimental
 and extended-$t_0$ 
definitions of the $\chi^{2}$ function. The cross sections
are computed at NLO using the specified NLO PDFs, $\alpha_s$ values,
and the following codes: {\tt FastNLO}, {\tt MEKS} with $\mu_{F,R}$ equal
to the individual jet $p_{T}$ ({\tt MEKS1}) or $p_{T}$ of the hardest
jet ({\tt MEKS2}), and {\tt APPLgrid}. 
\label{tab:ATL7jtR4}}
\end{center}
\end{table}

The comparisons of the three covariance matrix definitions 
in the two tables indicate that, for the ATLAS jet data, the difference 
in the corresponding $\chi^2$ values is quite large. Note that in 
this comparison, the $t_0$ covariance matrix treats only the normalization 
of these data as multiplicative, whereas the extended-$t_0$ treats all 
systematic uncertainties as multiplicative. Hence, it is always 
important to know when performing a fit whether a 
correlated error as determined by the experimentalists 
is multiplicative (hence, susceptible to the d'Agostini bias) 
or additive, since this will affect the impact of that data on the fit.

\clearpage

%\bibliography{PDFbench}

\end{document}